\documentclass[letterpaper,onecolumn,12pt,accepted=2020-10-07]{quantumarticle} 
\pdfoutput=1

\usepackage{amsfonts, amsmath, amssymb, amsthm, xcolor}
\usepackage{graphicx}
\usepackage{
hyperref, doi}
\usepackage{xspace}

\newcommand{\hb}[1]{{\color{black} #1}} 

\newcommand{\cc}{{\cal C}}
\newcommand{\Cu}{C^{\ast}} 
\newcommand{\Tr}{\mbox{Tr}}
\newcommand{\tr}{\mbox{tr}}

\newcommand{\R}{\mathbb{R}}

\newcommand{\D}{\mathbb{D}}
\renewcommand{\L}{\mathcal{L}}
\newcommand{\cL}{\mathcal{L}}
\newcommand{\M}{\mathbf{M}}
\newcommand{\N}{\mathbf{N}}
\renewcommand{\O}{\mathbb{O}}
\newcommand{\bbN}{\mathbb{N}}
\newcommand{\C}{\mathbb{C}}
\newcommand{\Q}{\mathbb{H}}
\newcommand{\Cat}{\mathcal{C}} 
\renewcommand{\H}{\mathbb{H}}
\newcommand{\cH}{\mathcal{H}}
\newcommand{\Hilb}{\mathbf{H}}

\newcommand{\Cliff}{\mbox{Cliff}}

\newcommand{\iso}{\simeq}
\newcommand{\St}{\mbox{St}}
\newcommand{\beq}{\begin{equation}}
\newcommand{\eeq}{\end{equation}}

\newcommand{\EJC}{\mbox{\bf EJC}}

\newcommand{\CJP}{\mbox{\bf CJP}}

\newcommand{\RSE}{\mbox{\bf RSE}}

\newcommand{\URUE}{\mbox{\bf URUE}}
\newcommand{\URSE}{\mbox{\bf URSE}}
\newcommand{\FinVec}{\mbox{\bf FinVec}}
\newcommand{\CQM}{\mbox{$\mathbb C$QM}}
\newcommand{\stalg}{\mbox{\bf $\ast$-Alg}}
\newcommand{\EJA}{\mbox{\bf EJA}}

\newcommand{\Cstar}{C^{*}}
\newcommand{\InvQM}{\mbox{\red \bf InvQM}}
\newcommand{\InvCQM}{\mbox{\red \bf InvCQM}}
\newcommand{\IP}{\langle \  | \ \rangle}

\renewcommand{\bar}{\overline}
\renewcommand{\hat}{\widehat}
\newcommand*{\jProd}{\raisebox{-0.88ex}{\scalebox{2.5}{$\cdot$}}}
\renewcommand{\dot}{\jProd}
\newcommand{\hotimes}{\widetilde{\otimes}}
\newcommand{\otilde}{\widetilde{\otimes}}

\newcommand{\id}{\mbox{id}}
\newcommand{\1}{\mathbf{1}}
\newcommand{\0}{\mathbf{0}}
\newcommand{\sa}{{sa}}
\newcommand{\exc}{\mbox{\em ex}}
\newcommand{\ex}{\mbox{\em ex}}
\renewcommand{\sp}{\mbox{\em sp}}

\newcommand{\Aut}{\mbox{Aut}}

\newcommand{\potimes}{\circledcirc}
\newcommand{\tensor}{\otimes}
\newcommand{\intersect}{\cap}

\newcommand{\ket}[1]{|#1\rangle}
\newcommand{\tempout}[1]{{}}
\newcommand{\ffootnote}{\tempout}

\definecolor{darktangerine}{rgb}{1.0, 0.66, 0.07}
\definecolor{kellygreen}{rgb}{0.3, 0.73, 0.09}
\definecolor{indiagreen}{rgb}{0.07, 0.53, 0.03}
\definecolor{darkgreen}{rgb}{0.0,0.5, 0.1}
\definecolor{umbre}{rgb}{.7,.3,.2}
\definecolor{bluegrey}{rgb}{.2,.4,.7}
\newcommand{\hblue}{\color{black}}
\newcommand{\bluegrey}{\color{black}}

\newcommand{\awedit}{\color{black}}

\newcommand{\awcomment}{\color{black}}
\newcommand{\newawe}{\color{black}}

\newcommand{\bblue}{}
\newcommand{\blue}{}
\newcommand{\redd}{}
\newcommand{\red}{}

\newcommand{\magenta}{}
\newcommand{\mmagenta}{\color{black}}

\newtheorem{theorem}{Theorem}[section]

\newtheorem{lemma}[theorem]{Lemma}
\newtheorem{proposition}[theorem]{Proposition}
\newtheorem{corollary}[theorem]{Corollary}

\newtheorem{example}[theorem]{Example}

\newtheorem{definition}[theorem]{Definition}

\title{Composites and Categories of Euclidean Jordan Algebras}
\author{Howard Barnum}
\affiliation{Riemann Center for Geometry and Physics, Institute for Theoretical Physics, Leibniz Universit{\"a}t Hannover}
\affiliation{University of New Mexico}
\affiliation{Currently unaffiliated \texttt{hnbarnum@aol.com}}

\author{Matthew A. Graydon}
\affiliation{Department of Applied Mathematics, University of Waterloo}
\affiliation{Institute for Quantum Computing, University of Waterloo \texttt{m3graydo@uwaterloo.ca}}

\author{Alexander Wilce}
\affiliation{Department of Mathematical Sciences, Susquehanna University \texttt{wilce@susqu.edu}}

\begin{document}

\begin{abstract}\hspace{\parindent}We consider possible non-signaling composites of probabilistic models
based on euclidean Jordan algebras (EJAs), satisfying some reasonable
additional constraints motivated by the desire to construct
dagger-compact categories of such models. We show that no such composite
 has the exceptional Jordan algebra as a direct summand, nor
does any such composite exist if one factor has an exceptional summand, unless
the other factor is a direct sum of one-dimensional Jordan algebras (representing
essentially a classical system). Moreover, we show that any composite of simple,
non-exceptional EJAs is  a direct summand of their
universal tensor product, sharply limiting the possibilities.

These results warrant our focussing on concrete Jordan algebras of
hermitian matrices,  i.e., euclidean Jordan algebras with a preferred
embedding in a complex matrix algebra. We show that these can be organized in a natural
way as  a symmetric monoidal category, albeit one that is not compact closed. 
We then construct a related category $\InvQM$ of embedded euclidean Jordan algebras, having fewer objects but more morphisms, that is not only compact closed but dagger-compact. This 
category unifies finite-dimensional real, complex and quaternionic mixed-state quantum mechanics, except that the composite  of two complex quantum systems comes with an extra classical bit.  

Our notion of composite requires neither tomographic locality, nor
preservation of purity under tensor product.  The categories we
construct include examples in which both of these conditions
fail. {\redd In such cases, the information capacity (the maximum
number of mutually distinguishable states) of a composite is
greater than the product of the capacities of its constituents.}
\end{abstract}
\section{Introduction} 
\label{sec: introduction}

Formally real Jordan algebras were first proposed as models of 
quantum systems by P. Jordan in 1933 \cite{Jordan}. Abstractly, a Jordan algebra is a real vector space $A$ equipped with a commutative bilinear product $\dot$ satisfying the {\em Jordan identity} 
$a^2\dot (a \dot b) = a\dot (a^2 \dot b)$ for all $a,b \in A$ (where $a^2 = a \dot a$). 
We also assume  that $A$ has a unit element, which we denote by $u_{A}$. 
$A$ is {\em formally 
real} if, for $a_1,...,a_n \in A$, 
$\sum_{i=1}^{n} a_{i}^2 = 0$ only when $a_i = 0$ for all $i$. If $A$ is finite-dimensional, this is equivalent to 
$A$'s being {\em euclidean}, meaning that 
it carries an inner product such that 
$\langle a \dot b , c \rangle = \langle a , b \dot c \rangle$ for all $a,b,c \in A$.  {\bluegrey It will be convenient 
in what follows  to compress the phrase {\em euclidean Jordan algebra}, which will occur very frequently, to the acronym EJA.}

 The standard example is the {\awedit space} $\L(\Hilb)$ of 
self-adjoint operators on a finite-dimensional Hilbert space $\Hilb$, with $a \dot b = (ab + ba)/2$, and 
with $\langle a,b \rangle = \Tr(ab)$. 
In 1934, Jordan, von Neumann and Wigner \cite{JNW} showed that 
all finite-dimensional formally real --- equivalently, euclidean ---  Jordan algebras are direct sums 
of irreducible, or {\em simple}, such algebras, and that the latter are of just five kinds:
self-adjoint parts of real, complex or quaternionic matrix algebras
(which we can regard as real, complex or quaternionic quantum systems)
{\em spin factors} (which are analogues of qubits in which the ``Bloch
sphere" can have arbitrary finite dimension), and the {\em exceptional
  Jordan algebra} of $3 \times 3$ self-adjoint octonionic matrices.

A reasonable objection is that the physical meaning of the Jordan
product is obscure. (Indeed, it is not obvious why the observables of
a physical system should carry any physically meaningful bilinear
product at all.) Happily, there are alternative characterizations of
euclidean Jordan algebras in terms of ordered vector spaces and
related concepts having readier physical, probabilistic, or
operational interpretations.
The Koecher-Vinberg Theorem   
(\cite{Koecher, Vinberg}; see also \cite{KoecherBook}, 
\cite{FK}, Chapter III or
\cite{Satake}, Chapter I, \S8) identifies 
euclidean Jordan algebras with finite-dimensional ordered vector spaces 
having homogeneous, self-dual cones; work of Alfsen and Shultz 
\cite{ASPaper, AS} 
characterizes EJAs in terms of certain projections associated 
with closed faces of the positive cone.\footnote{Alfsen and Shultz' 
results apply, more generally, to JB-algebras, which in the context of 
finite dimension are the formally real Jordan algebras.}
 Exploiting these results, 
several recent papers \cite{BMU, Barnum-Hilgert, MU, Wilce09, Wilce11, Wilce12, Wilce18} have shown that physically reasonable postulates 
force a finite-dimensional physical system to have the structure of a euclidean Jordan algebra. To this extent, 
euclidean Jordan algebras are a natural class of models for physical systems.

A physical theory, however, is more than a collection of models of
physical systems. It must also describe how systems change
and how they interact. It is natural, therefore, to represent a physical 
theory as a {\em category}, in which objects represent physical systems and
morphisms represent physical processes.  To accommodate
composite systems, one wants the category to be monoidal, i.e., to be
equipped with an associative ``tensor product''.   This point of view
has been developed very fruitfully in \cite{Abramsky-Coecke,
  BaezCategories} and elsewhere,  where it is shown that
many features of finite-dimensional quantum mechanics can be recovered
if the category in question is compact closed, or, better still,
dagger-compact (terms we explain in Section \ref{sec: Euclidean JC}).

In this paper, building on work of Hanche-Olsen \cite{HO} and Jamjoom
\cite{JamjoomTensorJW} on tensor products of  JC-algebras,
  we classify the possible composites
  of euclidean Jordan-algebraic systems, subject to {\awedit a standard 
``no-signaling" condition and}  a few reasonable {\awedit additional} 
  constraints.  In particular, we show that no such composite exists
if either factor is exceptional, unless the other factor is a direct sum of trivial (\textit{i.e}.\ 1-dimensional) Jordan algebras. Furthermore, we show that a composite  of simple, nontrivial Jordan algebras is always {\red
  a special Jordan algebra, and, indeed, a direct summand of the
  universal tensor product defined by Hanche-Olsen}

Restricting attention further to Jordan algebras corresponding to
real, complex and quaternionic quantum systems {\awedit --- equivalently, self-adjoint parts of real, complex and quaternionic matrix algebras ---} we then identify two
different monoidal sub-categories extending the
category of finite-dimensional complex matrix algebras and CP maps.
One of these, {\mmagenta which we call $\RSE$}, unifies real, complex and quaternionic quantum mechanics,
but lacks certain desirable features. In particular, in 
this category, {\em states} are not represented by morphisms; 
hence, the category is far from being compact closed.
\tempout{\mmagenta which we call $\URUE$, unifies real, complex and quaternionic 
quantum mechanics, albeit excluding the quaternionic bit. 
In this theory, real and 
quaternionic systems compose in the expected way (in particular, composites of quaternionic 
systems are real); however, the composite of two complex quantum systems, splits into a direct 
sum of two copies of the usual composite; put another way, such a composite comes equipped with 
an extra classical bit. This theory also allows certain non-CP maps as processes.}
The other category, which we call $\InvQM$, and which {\em is} compact closed, also embraces real, complex and quaternionic quantum systems and processes (CP maps), 
{\em except} that its rule for composing standard complex quantum systems yields an extra classical
bit.  

These results, combined with the those of (any of) the papers cited
above, in which a euclidean Jordan structure emerges from
information-theoretically, physically or operationally natural
assumptions, {\red lend support to the idea of}
unified quantum theory that
embraces real, complex and quaternionic quantum systems, and
permits the formation of composites of these.  Consistent with the
results of \cite{BarnumWilceLocalTomography}, the composites that
arise in these constructions do not in general have the property of
``tomographic locality'': states on the composite of two Jordan-algebraic 
systems are not, in general, determined by the joint probabilities they 
assign to measurement outcomes associated with the two component systems. 
Equivalently, 
the Jordan algebra $AB$ corresponding to a composite of two formally
real Jordan-algebraic systems $A$ and $B$, will generally be 
larger than the algebraic (i.e., vector-space) tensor product $A \otimes B$.\\[0.3cm]
\noindent{\em Remark:} 
A related proposal is advanced by Baez \cite{BaezDivAlg}, who points out that one can regard real and quaternionic quantum systems as pairs $(\Hilb,J)$, where $\Hilb$ is a complex Hilbert space and $J$ is an anti-unitary satisfying $J^2 = \1$ (the real case) or $J^2 = -\1$ (the quaternionic case). Such pairs can be organized into a dagger-compact symmetric monoidal category, {\awedit taking} morphisms $(\Hilb_1, J_1) \rightarrow (\Hilb_2, J_2)$ {\awedit to be} linear mappings intertwining $J_1$ and $J_2$, and $(\Hilb_1, J_1) \otimes (\Hilb_2, J_2) = (\Hilb_1 \otimes \Hilb_2, J_1 \otimes J_2)$. This provides a unification of real and quaternionic quantum mechanics at the level of pure states and linear mappings between the relevant Hilbert spaces, whereas our approach takes in quantum systems over all three of the  associative division algebras $\R, \C$ and $\Q$, at the level of mixed states, observables and completely positive maps. While the precise connection between Baez' approach and ours is {\awedit not yet entirely clear},  it seems to us likely that an application of Selinger's CPM construction \cite{Selinger} to Baez' category will yield a category of the type we consider here.

Our results rest on a mixture of standard facts about ordered vector spaces, the order structure and the representation theory of euclidean Jordan algebras, and the pioneering work of Hanche-Olsen \cite{HO} on 
universal representations and tensor products of JC-algebras. 
Since much of this  
will be unfamiliar to many readers, we have 
included a 
good amount of purely expository material. 
Section \ref{sec: background probabilistic} provides background on 
order unit spaces and 
their interpretation as general probabilistic
models, including a fairly general notion of composite for such model. This material 
will be more familiar to many readers, but some of our 
notation and terminology, {\redd and some notions specific to our present purposes,} may not be. 
Section \ref{sec:
  background Jordan} collects background material on euclidean Jordan
algebras, their universal representations, and Hanche-Olsen's universal tensor
product.  

The balance of the paper is organized as follows. 
In Section \ref{sec: composites Jordan}, we introduce a general
definition for the composite of two euclidean Jordan-algebraic probabilistic
models, and establish some basic properties of any such
composite. Along the way, we see that the composite of two simple,  nontrivial EJAs
must be embeddable in a complex matrix algebra, i.e. it must
be \emph{special} 
(Theorem \ref{thm: composites special}). 
 From this it follows that no simple,  nontrivial
EJA has any composite with an exceptional EJA (Corollary 
\ref{cor: no composite with exceptional}), and
that if $A$ and $B$ are simple, special EJAs, then any composite of $A$ and $B$
must be an ideal  --- that is, a direct summand --- in their universal tensor product (Theorem \ref{thm: simple composites ideals}). 

These results warrant our focusing on special EJAs. Section 
\ref{sec: Euclidean JC} 
develops a canonical, and naturally associative, tensor product of embedded EJAs, that is, pairs
$(A,\M_A)$ where $\M_A$ is a  finite-dimensional complex $\ast$-algebra and $A$ is a Jordan
subalgebra of the self-adjoint part of $\M_A$. (The universal tensor
product is the special case in which $\M_A$ and $\M_B$ are the 
universal complex enveloping algebras of $A$ and $B$.) 
In Section \ref{sec: categories EJC} we introduce a class of mappings we call
{\em completely Jordan-preserving} and use these to construct 
symmetric monoidal categories of embedded EJAs, some of which we then show are
compact closed {\redd or, indeed, dagger-compact}.  
Section \ref{sec: conclusion} concludes with further discussion of these 
categories and their physical and information-processing significance. 
To avoid obstructing the flow of the main arguments, we 
have removed some technical details to a series of appendices.\\[0.2cm]
\noindent{\bf Acknowledgments} {\red Some of our results have previously been announced, without proof, in 
\cite{BGraW}.\footnote{There, it is erroneously claimed that $\RSE$ is compact closed. This error is corrected 
in Section 6 of the present paper; see especially Examples 6.3 and 6.11} } HB and AW wish to thank C. M. Edwards
for introducing them to the paper \cite{HO} of Hanche-Olsen.  HB and MG 
thank Cozmin Ududec for valuable discussions. AW {\awedit was} 
supported by a grant from the FQXi foundation (FQXi-RFP3-1348). This research was supported in part by Perimeter Institute for Theoretical Physics. Research at Perimeter Institute is supported by the Government of Canada through the Department of Innovation, Science and Economic Development Canada and by the Province of Ontario through the Ministry of Research, Innovation and Science.

\section{Ordered vector spaces and probabilistic models}
\label{sec: background probabilistic}

In this section and the next, we present enough background material to
make this paper reasonably self-contained. This section summarizes
basic information about ordered vector spaces and the 
  ``convex operational" (or ``generalized probabilistic theories")
framework for discussing probabilistic physical theories. 

A good general reference for ordered vector spaces is Chapter 1 of 
\cite{ASbasic} (or the summary
in Appendix \ref{appendix: direct sums} of \cite{AS}).  The use of
ordered vector spaces with 
an order unit as probabilistic models goes back at least to the work of
Ludwig \cite{Ludwig, LudwigAx}; 
see also \cite{Davies-Lewis, Edwards, Holevo}. 
\tempout
{\newawe A survey of more recent literature in this tradition can 
be found in \cite{BarnumWilceFoils}. This has largely been 
concerned with generalizing finite-dimensional quantum theory, 
and accordingly makes use only of finite-dimensional ordered 
vector spaces. This will also be the case in this paper. Hence, in order to avoid constant repetition of 
the qualifier, we adopt the convention that {\em all vector spaces 
are finite dimensional unless otherwise indicated.} All topological 
statements concerning such a space should be understood as referring 
to its unique Hausdorff linear topology.} 
{\newawe The more recent literature in this tradition 
(for a survey of which, see \cite{BarnumWilceFoils}) has focussed 
on finite-dimensional quantum theory, and accordingly makes 
use only of finite-dimensional ordered spaces. This will also be the case 
in this paper. Hence, in order to avoid constant repetition of 
the qualifier, we adopt the convention that {\em all vector spaces 
are finite dimensional unless otherwise indicated.} All topological 
statements concerning such a space\hb{, for example references to the interior of a subset,} should be understood as referring 
to its unique linear topology.}

\subsection{Ordered vector spaces} 

Let $A$ be a real vector space. A (convex, pointed) {\em
  cone} in $A$ is a convex set $K
\subseteq A$ such that $a \in K$ implies $ta \in K$ for all $t \in
\R_+$, and $K \cap -K = \{0\}$. A cone $K$ is \emph{generating}
  iff it spans $A$, i.e., if every $a \in A$ can be expressed as a difference of elements of $K$.  
 Any cone (generating or not) induces
  a partial ordering of $A$, given by $a \leq b$ iff $b - a \in K$;
  this is translation-invariant, i.e, $a \leq b$ implies $a + c \leq b
  + c$ for all $a, b, c \in A$, and homogeneous, i.e., $a \leq b$
  implies $ta \leq tb$ for all $t \in \R_+$. Conversely, such an
  ordering determines a cone, namely $K = \{ a | a \geq 0\}$.
  Accordingly, an {\em ordered vector space} is a real vector space
  $A$ equipped with a designated {\em positive cone} cone $A_+$.  
  It  is common to assume, and we shall assume here, that $A_+$ is closed and generating.


If $A$ and $B$ are ordered vector spaces, a linear mapping $f : A
\rightarrow B$ is {\em positive} iff $f(A_+) \subseteq B_+$. If $f$ is
bijective and $f(A_+) = B_+$, then $f^{-1}(B_+) = A_+$, so that
$f^{-1}$ is also positive. In this case, we say that $f$ is an {\em
  order isomorphism}.  An {\em order automorphism} of $A$ is an order
isomorphism from $A$ to itself.

Denoting the dual space of $A$ by $A^{\ast}$, the {\em dual cone},
$A^{\ast}_+$, is the set of positive linear functionals on $A$. Since
we are assuming that $A_+$ is generating, it is easy to see that
$A^{\ast}_+ \cap -A^{\ast}_{+} = \{0\}$. In our finite-dimensional
setting, $A^{\ast}_{+}$ is also generating.  Note that if
$B$ is another ordered vector space and $\phi : A \rightarrow B$ is a
positive linear mapping, then the dual mapping $\phi^{\ast} : B^{\ast}
\rightarrow A^{\ast}$ is also positive.

\subsection{Order units and probabilistic models}  An {\em order unit} in an ordered vector space $A$ is an element $u \in A_+$ such that, 
for all $a \in A$, $a \leq tu$ for some $t \in \R_+$. In finite
dimensions, this is equivalent to $u$ belonging to the interior 
of $A_+$ (cf. \cite{AT}, Theorem 2.8). An {\em order unit space} is a pair $(A,u)$ where $A$ is an
ordered vector space and $u$ is a designated order unit.

An order unit space provides the machinery to discuss probabilistic
concepts. A {\em state} on $(A,u)$ is a positive linear functional
$\alpha \in A^{\ast}$ with $\alpha(u) = 1$. An {\em effect} is an
element $a \in A_+$ with $a \leq u$.  If $\alpha$ is a state and $a$
is an effect, we have $0 \leq \alpha(a) \leq 1$: we interpret this as
the {\em probability} of the given effect on the given state.
  
A {\em discrete  observable} on $A$ with values $\lambda \in \Lambda$ is
represented by an indexed family $\{a_{\lambda} | \lambda \in
\Lambda\}$ of effects summing to $u$, the effect $a_{\lambda}$
representing the event of obtaining value $\lambda$ in a measurement
of the observable. {\redd Thus, if $\alpha$ is a state, $\lambda \mapsto \alpha(a_{\lambda})$ 
gives a probability weight on $\Lambda$.}
  (One can extend this discussion to include more
general observables by considering effect-valued
measures (cf. \cite{Edwards}), but we will not need this extra
generality here.)

We denote the set of all states of $A$ by $\Omega$; the set of all
effects {\magenta --- the interval between $0$ and $u$ --- is} denoted
$[0,u]$. In our present finite-dimensional setting, both are compact convex sets. 
Extreme points of $\Omega$ are called {\em pure
  states}.\\[0.3cm]
\noindent{\bf Examples} 

(1) {\em Discrete Classical Probability Theory:} If $S$ is a finite
set, regarded as the {\redd outcome} space of some classical experiment, let
$A(S) = \R^{S}$, ordered pointwise.  Elements of $A(S)_+$ are then
non-negative random variables, and effects are random variables with
values between $0$ and $1$. 
We turn this into an order unit space by taking $u \in A(S)_+$ to be the constant
function with value $1$. It is then easy to show that normalized states on
$A(S)$ correspond exactly to probability weights on $S$; discrete
observables correspond in a natural way to discrete ``fuzzy" random
variables. Extreme effects, i.e, extreme points of $[0,u]$, are easily
seen to be characteristic functions of subsets of $S$; hence, an
observable $\{a_{\lambda}\}$ with $a_{\lambda}$ extreme for each
$\lambda$, corresponds to an ordinary ``sharp" random variable.

(2) {\em Discrete Quantum Probability Theory:} If $\Hilb$ is a
finite-dimensional Hilbert space, let $A(\Hilb) = \L(\Hilb)$, the
space of self-adjoint operators on $\Hilb$, ordered by the cone of
positive semidefinite operators; let $u = \1$, the identity operator
on $\Hilb$. Then each normalized state $\alpha$ has the form
$\alpha(a) = \Tr(\rho a)$ where $\rho$ is a density operator, and,
conversely, every density operator determines in this way a normalized
state,.  Observables correspond to discrete POVMs; thus, we recover
orthodox finite-dimensional quantum probability theory.

(3) A class of examples that embraces both (1) and (2) is the
following. Let $\M$ be a unital complex $\ast$-algebra; define $\M_+$ to
consist of all $a \in \M$ with $a = b^{\ast} b$ for some $b \in \M$:
then $(\M_{\sa}, \M_{+})$ is an ordered vector space, in which the
unit element of $\M$ serves as an order unit.  If $\M$ is
finite-dimensional and commutative, one essentially recovers example
(1); if $\M$ is the algebra $M_n(\C)$ of $n \times n$ complex
matrices, one essentially recovers example (2).  More generally, if
$\M$ is finite-dimensional, Wedderburn's theorem tells us that $\M$ is
a direct sum of matrix algebras, so one has finite-dimensional quantum
theory with superselection rules (classical discrete probability
theory being the special case in which all superselection sectors
--- that is, direct summands --- are one-dimensional).\\[0.3cm]
\noindent One {\bluegrey might} wish to privilege certain states and/or certain effects of a
probabilistic model as being ``physically possible".  One way of doing
so is to consider ordered subspaces {\bluegrey $V \leq A^{\ast}$ and $E \leq A$, with $u \in E$:} 
this picks out the set of states $\alpha \in V \cap
A^{\ast}_+$ and the set of effects $a \in E \cap A_+$, $a \leq u$. The pair
$(E,V)$ then serves as a probabilistic model for a system having these
allowed states and effects.  
However, in the three
examples above, and in those that concern us in the rest of this
paper, it is always possible to regard {\em all} states in $A^{\ast}$,
and {\em all} effects in $A$, as allowed. Henceforth, then, when we
speak of a {\em probabilistic model} --- or, more briefly, a {\em
  model} --- we simply mean an order unit space $(A,u)$. It will be
convenient to adopt the shorthand $A$ for such a pair, writing $u_A$
for the order unit where necessary.\\[0.3cm]
\noindent \textbf{Processes, Symmetries and Dynamics} By a {\em process}
affecting a system represented by a probabilistic model $A$, we mean a
positive linear mapping $\phi : A \rightarrow A$, subject to the
condition that $\phi(u_A) \leq u_A$. 
The probability of observing an effect $a$ after the system has been
prepared in a state $\alpha$ and then subjected to a process $\phi$ is
$\alpha(\phi(a))$. One can regard $\alpha(\phi(u))$ as the probability
that the system is not destroyed by the process.  We can, of course,
replace $\phi : A \rightarrow A$ with the adjoint mapping $\phi^{\ast}
: A^{\ast} \rightarrow A^{\ast}$ given by $\phi^{\ast}(\alpha) =
\alpha \circ \phi$, so as to think of a process as a mapping from
states to possibly sub-normalized states. Thus, we can view processes
either as acting on effects (the ``Heisenberg picture"), or on states
(the ``Schr\"{o}dinger picture").

Any non-zero positive linear mapping $\phi : A \rightarrow A$
is a non-negative scalar multiple of a process in the above sense:
since $\Omega(A)$ is compact, {\awedit the function $\alpha \mapsto 
\alpha(\phi(u_A))$ attains a maximum value $m > 0$ on $\Omega(A)$;
$m^{-1}\phi$} is then a
process. For this reason, we make little further distinction here
between processes and positive mappings. In particular, if $\phi$ is
an order automorphism of $A$, then both $\phi$ and $\phi^{-1}$ are
scalar multiples of processes in the above sense: each of these
processes ``undoes" the other, {\em up to normalization}, i.e.,
with nonzero probability.  A process that can be reversed with
probability one is represented by an order-automorphism $\phi$ such
that $\phi(u_A) = u_A$, in which case $\phi^{\ast}$ takes normalized
states to normalized states. Such an order-automorphism is called a
{\em symmetry} of $A$.

We denote the group of all order-automorphisms of $A$ by $\Aut(A)$.\footnote{Here our usage diverges from that of \cite{AS} and
  \cite{FK}, who use $\Aut(A)$ to denote the group of Jordan
  automorphisms of a Jordan algebra $A$.}  This is a Lie group 
{\magenta (see e.g. \cite{HilgertHoffmanLawson}, pp. 182-183);} 
  its connected identity component (consisting of
those processes that can be obtained by continuously deforming the
identity map) is denoted $\Aut_{0}(A)$.  A possible
(probabilistically) reversible {\em dynamics} for a system modelled by
$A$ is a homomorphism $t \mapsto \phi_t$ from $(\R,+)$ to $\Aut(A)$,
i.e., a one-parameter subgroup of $\Aut(A)$.  

One might wish to privilege certain processes as reflecting physically
possible motions or evolutions of the system. In that case, one might
add to the basic data $(A,u)$ a preferred 
{\bblue group $G(A) \leq \Aut(A)$} of order automorphisms.  We refer to such a structure
as a {\em dynamical} probabilistic model, since the choice of $G(A)$
constrains the permitted probabilistically reversible dynamics
of the model.\footnote{A more general definition of a dynamical model would require only that the set of possible evolutions form a semigroup of positive maps. This level of generality will emerge naturally later in this paper, when we consider categories of systems. The definition above is sufficient for our immediate purposes.}  
\interfootnotelinepenalty=200 \\[0.3cm]

{\newawe When $\Aut(A)$ acts transitively on the interior of $A_+$, the cone $A_+$ (or 
the ordered space $A$) is said to be {\em homogeneous}. The positive cone of a euclidean Jordan algebra is always homogeneous, as we will see below. \footnote{Although we do not need this fact, we note that for a homogeneous cone in an ordered 
vector space $A$, the connected identity component $\Aut_o(A)$ of the $\Aut(A)$ also acts transitively on the interior of $A_+$; see e.g. 
\cite{FK}, pp. 5-6.  
}
 }
\\
[0.3cm]
\noindent{\bf Self-Duality}
  An inner product $\IP$ on an
ordered vector space $A$ is {\em positive} iff the associated mapping
$A \rightarrow A^{\ast}$, $a \mapsto \langle a|$, is positive,
i.e,. if $\langle a , b \rangle \geq 0$ for all $a, b \in A_+$. We say that $\IP$ is {\em self-dualizing} if $a \mapsto \langle a |$ maps
$A_+$ {\em onto} $A^{\ast}_+$, so that $a \in A_+$ if and only if
$\langle a , b \rangle \geq 0$ for all $b \in B$.  We say that $A$ (or
its positive cone) is {\em self-dual} if $A$ admits a self-dualizing
inner product.  
 In this case, we can represent states of $A$ {\em internally}: if $\alpha \in
A^{\ast}_+$ with $\alpha(u) = 1$, there is a unique $a \in A_+$ with
$\langle a | b \rangle = \alpha(b)$ for all $b \in A_+$. Conversely,
if $a \in A_+$ with $\langle a , u \rangle = 1$, then $\langle a |$ is
a state.  We will also use the notation $\langle ~ , ~ \rangle$ for an inner product.  

The probabilistic models associated with classical and quantum
systems, as discussed above, are self-dual.
Indeed, in non-relativistic quantum theory, where $A = \L(\Hilb)$, the standard trace inner product 
$\langle a , b \rangle = \Tr(ab)$ is self-dualizing. Here it is {\em usual} to identify states internally, 
i.e., as density operators. 

If $A$ and $B$ are both self-dual and $\phi : A \rightarrow B$ is a positive linear mapping, we can use  
self-dualizing inner products on $A$ and $B$ to represent the mapping $\phi^{\ast} : B^{\ast} \rightarrow A^{\ast}$ as a positive linear mapping $\phi^{\dagger} : B \rightarrow A$, setting $\langle a , \phi^{\dagger}(b) \rangle = \langle \phi(a) , b \rangle$ for all $a \in A$ and $b \in B$. If $\phi : A \rightarrow A$ is an order-automorphism, then so is $\phi^{\dagger}$. 

\subsection{Composites of probabilistic models} If $A$ and $B$ are probabilistic models of two physical systems, one may want to construct a model of  the pair of systems considered together. In quantum mechanics, 
where $A = \L(\Hilb_1)$ and $B = \L(\Hilb_2)$, one would form the
model $AB = \L(\Hilb_1 \otimes \Hilb_2)$ associated with the tensor
product of the two Hilbert spaces. In the framework of general
probabilistic models, there is no such canonical choice for a model of
a composite system. However, one can at least say what one {\em means}
by a composite of two probabilistic models: at a minimum, one should
be able to perform measurements on the two systems separately, and
compare the results. More formally, there should be a mapping $\pi : A
\times B \rightarrow AB$ taking each pair of effects $(a,b) \in A
\times B$ to an effect $\pi(a,b) \in AB$. One would like this to be
       {\em non-signaling}, meaning that the probability of obtaining
       a particular effect on one of the component systems in a state
       $\omega \in \Omega(AB)$ should be independent of what
       observable is measured on the other system. One can show that
       this is equivalent to $\pi$'s being bilinear, {\redd with $\pi(u_A, u_B) = u_{AB}$} 
       \cite{BarnumWilceFoils}.  Finally, one would like to be able to
       prepare $A$ and $B$ separately in arbitrary
       states. Summarizing:

\begin{definition} \label{def: composites} 
{\em 
A (non-signaling) {\em composite} of probabilistic models $A$ and $B$, is a pair $(AB,\pi)$ where 
$AB$ is a probabilistic model, and $\pi : A \times B \rightarrow AB$ is a bilinear mapping such that 
\begin{itemize} 
\item[(a)] $\pi(a,b) \in (AB)_+$ for all $a \in A_+$ and $b \in B_+$; 
\item[(b)] $\pi(u_A, u_B) = u_{AB}$; 
\item[(c)] For every pair of states $\alpha \in \Omega(A)$ and $\beta \in \Omega(B)$, there exists a 
state $\gamma \in \Omega(AB)$ such that for every pair of effects $a \in A$ and $b \in B$,  $\gamma(\pi(a,b)) = \alpha(a)\beta(b)$.\footnote{$\gamma$ is unique if $AB$ is locally tomographic.}
\end{itemize} 
}
\end{definition}
Since $\pi$ is bilinear, it extends uniquely to a linear mapping $A \otimes B \rightarrow AB$. 
{\newawe In what follows, we abuse notation slightly to denote 
this unique extension also by $\pi$, so that, 
for instance, $\pi(a \otimes b) = \pi(a,b)$ for $a \in A, b \in B$.}

\begin{lemma}\label{lemma: pi injective} $\pi$ is injective. \end{lemma}

\noindent{\em Proof:} If $\pi(T) = 0$ for some $T \in A \otimes B$,
then for {\bblue each pair of states $\alpha \in \Omega(A)$, $\beta \in \Omega(B)$, we have by (c) of 
Definition \ref{def: composites}} a state
$\gamma \in \Omega(AB)$ with $(\alpha \otimes \beta)(T) = \gamma(\pi(T)) =
0$. But then $T = 0$ since product states span $(A \otimes B)^*$. $\Box$
\\[0.3cm]
This warrants our treating $A \otimes B$ as a subspace of $AB$ and writing $a \otimes b$ for $\pi(a,b)$. 
Note that if $\omega$ is a state on $AB$, then $\pi^{\ast}(\omega) := \omega \circ \pi$ defines 
a joint probability assignment on effects of $A$ and $B$: 
\[\pi^{\ast}(\omega)(a,b) = \omega(a \otimes b).\]
This gives us marginal states $\omega_A = \omega(u_A \otimes - )$ and $\omega_B = \omega(- \otimes u_B)$. Where 
these are non-zero, we can also define conditional states $\omega_{1|b}(a) := \omega(a \otimes b)/\omega_{B}(b)$ 
and $\omega_{2|a}(b) = \omega(a \otimes b)/\omega_{A}(a)$.\footnote{In the context of a more general definition 
of probabilistic models, in which the cone generated by allowed states might not be the full dual cone 
$A^{\ast}_+$, we would need to modify this definition 
to enforce that these conditional states belong to the allowed state-space. See \cite{BarnumWilceFoils} for 
details.} 

When the mapping $\pi : A \otimes B \rightarrow AB$ is {\awedit also} surjective, we can identify $AB$ with $A \otimes B$. The joint probability assigment $\pi^{\ast}(\omega)$ then completely determines $\omega$, so that states on $AB$ 
{\em are} such joint probability assignments. In this case, we say that $AB$ is {\em locally tomographic}, 
since states of $AB$ can be determined by comparing the results of ``local" measurements, i.e., measurements carried out on $A$ and $B$ alone. In finite dimensions, both classical and {\em complex} quantum-mechanical composites have this feature, while composites of real quantum systems are not locally tomographic \cite{Araki80, Hardy-Wootters}. 

When dealing with dynamical probabilistic models, one needs to
supplement conditions (a), (b) and (c) with the further condition that
it should be possible for $A$ and $B$ to evolve independently within
the composite $AB$. That is:

\begin{definition}\label{def: dynamical composites}
{\em A {\em dynamical composite} of dynamical probabilistic models $A$ and $B$ is a composite $AB$, in the sense of 
Definition \ref{def: composites}, plus a  homomorphism 
\[\otimes : G(A) \times G(B) \rightarrow G(AB)\]
such that 
$(g \otimes h)(a \otimes b) = ga \otimes hb$
 for all $g \in G(A)$, $h \in G(B)$, $a \in A$ and $b \in B$.
}
\end{definition}
(Note that since $AB$ may be larger than the algebraic tensor product $A \otimes B$, the  order automorphism  
$(g \otimes h)$ need not be uniquely 
 determined by the aforementioned condition.)

\subsection{Probabilistic theories as categories} \label{subsec: probabilistic theories as categories}  A physical theory is more than a collection of models. 
At a minimum, one also needs the means to describe interactions between physical systems. 
A natural way of accomplishing this is to treat physical theories as {\em categories}, in which 
objects represent physical systems, and morphisms represent processes. 
In the setting of this paper, then, it's natural to regard a probabilistic {\em theory} as 
a category $\Cat$ in which objects are probabilistic models, i.e., order unit spaces, and in which 
morphisms give rise to positive linear mappings between these.  

The reason for this phrasing --- morphisms {\em giving rise to}, as opposed to simply {\em being},  
positive linear mappings --- is to allow for the possibility that two abstract processes
that behave the same way on effects of their source system, may differ
in other ways---even in detectable ways, such as their effect on
composite systems of which the source and target systems are
components.  If  distinct morphisms between the same two
objects always induce  {\hblue distinct positive maps}, we say the category, and
the set of morphisms, {\em has} \emph{local process tomography}\footnote{One way to make this more precise is to require that $\Cat$ contain $\R$, ordered as usual and with order unit $1$, 
and that $\Cat(I,A)$ {\em be} the cone of positive linear maps $\R \rightarrow A$, so that $\Cat(I,A) \simeq A_+$. Any morphism $\phi \in \Cat(A,B)$ then gives rise to a mapping $\hat{\phi} : \Cat(I,A) \rightarrow \Cat(I,B)$ by $\hat{\phi}(a) = \phi \circ a$ for every $a \in \Cat(I,A)$; this extends to a positive linear mapping $A \rightarrow B$. We shall not pursue this further here; see \cite{BarnumWilce11} for more on these lines. }.
Notice that invertible morphisms $A \rightarrow A$ that preserve the order unit then induce processes in the sense given above, so that every model $A \in \Cat$ carries a distinguished group of reversible processes: models in 
$\Cat$, in other words, are automatically {\em dynamical} models.  

In order to allow for the formation of composite systems, it is
natural to ask that $\Cat$ be a symmetric monoidal category. That is,
we wish to equip $\Cat$ with a bifunctorial product $\otimes : \Cat
\times \Cat \rightarrow \Cat$ that is naturally associative and
commutative, and for which there is a unit object $I$ with $I \otimes
A \simeq A \simeq A \otimes I$ for objects $A \in \Cat$.  Of course,
we want to take $I = \R$. Moreover, for objects $A, B \in \Cat$, we
want $A \otimes B$ to be a composite in the sense of Definitions
\ref{def: composites} and \ref{def: dynamical composites} above. 
In fact, though, every part of those definitions 
simply {\em follows from} the monoidality
    of $\Cat$, except for part (b) of \ref{def: composites}; we must
    add ``by hand" the requirement that $u_A \otimes u_B = u_{A
      \otimes B}$.  The category will also pick out, for each object
    $A$, a preferred group $G(A)$, namely, the group of invertible
    morphisms in $\Cat(A,A)$.  The monoidal structure then picks out,
    for $g \in g(A)$ and $h \in G(B)$, a preferred $g \otimes h \in
    G(AB)$.  {\bblue The bifunctoriality of $\otimes$ guarantees that this will satisfy condition (b) of Definition \ref{def: dynamical composites}; it will also satisfy condition (a) as long as $\cc(I,A) = \cal L_{+}(\R,A) \simeq A_{+}$.}

\section{Background on Euclidean Jordan algebras}  
\label{sec: background Jordan}

In this section, we summarize the essential background information 
on euclidean Jordan algebras and their universal tensor products that 
will be used in the sequel. General references for this material are
the monographs \cite{AS} of Alfsen and Shultz and \cite{FK}
of Faraut and Koranyi and \cite{HO-Stormer} of Hanche-Olsen and
St{\o}rmer, plus the paper \cite{HO} of Hanche-Olsen.

\subsection{Euclidean Jordan algebras} 

As  we have already mentioned, a {\em euclidean Jordan algebra} (hereafter:
EJA) is a finite-dimensional commutative (but not necessarily
associative) real algebra $(A,\dot)$ with a multiplicative unit
element $u$, satisfying the {\em Jordan identity} $a^2 \dot (a \dot
b) = a \dot (a^2 \dot b)$ for all $a,b \in A$, and equipped with an
inner product satisfying $\langle a \dot b , c \rangle = \langle b ,
a \dot c \rangle$ for all $a,b,c \in A$.  Obviously, any commutative,
associative real algebra provides an example, but not a very interesting
one from the algebraic point of view. The basic nonassociative example is the self-adjoint part $M_{sa}$ of a
complex matrix algebra $M$, with $a \dot b = (ab + ba)/2$ and with
$\langle a, b \rangle = \Tr(ab)$.   A \emph{Jordan subalgebra} of a Jordan algebra $A$ is 
a subspace $B$ of $A$ that is closed under inclusion of Jordan products, and hence is a Jordan algebra
with Jordan product given by the restriction of $A$'s.   Any Jordan subalgebra of an EJA is
also an EJA.  Since real and quaternionic matrix algebras have
representations as subalgebras of complex matrix algebras, their
self-adjoint parts are EJAs.  So, too, is the {\em spin factor}
$V_n = \R \times \R^n$, with an inner product given by the usual vector dot product, and with a Jordan product given by
\[(t,x) \dot (s,y) = (ts + \langle x, y \rangle, ty + sx);\]
this can be embedded in $M_{2^{k}}(\C)$ if $n = 2k$ or $2k+1$, as discussed in more detail in Appendix \ref{appendix: spin factors}.

As we shall see in the next section, each EJA has an associated probabilistic model.  
In the case of the EJAs $M_n(\C)_{sa}, M_n(\R)_{sa}$, and $M_n(\Q)_{sa}$, the associated state spaces $\Omega$ are the positive 
semidefinite trace-1 
self-adjoint matrices, which can be viewed as the density matrices associated with $n$-dimensional
complex, real, or quaternionic Hilbert spaces respectively.  
{\newawe In each case, the} 
maximal number of mutually perfectly distinguishable states (a quantity
we call  the \emph{information capacity} of a state space) is $n$.
The state space of $V_n$ is an $n$-dimensional Euclidean ball, which has information capacity $2$, and 
so constitutes a ``generalized bit".  The \emph{associative} EJA of dimension $n$ has a simplex with $n$ vertices as its
state space, so it can be viewed as a classical system of information capacity $n$.
\\[0.3cm]
\noindent{\bf Classification} {\awedit If $A$ and $B$ are EJAs, their vector-space direct sum $A \oplus B$ 
is also an EJA 
under the obvious inner product and slot-wise Jordan product. In this case, $A$ and $B$ can be regarded as 
subalgebras of $A \oplus B$. More than that, they are {\em Jordan ideals}, that is, if $a \in A$ and 
$x \in A \oplus B$, then $a \dot x \in A$ as well, and similarly for $B$.  Conversely, if 
$A$ is any EJA and $A_0 \leq A$ is a Jordan ideal, then so is $A_{1} \hb{ :=} A_0^{\perp} 
\hb{ :=} \{ b \in A | \forall a\in A_0\;a \dot b = 0\}$, and $A \simeq A_0 \oplus A_1$.  Since $A$ is finite-dimensional, one can repeat this 
process so as to decompose $A$ into a direct sum $A = \bigoplus_i A_i$ of finitely many 
irreducible or {\em simple} EJAs $A_i$.  (See Appendix \ref{appendix: direct sums} for further details.)

The EJAs $M_n(\D)_{\sa}$ for $\D = \R, \C, \Q$, and the spin factors $V_n$, discussed above, are 
all simple.\footnote{The $n$-dimensional associative EJA is not simple, rather it is a direct sum of $n$ copies of the one-dimensional EJA.} The {\em Jordan-von Neuman-Wigner Classification Theorem} \cite{JNW}  provides 
a near converse: every simple EJA is isomorphic to one of these types, {\em i.e}.\ isomorphic to a spin factor $V_n$, or}
to the self-adjoint part of a
matrix algebra $M_n({\D})$ where $\mathbb{\D}$ is one of the classical division
 algebras $\R, \C$ or $\Q$, {\em or}, if $n = 3$,  the {\em octonions},
$\O$.  This last example, which is not embeddable into a complex matrix
algebra, is called the {\em exceptional Jordan algebra}, or the {\em
  Albert algebra}. 

A Jordan algebra that {\em is} embeddable in the self-adjoint part
of a complex matrix algebra is said to be {\em special}.   In addition to 
$M_{n}(\C)_{\sa}$, the simple EJAs $M_n(\R)_{\sa}$, $M_n(\Q)_{\sa}$ and $V_n$ are 
all special. It follows
from the classification theorem that any EJA decomposes as a direct
sum $A_{\sp} \oplus A_{\ex}$ where $A_{\sp}$ is special and $A_{\ex}$
is a direct sum of copies of the exceptional Jordan algebra. 
\\[0.3cm]
\noindent{\bf Operator commutation} For each $a \in A$, define $L_a : A \rightarrow A$ to be 
the operation of Jordan multiplication by $a$: $L_a(x) = a \dot x$ for all $x \in A$. Elements 
$a, b \in A$ are said to {\em operator commute} iff $L_a \circ L_b = L_b \circ L_a$. If 
$A$ is a Jordan subalgebra of $M_{\sa}$, where $M$ is a complex $\ast$-algebra, then 
for all $x \in A$, 
\[4 L_a (L_b x)  = a(bx + xb) + (bx + xb)a = abx + axb + bxa + xba\]
and similarly
\[4 L_b(L_a x) = bax + bxa + axb + xab.\]
If $a$ and $b$ operator commute, the left-hand sides are equal. Subtracting, we have 
\[abx + xba - bax - xab = 0\]
or 
\[[a,b]x + x[b,a] = 0\]
which is to say, $[a,b]x - x[a,b] = 0$. If $M$ is unital and $A$ is a unital subalgebra, so that $u_A = 1_{\M}$, 
then setting $x = 1_{M} \in A$ gives us $[a,b] = -[a,b]$, i.e., $a$ and $b$ commute in $M$. 
\\[0.3cm]
\noindent
{\bf Projections and the Spectral Theorem} 
A {\em projection}
in an EJA $A$ is an element $a \in A$ with $a^2 = a$. If $p, q$ are
projections with $p \dot q = 0$, we say that $p$ and $q$ are {\em orthogonal. } 
This implies that $\langle p, q \rangle = \langle p, p \dot q \rangle = 0$.\footnote{ In fact the converse is also true, cf. Ch. II Exercise 3 or Ch. III Exercise 7 in \cite{FK}.}
In this case, $p + q$ is another projection. A projection
not representable as a sum of other projections is said to be {\em
  minimal} or {\em primitive}. A {\em Jordan frame} is a set $E
\subseteq A$ of pairwise orthogonal minimal projections that sum to
the Jordan unit.  The {\em Spectral Theorem} for EJAs (see e.g.
\cite{FK}, Theorem III.1.1, or \cite{AS}, Theorem 2.20 for 
an infinite-dimensional version) asserts that 
every element $a \in A$
can be expanded as a linear combination $a = \sum_{x \in E} t_x x$
where $E$ is some Jordan frame {\red and $t_x$ is a coefficient in $\R$ for each $x \in E$.} 
{\bblue If $a$ has spectral decomposition $a = \sum_{x \in E} t_{x} x$, where $E$ is a Jordan frame, then $a \geq 0$ iff $t_{x} \geq 0$ for all $x \in E$. (The ``if" direction is trivial; for the converse, notice that since $x, a \in A_+$, 
$\langle x, a \rangle \ = \ t_x \|x\|^2 \geq 0$, whence, $t_x \geq 0$.

{\newawe By summing over those $x \in E$ for which $t_x$ has a given value, 
we obtain a decomposition $\sum_{i} t_i p_i$ 
where $p_i$ are pairwise Jordan-orthogonal projections and the coefficients 
$t_i$ are distinct. In this form, the spectral decomposition is unique \cite{FK}. This gives us a functional calculus, as we can now define $f(a) = \sum_i f(t_i)p_i$ 
for any real-valued function $f$ defined on the set of coefficients $t_i$. 
In particular, every $a \in A_+$ has a square root 
$\sqrt{a} = \sum_{i} \lambda_{i}^{1/2} p_i$ also in $A_+$. }

{\bblue Decomposing $A$ as an orthogonal direct sum $A = \bigoplus_i A_i$ of simple Jordan ideals $A_i$, 
the Jordan unit $u_A$ is the sum 
of the Jordan units $u_{A_{i}}$ of these ideals. It follows that each Jordan frame $E$ of $A$ is the disjoint union of 
Jordan frames $E_i$ belonging to the various simple ideals $A_i$.  If $A$ is simple, the group of Jordan automorphisms acts transitively on the set of
Jordan frames (\cite{FK}, Theorem IV.2.5). It follows that} all
Jordan frames for a given euclidean Jordan algebra $A$ have the same
number of elements.  This number is called the {\em rank} of $A$. By
the Classification Theorem, all simple Jordan algebras having rank $4$
or higher are special.

\subsection[Euclidean Jordan algebras as probabilistic models]{Euclidean Jordan algebras as {\awedit probabilistic models}} 

\tempout{Any euclidean Jordan algebra $A$ can be regarded as an ordered real
vector space, with positive cone $A_+ = \{ a^2 | a \in A\}$. (That this 
{\em is} a cone is a non-trivial fact; see \cite{FK}, Theorem III.2.1, or \cite{ASbasic}, pp. 36-28.) By the
spectral theorem, $a = b^2$ for some $b \in A$ iff $a$ has a spectral
decomposition $a = \sum_{i} \lambda_i x_i$ in which all the
coefficients $\lambda_i$ are non-negative. }

{\newawe As remarked earlier, any} euclidean Jordan algebra $A$ can be regarded as an ordered real
vector space, with positive cone $A_+ = \{ a^2 | a \in A\}$. (That this 
{\em is} a cone is a non-trivial fact (see \cite{FK}, Theorem III.2.1, or \cite{ASbasic}, pp. 36-28).) By the
spectral theorem, $a = b^2$ for some $b \in A$ iff $a$ has a spectral
decomposition $a = \sum_{i} \lambda_i x_i$ in which all the
coefficients $\lambda_i$ are non-negative. {\newawe It can also be shown 
(see \cite{FK}, Proposition I.1.4) that $a$ belongs to the interior 
of $A_+$ iff $\langle a, b \rangle > 0$ for all nonzero $b \in A_+$. 
Using this, it follows that $a$ belongs to the interior of $A_+$ iff it has 
a spectral decomposition $a = \sum_{x \in E} t_{x} x$ with 
all coefficients $t_{x}$ strictly positive. Hence, if $a$ belongs to 
the interior of $A_+$, so does $\sqrt{a}$. }

The Jordan unit $u$ is also an order unit. Thus, any EJA can serve as a probabilistic 
model, as defined in Section 2:  physical states correspond to states {\em qua} normalized positive linear
functionals on $A$, while 
measurement outcomes are represented by \emph{effects}, i.e,
elements $a \in A_+$ with $0 \leq a \leq u$, and (discrete) observables, 
by sets $\{e_i\}$ of events with $\sum_i e_i = u$.  {\awedit Note that if 
$A = \bigoplus_{i=1}^{n} A_i$ where each $A_i$ is a copy of the one-dimensional Jordan algebra $\R$, 
then $A \simeq \R^{n}$, regarded as a commutative algebra (in particular, $a \dot b = ab$). As a probabilistic system, this is {\em classical}, in the sense that it is simply the space of random variables on a finite sample space. From now on, when we speak of a classical system, this is 
what we have in mind.}

 As discussed earlier, the inner product on $A$ allows us to represent states internally, i.e., 
for every state $\alpha$ there exists a unique $a \in A_+$ with $\alpha(x) = \langle a | x \rangle$ 
for all $x \in A$; conversely, every vector $a \in A_{+}$ with $\langle a | u \rangle = 1$ defines 
a state in this way. Now, if $a$ is a projection, i.e., $a^2 = a$, let $\hat{a} = \|a\|^{-2} a$: 
then 
\[\langle \hat{a} | u_{A} \rangle = \frac{1}{\|a\|^2} \langle a | u_{A} \rangle = 
\frac{1}{\|a\|^2} \langle a^2 | u_{A} \rangle =
\frac{1}{\|a\|^2} \langle a | a \rangle = 1.\]
Thus, $\hat{a}$ represents a state. A similar computation shows that $\langle \hat{a} | a \rangle = 1$. 
Thus, every projection, regarded as an effect, has probability $1$ in some state.  

{\bblue 
\begin{lemma}\label{lemma: projections}
Let $A$ be an EJA, and let $a$ be an effect, i.e. $a \in [0,u_A] \subseteq A_+$.
Then $a$ is a projection iff $\langle u_A, a \rangle = \langle a, a \rangle$. 
\end{lemma} 

\noindent{\em Proof:} If $a$ is a projection, then 
$\langle a, a \rangle = \langle u_{A} \dot a, a \rangle = \langle u_A, a \dot a \rangle = \langle u_A, a \rangle$. Conversely, 
suppose $\langle u_A, a \rangle = \langle a, a \rangle$. Let $a$ have spectral decomposition $a = \sum_{x \in E} t_{x} x$ where $E$ is a Jordan frame. We then have 
$\langle u_A, a \rangle = \sum_{x \in E} t_{x} \|x\|^2$, while $\langle a, a \rangle = \sum_{x \in E} t_{x}^{2}\|x\|^2$.  
Since $a$ is an effect, $0 \leq t_x \leq 1$ for every $x \in E$, so that $t_{x}^{2}\|x\|^2 < t_{x}\|x\|^2$ unless $t_{x} = 0$ or $t_{x} = 1$.  In order for the two sums
 above 
to be equal, therefore, we must have  
that $t_x = 0$ or $t_x = 1$ for every $x$. 
Setting $B = \{ x \in E | t_{x} > 0\}$, we have $a = \sum_{x \in B} x$, a projection.} $\Box$\\[0.3cm]

When $A$ is special, i.e., a Jordan subalgebra of a matrix algebra, its order structure is inherited from that of the latter.  

\begin{proposition}\label{prop: embedded order} Let $A \leq \M_{\sa}$, i.e., $A$ is a Jordan subalgebra of a finite-dimensional complex matrix algebra $\M$. 
Then $A_{+} = A \cap \M_{+}$.  
\end{proposition} 

\noindent{\em Proof:}  $A_{+} \subseteq \M_{+}$ because squares in $A$ are squares in $\M_+$. For the converse, 
let $a \in A \cap \M_{+}$. By the spectral theorem for EJAs, we can express $a$ as a sum 
$a = \sum_{i} \lambda_i e_i$ where the $e_i$ are pairwise Jordan-orthogonal idempotents, i.e, $e_i \dot e_j = 0$ 
for $i \not = j$. Jordan-orthogonal idempotents in $A$ are again Jordan-orthogonal idempotents in $\M_{\sa}$. Since 
Jordan-orthogonal idempotents in $\M_{\sa}$ are orthogonal in the usual sense and $a \in \M_+$, it follows that the coefficients $\lambda_i$ are all non-negative, whence, $a \in A_+$.
 $\Box$ \\

 \noindent{\bf  Order-automorphisms} The order structure of an EJA $A$, together with its inner product and order unit, entirely determines its Jordan structure, as a consequence of the Koecher-Vinberg theorem \cite{Koecher, Vinberg} (discussed in more detail below).  One manifestation of this is that 
a \emph{symmetry} of $A$ --- that is, an order-automorphism preserving the unit $u_A$ --- is the same thing as a 
Jordan automorphism (\cite{AS}, Theorem 2.80).

Another class of order automorphisms is given by the {\em quadratic representations} of certain elements of $A$. 
The quadratic representation of $a \in A$ is 
the mapping $U_a : A \rightarrow A$ given by 
\[U_a = 2L_{a}^{2} - L_{a^2}\]
i.e., $U_a(x) = 2a \dot (a \dot x) - (a \dot a) \dot x $. 
These mappings have direct physical interpretations as {\em filters} in the sense of \cite{Wilce12}. 
The following non-trivial facts will be used repeatedly in what follows:

\begin{proposition}\label{prop: quadratic representation} Let $a \in A$. Then 
\begin{itemize} 
\item[(a)] $U_a$ is a positive mapping; 
\item[(b)] If $a$ lies in the interior of $A_+$, $U_a$ is invertible,  with 
inverse given by $U_{a^{-1}}$;
\item[(c)] $e^{L_{a}} = U_{e^{a/2}}$
\end{itemize} 
\end{proposition}

\noindent{\em Proof:} For (a), 
see  Theorem 1.25 of \cite{AS}; for (b), \cite{AS} Lemma 1.23 or \cite{FK}, Proposition II.3.1. 
Part (c) is Proposition II.3.4 in \cite{FK}.  $\Box$\\[0.3cm]
\indent Combining (a) and (b), $U_a$ is an order automorphism for every $a$ in the interior of $A_+$. 
Regarding (c), note that $e^{L_a}$ is the ordinary operator exponential.  Therefore if $a$ is in 
the interior of $A_+$, then 
$\phi_{t} := e^{tL_{a}} = U_{\frac{t}{2}a}$ is a one-parameter group of order-automorphisms 
with $\phi'(0) = L_{a}$.   Therefore, for all $a$ in the interior of $A_+$, $U_a$ is in the connected component {\newawe $\Aut_{0}(A)$} 
of the identity in the group {\newawe $\Aut(A)$} of 
order-automorphisms of $A$.  

{\newawe Notice that $U_a(u_A) = 2a^2 - a^2 = a^2$} {\newawe If 
$a$ belongs to the interior of $A_+$, then so does $\sqrt{a}$; thus,  
by the remarks above, $U_{\sqrt{a}} \in \Aut_{0}(A)$ and $U_{\sqrt{a}}(u_A) = a$. 
It follows that $\Aut_{0}(A)$, and hence also the full group $\Aut(A)$ of order-automorphisms of $A$, act transitively on the interior of $A_+$: if $a, b$ belong to the interior of $A_+$, then $U_{\sqrt{b}} \circ U_{\sqrt{a}}^{-1}$ maps $a$ to $b$. In other words, $A_+$ is homogeneous. }\\[.03cm]

\tempout{Since the square-root of an element of the interior of $A_+$ is also 
in the interior of $A_+$, it follows that 
the map $a \mapsto a^2$ on the interior of $A_+$ is surjective,
 it follows that {\newawe $\Aut_{0}(A)$}, 
 and hence the full group {\newawe $\Aut(A)$}
of order-automorphisms of $A$,  
acts  transitively on
 the interior of $A_+$.   {\newawe In other words, $A_+$ is homogeneous.}
 \\[.03cm]
 }
 
{\bluegrey
\noindent{\bf EJAs as dynamical models}  {\newawe Henceforth, we will write $G(A)$ for the identity component 
$\Aut_{0}(A)$ of an EJA $A$. }
As {\newawe this} notation 
suggests, we henceforth 
regard an EJA $A$ as a {\em dynamical} probabilistic model with $G(A)$ as its  dynamical group. }
{\hblue 
This is 
a reasonable choice. First, elements of $G(A)$ are exactly those automorphisms of $A_+$ that figure 
in the system's possible dynamics, as elements of 
one-parameter groups of automorphisms. 
This suggests that the ``physical" dynamical group of a dynamical 
model based on $A$ should 
at least be a subgroup of $G(A)$, so that the latter is the least constrained choice. 
\ffootnote{\blue Though we might want to think about whether we actually {\em need} all of $G(A)$...} 
Moreover, $G(A)$, like the full group of order-automorphisms, acts transitively on the interior of the 
cone $A_+$, and its unit-preserving subgroup acts transitively on the set of {\em Jordan frames}, i.e., maximally informative sharp observables --- or, equivalently, on maximal lists $(\alpha_1,...,\alpha_n)$ of sharply-distinguishable  states. These transitivity properties, abstracted from the 
Jordan-algebraic setting, were among the postulates used (in somewhat different ways) in  
\cite{Wilce09, Wilce11} and \cite{BMU} to derive the Jordan structure of probabilistic models, so it 
is not unreasonable to require that the dynamical group enjoy them.
}

 If $\phi$ is any order-automorphism with $\phi(u_A) = a^2 \in A_+$, then $U_{a}^{-1} \circ \phi$ is a symmetry of $A$. Hence, every order-automorphism of $A$ decomposes as $\phi = U_{a} \circ g$ where $g$ is a symmetry. 
As we observed earlier, $a$ can be chosen to belong to the interior of $A_+$; it can be shown that with this 
 choice, the decomposition is unique (\cite{FK}, III.5.1)\\[0.3cm]
\noindent{\bf The {\awedit Koecher-Vinberg} Theorem} {\newawe As we have seen, 
the positive cone $A_+$ of an EJA is homogeneous, and also self-dual with respect to $A$'s inner product. }
\tempout{Recall that this 
means that an element $a \in A$ belongs to the cone $A_+$ iff $\langle a | b \rangle \geq 0$ for all elements $b$ of $A_+$ --- equivalently, iff $\langle a |$ belongs to the dual cone $A_{+}^{\ast}$.  
Thus, as an ordered vector space, $A$ is both self-dual and homogeneous. }
Conversely, {\awedit let $A$ be an homogenous  finite-dimensional ordered vector space with a self-dualizing inner product. Choosing any 
order-unit $u$ invariant under positive orthogonal transformations,  there exists a unique  bilinear 
product on $A$ making $A$, with the given inner product, a euclidean Jordan algebra, and the chosen  element $u$, the Jordan unit. This is the content of the {\em Koecher-Vinberg Theorem} (\cite{Koecher, Vinberg}; see also 
\cite{FK, Satake}). While we make no use of this result here, it is at the center of efforts to provide an operational motivation for euclidean Jordan algebras as models of physical systems, 
e.g., in \cite{Wilce12, BMU}.

\subsection{Representations of EJAs}

\label{subsec: reps of EJAs}

A {\em representation} of a Jordan algebra $A$ is a Jordan
homomorphism $\pi : A \rightarrow M_n(\C)_{\sa}$ for some $n$.\footnote{This is a finite dimensional, concrete representation.  For the finite-dimensional algebras we are concerned with, 
this definition suffices.}
\ffootnote{\blue AW: Do we want to allow range to be an arbitrary f.d. complex matrix algebra?}  
So a Jordan algebra is special iff it has an injective,  or {\em faithful}, representation. 
Recall that every EJA decomposes 
as a direct sum $A = A_{\sp}
\oplus A_{\exc}$, where $A_{\sp}$ is special and $A_{\exc}$ has no
nontrivial 
representations.  
The latter, in turn, is a direct sum of copies of the exceptional EJA $M_3(\O)_{\sa}$. {\redd See \cite{AS}, Theorem 4.3 for details.}\\[0.3cm]
\noindent {\bf Standard Representations} For the non-exceptional
simple EJAs, it will be useful to record what we will call their {\em
  standard} representations.
It will also prove helpful to adopt the following abbreviations: 
\[R_n = M_n(\R)_{\sa}; \ C_n = M_n(\C)_{\sa}, \ Q_n = M_n(\H)_{\sa}.\]
 As above, we write $V_n$ for the spin factor $\R \times \R^{n}$.  With this notation 
we have obvious embeddings $R_n \leq C_n = M_n(\C)_{\sa}$. For $Q_n$, note that a quaternion 
$a + bi + cj + dk$ can be written as $(a + bi) + (c + di)j$, and so, can be represented by 
the pair of complex numbers $(a + bi, c + di)$. 
Thus,  any $n \times n$ matrix of quaternions can be represented 
as a $2n \times 2n$ complex matrix having the form 
\begin{equation}
	\begin{pmatrix}
		 \hspace{0.23cm}\Gamma_1 & \Gamma_2 \\[0.02cm] -\bar{\Gamma}_2 & \bar{\Gamma}_1
	\end{pmatrix}\text{,}
\end{equation}
where the blocks $\Gamma_1$ and $\Gamma_2$ are $n \times n$ complex matrices. 
This gives us a faithful representation of $Q_n$ in $M_{2n}(\C)$, {\redd known as the {\em symplectic representation} 
\cite{Graydon11}.} 
There is also 
what we will call a {\em standard representation}
 of $V_n$ in $M_{2^k}(\C)$, where $n = 2k$ or $2k + 1$.  This is less obvious; the
details are given in Appendix \ref{appendix: spin factors}.\\[0.3cm]
\noindent{\bf Involutions and Reversibility} 
By an {\em involution} on a complex $\ast$-algebra $M$, we mean 
a  $\ast$-anti-automorphism of $M$ of period $2$ --- in more detail, a linear mapping 
$\Phi : M \rightarrow M$ such that $\Phi(a^{\ast}) = \Phi(a)^{\ast}$, $\Phi(ab) = \Phi(b)\Phi(a)$, and 
$\Phi(\Phi(a)) = a$ for all $a \in M$. A {\em real} involution is defined similarly, except that the 
mapping is only required to be real-linear.\footnote{Our usage is slightly nonstandard here: an involution 
on a complex $\ast$-algebra is more frequently defined to be a conjugate-linear. The only involution in this sense 
that will concern us is $a \mapsto a^{\ast}$.} It is straightforward that the set 
$M^{\Phi}_{\sa} = \{ a \in M | a = a^{\ast} = \Phi(a)\}$ of all self-adjoint fixed-points of $M$ under $\Phi$ is a Jordan subalgebra of $M_{\sa}$. 
Indeed, if $a, b \in M^{\phi}_{\sa}$, then $a \dot b = \tfrac{1}{2}(ab + ba) \in M_{\sa}$, and 
\[ \Phi(ab + ba) = \Phi(b)\Phi(a) + \Phi(a)\Phi(b) = ba + ab = ab + ba\]
so that $a \dot b \in M^{\Phi}_{\sa}$ as well. In fact, more is true: if $a_1,...,a_n \in M^{\Phi}_{\sa}$, then 
\[\Phi(a_1 \cdots a_n + a_n \cdots a_1) = a_n \cdots a_1 + a_1 \cdots a_n \]
so that $a_1 \cdots a_n + a_n \cdots a_1 \in M^{\Phi}_{\sa}$. 

\begin{definition}\label{def: reversible}{\em A Jordan subalgebra $A$ of $M_{\sa}$  is said to be {\em reversible}
{\hblue  if 
$a_1,...,a_n \in A \implies a_1 \cdots a_n + a_n \cdots a_1 \in A$.}  An 
abstract EJA $A$ is reversible iff it has a faithful (that is, injective) representation as a reversible Jordan subalgebra 
of some $\ast$-algebra $M$.  If {\em all} of $A$'s faithful representations are reversible, then $A$ is said to 
be {\em universally reversible} (hereafter: UR).}
\footnote{\hblue Some authors, for example Hanche-Olsen in \cite{HO}, define universal reversibility as reversibility in all 
representations, not just all faithful representations.  The two definitions are equivalent, as will become apparent below. }
\end{definition}

In this language, then, all Jordan algebras of the form $M^{\Phi}_{\sa}$ are reversible.
All self-adjoint parts of real and complex matrix algebras are universally reversible, as are
the quaternionic ones of rank 3 and higher {\redd (whence, all EJAs of rank $\geq 4$ are UR)}.  The quaternionic bit
$M_2(\Q)_{\sa}$, which is isomorphic to $V_5$, is not universally reversible, but 
it is reversible, since its
standard embedding into $M_4(\C)$ is reversible: indeed, it's the set fixed points 
of the involution  $\Phi(x) = (\sigma_{y}\otimes\1_{2})x^{T}(\sigma_{y}\otimes\1_{2})$ where $\sigma_{y}=-\overline{\sigma_{y}}$ is the usual Pauli $y$-matrix. Spin factors $V_n$ with $n=4$ or $n
\geq 6$ are not reversible at all. For details, see \cite{HO}. Thus, 
the reversible {\em simple} EJAs are  just those of the forms $M_n(\R)_{\sa}$, $M_{n}(\C)_{\sa}$ and $M_{n}(\Q)_{\sa}$.\\[0.3cm]
{\noindent{\bf The universal representation}}  In addition to the standard representations discussed above, 
every special EJA has a {\em universal} representation. 

\begin{definition}
{\em 
A \emph{universal $\Cu$ algebra} for a euclidean Jordan algebra $A$
is a complex $*$-algebra 
$\Cu(A)$, plus a Jordan {\newawe homomorphism} 
$\psi_A : A \rightarrow \Cu(A)_{\sa}$, 
such that for any $C^{\ast}$-algebra $\M$ and any Jordan 
homomorphism $\phi : A \rightarrow \M_{\sa}$, there exists a unique 
$\ast$-homomorphism $\hat{\phi} : \Cu(A) \rightarrow \M$ with 
$\phi = \hat{\phi} \circ \psi_A$. 
}
\end{definition}
 
{\newawe Note that the uniqueneness of the $\ast$-homomorphism $\hat{\phi} : \Cu(A) \rightarrow \M$ is equivalent to $\Cu(A)$ being generated by $\psi_{A}(A)$ as a $C^{\ast}$ algebra, which is how the definition is more usually presented.}
For the existence of $\Cu(A)$, see \cite{AS} or \cite{HO-Stormer}.  
Universal $C^*$-algebras for a given EJA $A$ are 
unique up to {\redd canonical $\ast$}-isomorphism, 
warranting our speaking of 
``the'' universal $C^*$-algebra of $A$.\footnote{In fact, by privileging 
any particular construction of $\Cu(A)$, we can take this literally.}
It is easy to see that any Jordan homorphism $A \rightarrow B$ lifts, via
  $\psi_A$ and $\psi_B$, to a $C^*$-homomorphism $\Cu(A) \rightarrow
  \Cu(B)$, and that this lifting respects composition of Jordan homomorphisms, 
  so that $\Cu(\ \cdot \ )$ defines a functor from the category of EJAs to the 
  category of  complex $\ast$-algebras.  It can be shown that this 
  functor is exact (\cite{HO}, Theorem 4.1). 
 It is an important fact that $A$ is exceptional iff $\Cu(A) = \{0\}$. 
  Otherwise, the representation $\psi_A$ of $A$ in $\Cu(A)$ is faithful, 
  and takes the unit of $A$ to the unit of $\Cu(A)$.  In
  this case, we will often identify $A$ with its image $\psi_{A}(A)$ in
  $C^*(A)$, referring to this (and to the homomorphism $\psi_{A}$) as 
  the {\em universal embedding} of $A$.

\bluegrey We can now see that reversibility in all faithful representations implies reversibility in all representations. 
 It is straightforward that any representation factoring through a reversible representation is also 
reversible. If $A$ is universally reversible in the sense of Definition \ref{def: reversible}, then the universal 
representation, which is faithful, is also reversible. Since every representation factors through this one, every 
representation of $A$ is reversible.
\\
[0.01cm]

\noindent
 {\hblue {\bf The canonical involution}}
  If $A$ is special, $\Cu(A)$ comes equipped with a unique
    involution that fixes every point of $A$.  To see
    this, note that that the opposite algebra $\Cu(A)^{\text{op}}$ (the same
    vector space, equipped with reversed multiplication) is equally a
    universal $C^{\ast}$-algebra for $A$. Hence, there is a unique
    $\ast$-isomorphism $\Phi_{A} : \Cu(A)^{\text{op}} \rightarrow \Cu(A)$
    fixing all points of $A$. We can equally well regard $\Phi$ as a
    $\ast$-antiautomorphism of $\Cu(A)$; so regarded, $\Phi$ is
    self-inverse, {\magenta hence} an involution.

\begin{definition}\label{def: canonical involution}
We call the involution $\Phi: C^*(A) \rightarrow C^*(A)$ just described, the 
\emph{canonical involution}, $\Phi_A$, on $C^*(A)$.
\end{definition}

As discussed above, the self-adjoint fixed-points of an involution $\Phi$ on a complex $\ast$-algebra $M$    
  constitute a 
  Jordan subalgebra, $M^{\Phi}_{\sa}$, of $M_{\sa}$. With $\Phi_{A}$ the canonical 
  involution on $\Cu(A)_{\sa}$, we have a Jordan embedding $A \leq \Cu(A)^{\Phi}_{\sa}$. 
  
\begin{proposition}[\cite{HO}, Lemma 4.2 and Theorem 4.4] 
\label{prop: UR fixed points}
With notation as above, $A$ is UR iff $A = \Cu(A)^{\Phi}_{\sa}$. 
  More generally, {\redd if $A$ is UR and} there exist an embedding $A \leq
  M_{\sa}$ for a complex $\ast$-algebra $M$, and an involution $\Phi$
  on $M$ fixing points of $A$, then the $\ast$-subalgebra of
  $M$ generated by $A$ is isomorphic to $\Cu(A)$ and $\Phi$,
  restricted to this subalgebra, is the canonical involution.
\end{proposition}

This characterization allows the explicit computation of
  universal $C^{\ast}$-algebras of UR EJAs \cite{HO}. For
    instance, $R_n = M_n(\R)_{\sa}$ generates $M_n(\C)_{\sa}$ as a
    (complex) $\ast$-algebra, in which it sits as the set of symmetric
    self-adjoint matrices: in other words, the set of fixed points of
    the involution $\Phi(a) = a^{T}$. Thus, $\Cu(R_n) =
    M_n(\C)$. Similarly, the image of $M_n(\Q)_{\sa}$ under the
    standard (symplectic) embedding into $M_{2n}(\C) =
    M_2(M_n(\C))$ is fixed by the involution $\Phi(a) = -J a^T J$
    where $J$ is the block matrix 
		 \begin{equation}
			J=\begin{pmatrix}
				\hspace{0.23cm}0 & \1\\ 
				-\1 & 0
			\end{pmatrix}
			\label{bigJ}
		\end{equation}
		with $\1$ the $n \times n$ identity matrix. For a third example, consider the
    embedding $M_n(\C)_{\sa} \rightarrow (M_n(\C) \oplus
    M_n(\C))_{\sa}$ given by $\psi(a) = (a,a^T)$: the image of this
    embedding generates $M_n(\C) \oplus M_n(\C)$ as a $\ast$-algebra,
    and is exactly the set of fixed-points of the involution
    $\Phi(a,b) = (b^T, a^T)$. Thus, $\Cu(C_n) = M_n(\C) \oplus
    M_n(\C)$.  Note, in passing, that this also
    shows that there is no involution on $M_n(\C)$ fixing points of
    $C_n = M_n(\C)_{\sa}$. {\redd (Otherwise, $M_n(\C)$ would be the universal 
    $\ast$-algebra for $C_n$.)}  

The universal $C^{\ast}$-algebras for all simple, non-exceptional EJAs 
are are summarized in Table 1 below, {\redd along with the canonical involutions in the UR cases.} 
{\redd For a spin factor $V_n$, the universal $C^{\ast}$-algebra is the complex Clifford 
algebra $\Cliff_{\C}(n)$ on $n$ generators \cite{HO}; these are tabulated separately in Table 1(b).
 }    

For contrast, Table 
1(c) lists the complex matrix algebras into which $R_n, C_n$ and $Q_n$ are
{\em standardly} embedded.  We also list in Fig. 1 (c) algebras
supporting what we are calling standard embeddings of 
the spin factors $V_n$; these agree with the
universal ones for $n = 2k$, but for $n=2k+1$, embed $V_n$ in 
$M_{2^k}(\C)$ rather than the $M_{2^k}(\C) \oplus M_{2^k}(\C)$ of the
universal embedding.  Thus, all the targets of embeddings in Fig. 
1(c) are \emph{simple} complex matrix algebras.  
\[
\begin{array}{ccc}
\begin{array}{l|l|l} 
A  & \Cu(A) & \Phi  \\ 
\hline
R_n & M_n(\C) & a \mapsto a^T\\
C_n & M_n(\C) \oplus M_n(\C) & (a,b)  \mapsto (b^T,a^T) \\ 
Q_n & M_{2n}(\C) \ \mbox{if} \ n > 2 & a \mapsto -J a^T J  \\
V_n & \Cliff_{\C}(n) & 
 \\
\vspace{.05in} & 
\end{array}
& & 
\begin{array}{l|l}
n ~(k \in \bbN_+ )  & { \Cliff_{\C}(n)}  \\
\hline
2k & M_{2^k}(\C) \\
2k+1 & M_{2^k}(\C) \oplus M_{2^k}(\C)\\
\vspace{.4in}
\end{array}\\
\mbox{(a)} & & \mbox{(b)} \\
 \mbox{ Universal embeddings } & & \mbox{ Clifford algebras}
\end{array}
\]
\[
\begin{array}{c}
\begin{array}{l|l} 
A  & \M \\ 
\hline
R_n & M_n(\C) \\
C_n & M_n(\C) \\ 
Q_n & M_{2n}(\C)   \\
V_{2k} & M_{2^k}(\C) \\
V_{2k+1} & M_{2^k}(\C)\\
\end{array} \\
\mbox{(c)}\\
\mbox{ Standard embeddings}\\
\\
\mbox{{\bf Table 1:} Universal and standard embeddings} 
\end{array}\]
\vspace{.3in}

Note that the spin factors $V_2, V_3, V_5$ correspond to the three
types of quantum bits: $V_2 \simeq R_2$; $V_2 \simeq C_2$ and $V_5 \simeq Q_2$. 
This last, together with line 2 of Fig. 1 table (b), gives us the missing item in table (a):
\[\Cu(Q_2) \simeq M_4(\C) \oplus M_4(\C).\]

It will be helpful to record here two further facts about
  universal $C^{\ast}$ algebras.  First, $\Cu(A \oplus B) = \Cu(A)
  \oplus \Cu(B)$. This follows from the exactness of the $\Cu(
  \ )$-functor.  Combining this with Proposition 
\ref{prop: UR fixed points}, it follows that
  if $A$ and $B$ are both UR, so is $A \oplus B$. Details can be found
  in Appendix \ref{appendix: direct sums}.

\subsection{The universal tensor product} 

The universal representation allows one to define a natural tensor product of EJAs,  first studied 
by H. Hanche-Olsen \cite{HO}:

\begin{definition}
{\em 
The {\em universal} tensor
product of two EJAs $A$ and $B$, denoted $A \hotimes B$,  
is the Jordan subalgebra of
$\Cu(A) \otimes \Cu(B)$ {\newawe (the tensor product of 
$\Cu(A)$ and $\Cu(B)$ as finite-dimensional $\ast$-algebras)} 
generated by {\newawe $\psi_A(A) \otimes \psi_B(B)$.}
}
\end{definition} 

{\awedit Since $\Cu(\R) = \C$, trivially we have $\R \hotimes A \simeq A$. Other examples are discussed below.} 
Some important {\awedit general} facts about the universal tensor product are collected 
in the following:

\begin{proposition} Let $A$, $B$ and $C$ denote EJAs. 
\label{prop: universal tensor product properties}
\begin{itemize}
\item[(a)]  If $\phi : A \rightarrow C$, $\psi : B \rightarrow C$ are unital Jordan homomorphisms with operator-commuting ranges, then 
there exists a unique Jordan homomorphism $A \hotimes B \rightarrow C$ taking $a
 \otimes b$ to $\phi(a) \dot \psi(b)$ for all $a \in A$, $b \in B$.
\item[(b)] $\Cu(A \hotimes B) = \Cu(A) \otimes \Cu(B) \ \ \mbox{and} \ \ \Phi_{A \hotimes B} = \Phi_{A} \otimes \Phi_{B}$. 
\item[(c)] $A \hotimes B$ is universally reversible unless one of the
  factors has a one-dimensional summand and the other has a
  representation onto a spin factor $V_n$ with $n=4$ or $n \ge 6$.
\item[(d)] If $A$ is universally reversible, then $A \hotimes
  M_n(\C)_{\sa} = (\Cu(A) \otimes M_{n}(\C))_{\sa}$.
\item[(e)] $u_{A \otilde B} = u_A \otimes u_B = u_{C^*(A \otilde B)}$.
\end{itemize}
\end{proposition}

\noindent{\em Proof:} (a), (c), and (d) are Propositions 5.2, 5.3 and
5.4, respectively, in \cite{HO}; (b) is observed in the
vicinity of these propositions in \cite{HO}, while (e) follows easily from 
the fact that $\psi^A(u_A) = u_{C^*(A)}$.  $\Box$\\[0.3cm]
Table 1 (c) shows that if $A$ and $B$ are simple and
nontrivial, $A \hotimes B$ will always be UR, and hence, by 
{\magenta Proposition \ref{prop: UR fixed points} and Proposition 
\ref{prop: universal tensor product properties} (b)},
the fixed-point set of $\Phi_A \otimes \Phi_B$.  Using this, one can compute $A
\hotimes B$ for simple $A$ and $B$ (with $n, k > 2$ in the case of
$Q_n$ and $Q_k$) { \cite{HO}}:
\[
\begin{array}{c}
\begin{array}{l|ccc} 
\hotimes    & R_k & C_k & Q_k \\
 \hline 
R_n & R_{nk} & C_{nk} & Q_{nk}    \\
C_n & C_{nk} & C_{nk} \oplus C_{nk} & C_{2nk}  \\
Q_n & Q_{nk} & C_{2nk} & R_{4nk} 
\end{array} \\
\\
\mbox{{\bf Table 2:} Universal tensor products of simple UR EJAs } 
\end{array}
\]
For $Q_2 \hotimes Q_2$, we obtain the direct sum of four copies of $R_{16} = M_{16}(\R)_{\sa}$. The details 
{\redd can be found in} \cite{Graydon16}.    
\ffootnote{What about $Q_2 \hotimes B$ for other $B$?? For example, what's $Q_2 \hotimes C_2$?}

{\bluegrey
In the next two sections, we are going to explore in some detail the possibilities for forming dynamical composites, 
in the sense of our Definition \ref{def: composites}, of Jordan-algebraic systems. As will emerge, the universal 
tensor product $A \hotimes B$ {\em is} such a composite of $A$ and $B$, though not the only one. 
{\newawe In this connection, notice that $A \hotimes B$ is in general a larger vector space than $A \tensor B$. In other words, as a composite of $A$ and $B$, $A \hotimes B$ is not necessarily locally tomographic. Indeed, this will typically be the 
case for composites of EJAs, as underlined by the following finite-dimensional case of a result of Hanche-Olsen (valid 
more generally for JC algebras), which was used in 
\cite{BarnumWilceLocalTomography} to show that the only EJAs having locally tomographic composites with a qubit are the 
complex quantum systems (those whose Jordan algebras are the self-adjoint parts of complex matrix algebras):}

\begin{proposition}[\cite{HO}, Theorem 5.5]\label{prop: HO no-go} Let $A$ be an EJA, and suppose 
that the vector-space tensor product $A \otimes M_{2}(\C)$ is a Jordan algebra with respect to a bilinear product 
$\dot$ such that 
\begin{itemize} 
\item[(i)] $(u_A \otimes b)^2 = u_{A} \otimes b^2$ and $(a \otimes \1)^2 = a^2 \otimes \1$, 
\item[(ii)] $(a \otimes \1) \dot (u_A \otimes b) = a \otimes b$,
\item[(iii)] $a \otimes \1$ and $u_A \otimes b$ operator commute for all $a \in A$ and $b \in M_2(\C)$, 
\end{itemize} 
where $\1$ is the identity in $M_2(\C)$. Then $A$  is the self-adjoint part of a complex matrix algebra.
\end{proposition}

\section{Composites of Jordan-Algebraic Systems} 
\label{sec: composites Jordan}

{\bluegrey As discussed in section 3.2, we can regard an EJA $A$ as a dynamical probabilistic model 
with dynamical group $G(A)$. If $A$ and $B$ are EJAs, thus regarded as models of probabilistic physical systems, 
we would like to know what possibilities exist for forming a Jordan-algebraic dynamical composite $AB$. We shall  
actually impose a further, but we think natural, condition on such composites, namely condition (b) in the definition 
below: }

\begin{definition}\label{def: composites of EJAs} 
{\em A {\em composite} of EJAs $A$ and $B$ is an EJA $AB$, 
plus a bilinear mapping $\pi : A \otimes B \rightarrow AB$, such that 
\begin{itemize} 
\item[(a)] $\pi$ makes $(AB,G(AB))$ a dynamical composite of $(A,G(A))$ and $(B,G(B))$, in the sense of Definition \ref{def: dynamical composites},  
\item[(b)] $(\phi \otimes \psi)^{\dagger} = \phi^{\dagger} \otimes \psi^{\dagger}$ for all $\phi, \psi \in G(A)$, and 
\item[(c)] $AB$ is generated, as a Jordan algebra, by (images of) pure tensors. 
\end{itemize}
{\redd By Lemma 2.2}, the 
mapping $\pi : A \otimes B \rightarrow AB$ is injective; 
hence, we can, and shall, identify $A \otimes B$ with its image in $AB$, writing 
$\pi(a,b)$ as $a \otimes b$.} 
\end{definition}

  Condition (b) is rather strong, but natural if we keep in mind that
  our ultimate aim is to construct dagger-compact categories of EJAs.
  Regarding condition (c), suppose $\pi : A \times B \rightarrow AB$
  satisfied only (a) and (b): letting $A \odot B$ denote the Jordan
  subalgebra of $AB$ generated by $\pi(A \otimes B)$, one can show
  (Appendix \ref{appendix: weak composites}) that the co-restriction of
  $\pi$ to $A \odot B$ also satisfies (a) and (b); thus, any composite
  in the weaker sense defined by (a) and (b) contains a composite
  satisfying all three conditions. 

In Section \ref{sec: Euclidean JC}, Proposition \ref{prop: canonical
  product composite}, we will show that $A \hotimes B$ is a composite
in the sense of Definition \ref{def: composites of EJAs}.  The main
result of the present section is to show that 
any such composite $AB$ 
is a direct summand of $A \hotimes B$.  In view of Table 2, this severely
limits the possibilities for $AB$.

\subsection[The Fundamental Identity]{The identity $\boldsymbol{(a \otimes u)} \bullet \boldsymbol{(x \otimes y) 
= \boldsymbol (a} \bullet \boldsymbol{ x ) \otimes y }$.} 

At this point, we have limited information
 about how the Jordan
structure of a composite $AB$ interacts with the Jordan structures of 
$A$ and $B$. However, we shall now establish, 
for
any $a \in A$ and any $x, y \in B$, the identity $(a \otimes u) \dot 
(x \otimes y) = (a \dot x) \otimes y$ --- in other words, that $L_{a
  \otimes u_B}$ acts on $A \otimes B \leq AB$ as $L_{a} \otimes \1_B$,
where $\1_B$ is the identity operator on $B$.\\[0.3cm]
\noindent{\bf One-parameter groups and exponentials} It will be
helpful first to recall some basic facts about operator exponentials,
or, equivalently, one-parameter groups of linear operators on
finite-dimensional spaces (see, e.g., \cite{Curtis}). Let $V$ be a finite-dimensional real vector
space, and $X$, a linear operator on $V$. Recall that 
$\phi(t) := e^{tX}$ is the unique function $\R \rightarrow {\cal
  L}(V)$ satisfying the initial-value problem
\[\phi'(t) = X \phi(t) ; \ \ \phi(0) = \1\]
(where $\1$ is the identity operator on $V$). In particular, $\phi'(0) =
X$. The function $\phi$ satisfies $\phi(t + s) = \phi(t)\phi(s)$ 
{\redd and hence, $\phi(t)\phi(-t) = \phi(0) = \1$,} 
 {\redd hence, as $\phi(0) = \1$,  
$\phi(t)$ is invertible, with $\phi(t)^{-1} = \phi(-t)$. In other 
words,} $\phi$ is a one-parameter group of linear operators on $V$.
Conversely, if $\phi : \R \rightarrow {\cal L}(V)$ is any continuous
one-parameter group of linear operators on $V$, then $\phi$ is
differentiable, and $\phi(t) = e^{tX}$ where $X = \phi'(0)$. Notice,
also, that in such a case we have
\[Xa = \frac{d}{dt} \phi(t)a|_{t = 0}\]
for any vector $a \in V$.

For later reference, the following lemma collects some standard facts: 

\begin{lemma}\label{lemma: operator commutation and one parameter groups}
Let $X, Y$ be linear operators on a finite-dimensional inner product space $V$. Then 
\begin{itemize} 
\item[(a)] $X$ commutes with $e^{tX}$ for all $t$; 
\item[(b)] If 
$e^{tX}$ commutes with $e^{sY}$ for all $t, s$, then $X$ commutes with $Y$; 
\item[(c)] $(e^{tX})^{\dagger} = e^{tX^{\dagger}}$.
\end{itemize}
\end{lemma}  

\tempout{
\noindent{\em Proof:} (a) is standard. For (b), define $F(s,t) =
e^{tX} e^{sY} = e^{sY} e^{tX}$. By the equality of mixed partials, we
have
\[XY F(s,t) = \frac{\partial^2}{\partial s \partial t} F(s,t) = \frac{\partial^2}{\partial s \partial t} F(s,t) 
= YX F(s,t)\] for all $s, t$. In particular, setting $s, t = 0$, so
that $F(s,t) = \1$ (the identity operator), we have $XY = YX$. For
(c), use the observation that if $\phi(t)$ is a one-parameter group,
then $\langle \phi'(0)x, y \rangle = \frac{d}{dt} \langle \phi(t) x, y
\rangle|_{t = 0}$. $\Box$\\}

Note that, by (c), if $\phi(t)$ is a one-parameter group with $\phi'(0) = X$ hermitian, then $\phi(t)$ is hermitian for all $t$, {\awedit and conversely.} 

Now let $A$ be an EJA. For $a \in A$, 
define 
\[\phi_{a}(t) := e^{t L_{a}} = e^{L_{ta}},\] 
i.e., $\phi_{a}$ is the solution to the initial-value problem
$\frac{d}{dt}\phi_{a} = L_{a} \phi_{a}, \ \ \phi_{a}(0) = \1$. 
By part (c) of Proposition \ref{prop: quadratic representation}, $\phi_{a}(t) = U_{e^{ta/2}}$; by part 
(b) of the same Proposition, this last is a positive mapping. Since 
$e^{tL_{a}}$ is invertible with inverse $e^{-tL_{a}} = e^{L_{-ta}}$, $\phi_{a}(t)$ is an order-automorphism 
belonging to $G(A)$. 
It follows that $L_a$ belongs to ${\mathfrak g}_{A}$, the Lie group of the identity component $G(A)$ of $A$. 
Note that $\langle L_{a} x, y \rangle = \langle a x, y \rangle = \langle x, ay \rangle = \langle x, L_{a} y\rangle$ 
for all $x, y \in A$; that is, $L_a$ is self-adjoint. One can show that, conversely, a self-adjoint 
element of ${\mathfrak g}_{A}$ has the form $L_{a}$ for a unique $a \in A$. (See \cite{FK}, pp. 6 and 49, for the details.)

We are now ready for the main result of this {\awedit sub-}section.

\begin{proposition}\label{prop: main equation} 
Let $AB$ be a composite (in the sense of Definition 
\ref{def: composites of EJAs}) of Jordan algebras $A$ and $B$. {\awedit Then the mapping 
$a \mapsto a \otimes u_B$ is a Jordan homomorphism from $A$ into $AB$. That is,} for 
all $a, x \in A$ and $b, y \in B$, 
{\awedit \begin{equation} 
(a \otimes u_B)\dot (x \otimes y) = (a \dot x) \otimes y \ \ \mbox{and} \ \ (u_A \otimes b)\dot (x \otimes y) = x \otimes( b \dot y ) 
\end{equation}}
\end{proposition}\vspace{-.1in}
{\awedit We shall refer to (1) as the {\em fundamental identity}.  }\\[0.3cm]
\noindent{\em Proof:} We prove the first identity; the second is 
handled similarly. Let $\phi(t)$ be a
one-parameter group {\awedit of order automorphisms of} $A$ with $\phi'(0) = L_a$, {\awedit that is, 
$\phi(t) = e^{tL_{a}}$.} Then $\psi(t) :=
\phi(t) \otimes \1$ is a one-parameter group of automorphisms on $AB$, by 
condition (b) of Definition \ref{def: dynamical composites}. 
 Let $Y = \psi'(0) {\redd \in {\mathfrak g}_{AB}}$; then, for all $x \in A$ and $y \in B$,  
\begin{eqnarray*} 
Y(x \otimes y) 
& = & \left[ \frac{d}{dt} \psi(t)\right]_{t=0}  (x \otimes y) \\
& = & \left[ \frac{d}{dt} (\psi(t)(x \otimes y))\right]_{t = 0}\\
& = & \left[ \frac{d}{dt} \left( \phi(t)x \otimes y \right) \right]_{t = 0} \\
& = & \left( \left[ \frac{d}{dt} \phi(t)\right ]_{t = 0} x \right) \otimes y 
\ = \ L_{a} x \otimes y \ = \ ax \otimes y.
\end{eqnarray*}
Using condition (b) of Definition 
\ref{def: composites of EJAs},  the hermiticity of $\phi'(0) = L_a$, and the consequent 
hermiticity of $\phi(t)$ (see the discussion following Lemma \ref{lemma: operator commutation and one parameter groups}), 
we have
\[(\phi_{a}(t) \otimes \1)^{\dagger} = \phi_{a}(t)^{\dagger} \otimes \1 = \phi_{a}(t) \otimes \1.\]
Hence, $Y$ is self-adjoint. 
As discussed above, it follows 
that 
there exists some $v \in AB$ with $Y = L_{v}$ on $A \otimes B$.
Thus, 
\[v  \dot  (x \otimes y)  = L_{v} (x \otimes y) = Y(x \otimes y) = (a \dot x) \otimes y.\]
Setting $x = u_A$ and $y = u_B$, we have 
 $v = v \dot u_{AB} = v \dot (u_{A} \otimes u_{B}) = (a \dot u_A) \otimes u_B = a \otimes u_B$, which gives the 
 advertised result. $\Box$ \\[0.3cm]
One immediate consequence of the fundamental identity (1) is that, for any composite $AB$ of EJAs, 
$A \otimes u_B = \{ a \otimes u_B |  a \in A\}$ is a Jordan subalgebra of $AB$ isomorphic to $A$. 
In particular,  since 
$A \otimes \R = A \otimes 1$, any composite of an EJA $A$ with the one-dimensional Jordan algebra $\R$ is canonically 
isomorphic to $A$. 

{\awedit The fundamental identity can be read as asserting that $L_a \otimes \id_B$ and $\id_A \otimes L_b$ act 
on $A \otimes B \leq AB$ in the expected way for all $a \in A, b \in B$. 
Recalling that, for $a \in A$, the mapping $U_a : A \rightarrow A$ is defined by 
$U_a = 2L_{a}^2 - L_{a^2}$, we have the} 

\begin{corollary}\label{cor: cor to main equation} In any composite $AB$ {\redd of EJAs $A$ and $B$}, and 
for any $a \in A$, $b \in B$, $U_{a \otimes u_B}$ and $U_{u_A \otimes b}$ act on $A \otimes B$ as 
$U_a \otimes \id_B$ and $\id_A \otimes U_b$, respectively. \end{corollary}

{\bblue Another, very important, consequence of Proposition \ref{prop: main equation} is the following: {\mmagenta 

\begin{proposition}\label{prop: product projections} 
Let $p \in A$ and $q \in B$ be projections. Then $p \otimes q$ is a
projection in $AB$, for any composite $AB$ of $A$ and $B$.
\end{proposition}

\noindent              
{\em Proof:} By Lemma \ref{lemma: projections}, if $A$ is an EJA, an effect $a \in A_+$ is a projection 
iff $\langle u_A |a \rangle = \langle a | a \rangle$. Certainly, $p \otimes q$ is an effect 
in $AB_{+}$. 
Now note that, by repeated application of Proposition \ref{prop: main equation}, plus the fact that $u_A \otimes u_B = u_{AB}$  and the associativity of the inner product, 
we have 
\begin{eqnarray*}
\langle u_A \tensor u_B , p \tensor q \rangle &  = & 
\langle u_A \tensor u_B , p \dot p \tensor q\rangle \\
& = & \langle u_A \tensor u_B , (p \tensor u_B) \dot (p \tensor q)\rangle \\
& = & \langle (u_A \tensor u_B)\dot (p \tensor u_B) , p \tensor q \rangle \\
& = & \langle p \tensor u_B , p \tensor q \rangle \\
& = & \langle p \tensor u_B , p \tensor q \dot q \rangle \\
& = & \langle p \tensor u_B , (u_A \tensor q) \dot (p \tensor q) \rangle \\
& = & \langle (p \tensor u_B) \dot (u_A \tensor q) , p \tensor q \rangle \\ 
& = & \langle p \tensor q , p \tensor q \rangle.
\end{eqnarray*}
It follows that $p \otimes q$ is a projection. $\Box$\\ 
}

{\mmagenta 
\begin{corollary}\label{cor: jordan-orthogonal projections} Let $p, q$ be Jordan-orthogonal projections in $A$. In any composite $AB$, and for any $b \in B_+$, 
$p \otimes b$ and $q \otimes b$ are Jordan-orthogonal.
\end{corollary}

\noindent{\em Proof:} Since $\otimes : A \times B \rightarrow AB$ is a {\em positive} bilinear map, we have for 
all $a_1, a_2 \in A$ and all $b \in B$ that $a_1 \leq a_2 \Rightarrow a_1 \otimes b \leq a_2 \otimes b$ 
(since $a_2 \otimes b - a_1 \otimes b = (a_2 - a_1) \otimes b$ is positive). Now suppose $b$ is an effect in $B$ 
and that $p$ and $q$ are Jordan orthogonal projections in $A$, so that $p + q \leq u_A$. It's enough to show that 
$p \otimes b + q \otimes b \leq u_{AB}$. But $p \otimes b + q \otimes b = (p + q) \otimes b$, and $p + q \leq u_{B}$, so 
we have $(p + q) \otimes b \leq u_A \otimes b \leq u_{A} \otimes u_{B} = u_{AB}$. Since the result holds for 
arbitrary effects $b$, it holds for arbitrary elements $b \in B$. $\Box$ \\}

{\awedit \begin{proposition}\label{prop: operator commutation} 
For all $a \in A$, $b \in B$, $a \otimes u_B$ and $u_A \otimes b$
operator commute in $AB$.
\end{proposition}

{\mmagenta Note that this would follow trivially from Proposition \ref{prop: main equation} if $AB$ were spanned 
by pure tensors,  that is, if $AB$ is locally tomographic.\\ }

\noindent{\em Proof:}  Suppose $p \in A$ and $q \in B$ are projections, and let $p' = u_A - p$ and $q' = u_B - q$.
Then we have 
\[u_{AB} = u_{A} \otimes u_{B} = (p + p') \otimes (q + q') = p \otimes q + p' \otimes q + p \otimes q' + p' \otimes q'.\]
{\bblue By Proposition \ref{prop: product projections}, the four terms on the right are projections. }
They are mutually orthogonal by Corollary \ref{cor: jordan-orthogonal projections}, and sum to the unit in $AB$.  Hence, $p \otimes
u_B = p \otimes q + p \otimes q'$ and $u_A \otimes q = p \otimes q +
p' \otimes q$ operator commute by \cite{AS}, Lemma 1.48.  Now let $a \in A$
and $b \in B$ be arbitrary: by the spectral theorem for EJAs, we have $a = \sum_i
t_i p_i$ and $b = \sum_j s_j q_j$ for pairwise orthogonal families of projections
$p_i$ and $q_j$ and scalars $t_i$ and $s_j$. Since $p_i \otimes u_B$ and $u_A \otimes q_j$
operator commute for all $i, j$, it follows that
$a \otimes u_B = \sum_i t_i p_i \otimes u_B$ and $ u_A \otimes b = \sum_j s_j u_A \otimes q_j$ 
also operator commute. $\Box$\\ 
}

{\awedit 
\subsection{Composites of Direct Sums} 
Let $A \simeq A_0 \oplus A_1$. Then there is a natural homomorphism $G(A_0) \rightarrow G(A)$ given by 
$\phi \mapsto \phi \oplus \id_{A_1}$ (where $(\phi \oplus \id_{A_1})(A_0, a_1) = (\phi(A_0), a_1)$).  If 
$AB$ is a dynamical composite of $A$ and $B$, then let $A_0B = J( \pi(A_0 \times B))$, the 
Jordan subalgebra of $AB$ generated by pure tensors $A_0 \otimes b$ with $A_0 \in A_0$ (equipped 
with the inner product inherited from $AB$). Note that we have a natural positive bilinear map $\pi_{0} : A_0 \times B \rightarrow A_0 B$, obtained by restricting and co-restricting $\pi : A \times B \rightarrow AB$ to $A_0 \times B$ and $A_0 B$. 

The following two results, taken together, will be very helpful below

\begin{proposition}\label{prop: subcomposites} $A_0 B$ is a composite of $A_0$ and $B$. \end{proposition}

\noindent{\em Proof:} We check the requirements of Definition \ref{def: composites of EJAs}. 
 First, note that $u_0 \otimes u_B$ is effective as the unit in $A_0B$. For all $a \in A_0$ and $b \in B$, we have 
$(u_0 \otimes u_B) \dot (a \otimes b) = (u_0 \dot a) \otimes b = a \otimes b$ by the fundamental identity and the fact that $u_0$ is the unit in $A_0$. Since this holds on all pure tensors in $\pi(A_0 \times B)$, it holds 
throughout $A_0B$.  

A  normalized state $\alpha$ on $A_0$ extends in a canonical way to a normalized state $\bar{\alpha}$ on $A = A_0 \oplus A_1$, given by $\bar{\alpha}(A_0, a_1) = \alpha(A_0)$. Thus, if $\beta$ is a normalized state on $B$, 
condition  (c) in Definition \ref{def: composites} tells us that there is a normalized state $\gamma$ on $AB$ with 
$\gamma(a,b) = \bar{\alpha}(a)\beta(b)$ for all $a \in A$ and $b \in B$. Restricting $\gamma$ to 
$A_0 B$ gives us a positive functional $\gamma_0$ with $\gamma_0(A_0 \otimes b) = \alpha(A_0) \beta(b)$ 
for all $A_0 \in A_0$ and $b \in B$. In particular, $\gamma_0 (u_0 \otimes u_B) = \alpha(u_0)\beta(u_B) = 1$, 
so $\gamma_0$ is a normalized state on $A_0 B$. 

Embedding $G(A_0)$ in  $G(A)$ in the obvious way (that is, $g \in G(A_0)$ acts on $(A_0, a_1)$ as $(gA_0, a_1)$), 
we have a homomorphism $\otimes_0 : G(A_0) \times G(B) \rightarrow G(A_0 B)$ obtained by restricting the domain of the given homomorphism $\otimes : G(A) \times G(B) \rightarrow G(AB)$. For any $h \in G(B)$ and any $A_0 \in A, b \in B$ we have  
$(g \otimes h)(A_0 \otimes b) = gA_0 \otimes hb$. Since this last lies in $\pi(A_0 \times B)$, $g \otimes h$ fixes $A_0B$. Thus, we have a natural homomorphism $G(A_0) \times G(B) \rightarrow G(A_0 B)$ satisfying the requirements for a dynamical composite. 
 Finally, note that $(g \otimes h)^{\dagger} = g^{\dagger} \otimes h^{\dagger}$ since this holds at the level of $AB$. 
$\Box$ \\
}

{\awedit 
Recall that the center of an EJA $A$ is the set of all elements in $A$ that operator commute with each element of $A$.  Such elements are called \textit{central}.  A \textit{central projection} is thus a projection $p\in A$ such that $L_{p}\circ L_{a}=L_{a}\circ L_{p}$ for all $a\in A$.
\begin{proposition}\label{prop: composites of sums} Let $AB$ be any composite of EJAs $A$ and $B$.  If $A = \bigoplus_{i} A_{i}$ and $B = \bigoplus_j B_j$ where 
$A_i = c_{i} \dot A$ and $B_j = d_j \dot B$ for mutually Jordan orthogonal families of central projections $c_i$ 
in $A$ and $d_j$ in $B$, then $AB = \bigoplus_{i,j} A_i B_j$. \end{proposition}

\noindent{\em Proof:} It is enough to prove this for the case in which $A = c \dot A \oplus c' \dot A$ for a central idempotent $c$, 
where $c' = u_A - c$.  By (\cite{AS}, Lemma 1.43), if $p$ is any idempotent in an EJA $A$, then $U_p(A)$ 
is a Jordan subalgebra of $A$. Moreover, if  $p$ and $q$ are Jordan-orthogonal idempotents, then 
$U_p(A)$ is Jordan-orthogonal to $U_p(B)$, by (\cite{AS}, Lemma 1.45).  Now let $c$ be a central projection in $A$, and write $A_c$ for $c \dot A$. Then  $c \otimes u_B$ is an idempotent in $AB$, and  Corollary 4.4 gives us 
\[x \otimes y = (c \dot x) \otimes y = U_{c}(x) \otimes y = U_{c \otimes y}(x \otimes y) \in U_{c \otimes y}(AB).\]
for all $x \in A_{c}$ and $y \in B$. As pure tensors $x \otimes y$ with $x \in A_c$ and $y \in B$ generate $A_{c}B$, 
we now have $A_{c}B \leq U_{c \otimes \id_B}(AB)$. By the same token $A_{c'}B \leq U_{c' \otimes \id_B}(AB)$. 
As noted above, the two larger subalgebras are pairwise Jordan-orthogonal; hence, so are the two smaller ones. In particular, the (internal) direct sum $A_{c}B \oplus A_{c'}B$ exists, and is a Jordan subalgebra, of $AB$. 
Every pure tensor in $A \otimes B$ has the form $(x_1 + x_2) \otimes y = x_1 \otimes y + x_2 \otimes y$, 
where $x_1 \in A_c$ and $x_2 \in A_{c'}$, and hence, belongs to $A_{c}B \otimes A_{c'}B$. 
Since such pure tensors generate $AB$ as a Jordan algebra, we have 
$AB = A_{c}B \oplus A_{c'}B$. 
The general case now follows by an easy induction. $\Box$ 

\begin{corollary}\label{cor: centrality} Let $c$ be a projection in an EJA $A$, and let $AB$ be a composite of 
$A$ with an EJA $B$. Then $c \otimes u_B$ is central in $AB$ if $c$ is central in $A$.
\end{corollary}
}

\subsection{Composites of simple EJAs} 

We now show that if $A$ and $B$ are nontrivial simple EJAs, 
then {\awedit any composite} $AB$ must be special, universally reversible, and an ideal (a direct summand) of the  universal tensor product $A \hotimes B$.  {\awedit It follows that $A$ and $B$ must be special; that is, 
no nontrivial composite exists if either factor is $M_3(\O)$.} The rough idea is that, since $A$ and $B$ have rank at least two, the fact that 
products of distinguishable effects are distinguishable will yield at least four distinguishable effects in $AB$. 
If the latter were simple, this would be the end of the story; but we know from the case of universal tensor products 
(which we will ultimately show are dynamical composites in our sense) that composites can have nontrivial direct 
summands. Therefore, we need to work a bit harder, and show that every irreducible direct summand of $AB$ has rank at least 4. 
 
{\awedit We will need some preliminaries. An element $s \in A$ is {\redd called} a {\em symmetry} iff $s^2 = u$.\footnote{\redd Not to be confused with a symmetry {\em qua} order-automorphism.} In this case
$U_s$ is a Jordan automorphism of $A$, with $U_{s}^{2} = \id$ 
(\cite{AS}, Prop. 2.34). Also
note that $p := \frac{1}{2}(s + u)$ is a projection, and, conversely,
if $p$ is a projection, then $s := 2p - u$ is a symmetry.
Two projections $p, q \in A$ are {\em exchanged by a symmetry} $s \in
A$ iff $U_s(p) = q$ (in which case, $p = U_{s}(q)$).  More
generally, $p$ and $q$ are {\em equivalent} iff there exists a finite
sequence of symmetries $s_1,...s_{\ell}$ with $q = (U_{s_{\ell}} \circ
\cdots \circ U_{s_1})(p)$. It will be important below that if $A$ and $B$ are simple, 
then any two atomic projections are exchanged by a symmetry (\cite{HO-Stormer}, Lemma 5.3.2).

\begin{lemma}\label{lemma: exchange} Let $s \in A$ be a symmetry exchanging projections 
$p_1, p_2 \in A$, and let $t \in B$ be a symmetry exchanging projections $q_1, q_2 \in B$.  
Then $s \otimes u_B$ and $u_A \otimes t$
  are symmetries in $AB$, and $U_{u_A \otimes t} U_{s \otimes u_B}
  (p_1 \otimes q_1) = p_2 \otimes q_2$. In particular, the projections
  $p_1 \otimes q_1$ and $p_2 \otimes q_2$ are equivalent.\end{lemma}

\noindent{\em Proof:} $(s \otimes u_B)^2 = s^2 \otimes u_B = u_A
\otimes u_B = u_{AB}$ by Proposition \ref{prop: main equation}. Similarly for $u_A \otimes
t$. Now by Corollary \ref{cor: cor to main equation}, we have 
\[\ \ U_{u_A \otimes t} U_{s \otimes u_B} (p_1 \otimes q_1) = U_{u_A \otimes t}(U_s (p_1) \otimes q_1) = U_{s}(p_1) \otimes U_{t}(q_1) = p_2 \otimes q_2. \ \Box\]
}

\begin{theorem}\label{thm: composites special}{ Let 
$AB$ be a composite of simple, {\red nontrivial} Jordan algebras $A$ and $B$.   
Then $AB$ is a special, {\red universally reversible} EJA.}
\end{theorem}

\noindent{\em Proof:} We shall show that every irreducible direct 
summand of $AB$ has rank $\geq 4$, from which the result follows. 
Decompose $AB$ as a direct sum of simple ideals,
say $AB = \bigoplus_{\alpha} M_{\alpha}$. Let $\pi_{\alpha} : AB 
\rightarrow M_{\alpha}$ be the corresponding projections, and let
$u_{\alpha} := \pi_{\alpha}(u_{AB})$ be the unit in $M_{\alpha}$.
Suppose now that $\{p_1,...,p_n\}$ is a Jordan frame in $A$ and
$\{q_1,...,q_m\}$ is a Jordan frame in $B$.  {\bluegrey By Proposition \ref{prop: product projections} and Corollary \ref{cor: jordan-orthogonal projections},  
$p_i \otimes p_j$ are pairwise orthogonal projections in $AB$.} {\awedit Since $A$ and $B$ are simple, 
 there are symmetries in $AB$ exchanging the $p_i$, and there are
symmetries in $B$ exchanging the $q_j$. By Lemma \ref{lemma: exchange}, therefore, the
projections $p_i \otimes q_j$ are pairwise equivalent. By \cite{AS} Lemma
3.9, therefore, these projections have the same central cover
$c$. This means that for each $\alpha$, the projection $\pi_{\alpha} :
AB \rightarrow M_{\alpha}$ takes  none of the projections $p_i \otimes
q_j$ to the zero projection in $M_{\alpha}$, or it takes  all of them
to zero --- the former case arising exactly when $u_{\alpha}
\leq c$, and the latter, when $u_{\alpha}c = c u_{\alpha} = 0$. If
$M_{\alpha}$ is of the first type, $\{\pi_{\alpha}(p_i \otimes q_j) |
i = 1,...,n, j = 1,...,m\}$ consists of $nm$ distinct orthogonal
projections in $M_{\alpha}$, summing to the unit $\pi_{\alpha}(u) =:
u_{\alpha}$. Hence, the rank of $M_{\alpha}$ is at least
$nm$. In particular, since $A$ and $B$
are nontrivial, $n, m \geq 2$, whence, $M_{\alpha}$ has rank at least
4, and hence, is special. }

Now let $p, q$ be arbitrary projections in $A$ and $B$, respectively:
extending each to a Jordan frame, as above, we see that for all
$\alpha$, if $\pi_{\alpha}(p \otimes q) \not = 0$, then $M_{\alpha}$
is special. Hence, $p \otimes q$ belongs to the direct sum of the
special summands of $AB$, i.e., to $M_{\mbox{sp}}$. 
Since projections $p \otimes q$ generate $AB$, the latter is 
special. 

{\red The argument also shows that each simple direct summand $M_{\alpha}$, in addition to being 
special, is not a spin factor, 
and hence, is UR. {\bblue Since  direct sums of universally reversible EJAs are 
again UR {\awedit (Appendix \ref{appendix: direct sums}, Proposition \ref{prop: UR composites}), it follows that $AB$ must be UR.} $\Box$ }

{\awedit 
\begin{proposition}\label{prop: composites UR}  Let $A$ and $B$ be EJAs having no 1-dimensional summands. If $AB$ is a composite of $A$ and $B$, then $AB$ is UR.\end{proposition} 

\noindent{\em Proof:} Decompose $A$ and $B$ into simple direct summands, say $A = \bigoplus_i A_i$ and 
$B = \bigoplus B_j$. By Proposition \ref{prop: composites of sums}, $AB = \bigoplus A_i B_j$. By 
Proposition \ref{prop: subcomposites}, each summand $A_i B_j$ is a composite, and hence, by the preceding result, UR. But direct sums of universally reversible EJAs are again UR. $\Box$ \\
}

We call an ideal \textit{trivial} if it is isomorphic to a direct sum of rank-one EJAs.
{\awedit 
\begin{proposition}\label{cor: no composite with exceptional} 
If $A$ contains an exceptional ideal and $B$ contains a nontrivial ideal,
there exists 
no composite $AB$ satisfying the 
conditions of Definition 1.\end{proposition}

\noindent{\em Proof:} Suppose $B$ has a nontrivial ideal $B_1$, and that a composite $AB$ exists. Let $A_0$ be any simple 
ideal in $A$. By Proposition \ref{prop: subcomposites}, 
$A_0 B$ is a composite of $A_0$ and $B$. Let $B = B_0 \oplus B_1$ where $B_1$ is a nontrivial simple ideal of $B$. 
Then, applying Proposition \ref{prop: subcomposites} again, $A_0 B_1 = J(A_0 \otimes B_1) \leq A_0 B$, is a composite. Since $A_0$ and $B_1$ are simple,  
Theorem \ref{thm: composites special} implies that $A_0 B_1$ is special.  The fundamental identity (1) 
implies that the mapping $A_0 \rightarrow A_0 B_1$, given by $a \mapsto a \otimes u_{1}$, where $u_1$ is 
the unit of $B_1$, is a faithful Jordan homomorphism. Therefore, $A_0$ is special. $\Box$ \\
}

{\awedit Another way to express Proposition \ref{cor: no composite with exceptional} is  that if 
$A$ is exceptional and $AB$ exists, then $B$ must be a direct sum of 1-dimensional EJAs --- in other words, a classical system.}

\begin{theorem}\label{thm: simple composites ideals} Let $A$ and $B$ be simple,  special EJAs. Then $AB$ is an ideal in $A \hotimes B$.\end{theorem}

\noindent{\em Proof:}  By {\awedit Propositions \ref{prop: main equation} and \ref{prop: operator commutation}},  we have Jordan homomorphisms $A, B \rightarrow AB$ with operator-commuting ranges. Since $AB$ is 
special, i.e., a JC-algebra, elements of $AB$ operator commute iff their images in $\Cu(AB)$ operator commute 
 (\cite{HO}, Lemma 5.1). Thus, we have Jordan homomorphisms $A, B \rightarrow \Cu(AB)$ with operator-commuting ranges. The universal 
property given by Proposition \ref{prop: universal tensor product properties} (a)  of $A \hotimes B$  yields a Jordan homomorphism $\phi : A \hotimes B \rightarrow \Cu(AB)$ taking 
(the image of) $a \otimes b$ in $A \hotimes B$ to (the image of) $a \otimes b$ in $\Cu(AB)$. Since both 
$A \hotimes B$ and $AB$ are generated by pure tensors, $\phi$ takes $A \hotimes B$ onto $AB$. Letting $K$ denote the kernel of $\phi$, an ideal of $A \hotimes B$, we have $A \hotimes B = K^{\awedit \perp} \oplus K$, where $K^{\awedit \perp}$ is the complementary ideal; 
the mapping $\phi$ factors through the projection $A \hotimes B \rightarrow K^{\awedit \perp}$ to give an isomorphism 
$\phi': K^{\awedit \perp} \simeq AB$, {\newawe since the restriction $\phi'$ of the surjection $\phi$ to $K^\perp$ is 
injective.}
$\Box$ \\[0.3cm]
{\awedit In appendix \ref{appendix: direct sums}, we show that $C^{\ast}(A \oplus B) = C^{\ast}(A) \oplus C^{\ast}(B)$ (Proposition 
\ref{prop: Cu additive}). Combining this 
fact with Proposition \ref{prop: composites of sums}, we can extend the preceding result as follows.

\begin{corollary} Let $AB$ be a composite of  special EJAs $A$ and $B$. 
Then $AB$ is a direct summand of $A \hotimes B$.
\end{corollary} 
}

Combined with Table 2, Theorem \ref{thm: simple composites ideals} sharply restricts the possibilities 
for composites of simple EJAS. In particular, it follows that {\em 
if $A \hotimes B$ is itself simple}, then $AB
\simeq A \hotimes B$. In other words, in this case the {\redd universal tensor product} 
is the only ``reasonable" tensor product (to the extent that we 
think the 
conditions of Definition \ref{def: composites of EJAs} 
constitute reasonableness, in this context). 
If $A = B = C_n$, 
so that $A \hotimes B = M_{n^2}({\mathbb C})_\sa \oplus M_{n^2}({\mathbb
  C})_\sa$, we have another candidate, i.e., the usual
quantum-mechanical composite $M_{n^2}({\mathbb C})_\sa$. If $A = B = M_2(\H)$ (that is, if $A$ and $B$ 
are two quaternionic bits), we have $A \hotimes B = M_{16}(\R)_{\sa} \oplus M_{16}(\R)_{\sa} \oplus M_{16}(\R)_{\sa} \oplus M_{16}(\R)_{\sa}$, giving us four possibilities for $AB$. {\em \awedit These exhaust the possibilities for 
composites of simple  real, complex 
and quaternionic quantum systems!}
\ffootnote{\blue [AW: What about spin factors?]}

\section{EJC-algebras} 
\label{sec: Euclidean JC}

In view of Theorem \ref{thm: composites special} and Corollary
\ref{cor: no composite with exceptional}, we now restrict our
attention to special EJAs.  A {\em JC-algebra} is variously defined as
a norm-closed Jordan subalgebra of $\L(\Hilb)$ for a real or complex
Hilbert space $\Hilb$, or as a Jordan algebra that is
Jordan-isomorphic to such an algebra. In finite dimensions, any JC
algebra is euclidean, and any special euclidean Jordan algebra is JC
(on the second definition).  Here, we consider EJAs
that are embedded, not necessarily in $\L(\Hilb)$ for a specific
Hilbert space, but in some definite complex $\ast$-algebra.

{\mmagenta \begin{definition}{\em An {\em embedded JC algebra} (EJC) is a triple $(A,\pi_A,\M_A)$ where $A$ is an EJA, $\M_A$ is a finite-dimensional complex $\ast$-algebra, and $\pi_A$ is a unital, injective Jordan homomorphism $A \rightarrow \M_{A_{\sa}}$. }
\end{definition} 
}

{\mmagenta As the notation suggests, we often simply write $A$ for $(A,\pi_A, \M_A)$.  Where no confusion is likely to arise, we shall usually identify $A$ with the Jordan subalgebra $\pi_A(A)$ of $\M_{A_{\sa}}$, denoting $(A,\pi,\M_A)$ by the pair $(A,\M_A)$. }  For aesthetic reasons, we abbreviate the phrase ``embedded euclidean JC" by EJC (rather than EEJC), letting the initial {\it E} stand simultaneously for {\em embedded} and {\em euclidean}.

{\hblue Notice that Definition 5.1 does not require that $A$ generate $\M_A$
as a complex $*$-algebra. While most of our examples satisfy this condition, few of our general results depend on it.}

In this section, we develop a canonical tensor product for EJC
algebras, {\redd and use this to construct several symmetric monoidal 
categories of EJC-algebras.} 
In one case, we obtain a category of reversible EJAs, with
a monoidal product extending that of ordinary complex matrix algebras;
{\redd in another, which we call $\InvQM$}, we 
restrict attention to EJAs that arise as fixed-point algebras of 
involutions on {\redd complex} $\ast$-algebras (a class that includes all 
universally reversible EJCs, but also includes the quaternionic bit, 
$M_{2}(\Q)_{\sa}$, {\redd as symplectically embedded in $M_{4}(\C)$}). Here, the monoidal structure agrees with the Hanche-Olsen tensor product 
{\redd in the cases in which the factors are UR.}

For any $\ast$-algebra $\M$ and any set $S \subseteq \M_{\sa}$, let
$J(S)$ denote the Jordan subalgebra of 
$\M_{\sa}$  generated by $S$. (We
should probably write this as $J_{\M}(S)$, but context will generally
make the usage unambiguous. See below for an example!)

\begin{definition} \label{def: canonical tensor product} 
{\em The {\em canonical tensor product} of $(A,\M_A)$ and
  $(B,\M_B)$ is $(A \odot B, \M_{A} \otimes \M_{B})$, where 
$A \odot B := J(A \otimes B) \subseteq \M_{A}
  \otimes \M_{B}$, the Jordan subalgebra of $(\M_A \otimes \M_B)_{\sa}$ generated 
  by $A \otimes B$.
\footnote{When $\M_A = M_n(\C)$ for some $n$ and similarly for $B$,
  this agrees with a construction by Jamjoom, Def. 2.1 in
  \cite{JamjoomTensorJW}.}}\end{definition}

Note that this makes it a matter of definition that $\M_{A \odot B} = \M_{A} \otimes \M_{B}$. 
In particular, if $A$ and $B$ are universally embedded, so that $\M_{A} = \Cu(A)$ and $\M_B = \Cu(B)$, then 
$A \odot B = A \hotimes B$, so that $\M_{A \hotimes B} = \Cu(A) \otimes \Cu(B)$ by definition; 
but the fact that this last is $\Cu(A \hotimes B)$ {\redd (whence, $A \odot B$ is UR)} is a theorem.
It will also be the case that $\odot$ preserves $C^*$-generation: if $A$ and $B$, as embedded, 
generate the $C^*$-algebras $\M_A$ and $\M_B$, then $A \odot B$ obviously generates $\M_{A \odot B}$.

Let us call an EJC $(A,\M_A)$ {\em reversible} iff $A$ is reversible in $\M_A$. Note that if $A$ is a 
simple EJC {\em standardly} 
embedded in $\M_A$, then $A$ is reversible iff $A = R_n, C_n$ or $Q_n$ for some $n$.

 We are going to show that the canonical composite of EJCs is a composite in the sense of Definition \ref{def: composites of EJAs}.  In order to do so, we will need the following Lemma on extensions 
of derivations.

\begin{lemma}\label{lemma: extension of derivations}
Let $D$ be a derivation of an EJA $A$, and $(A, \M_A)$ a faithful representation of $A$.  Then $D$ extends to a $*$-derivation of $\M_A$.  
\end{lemma}

The proof can be found in Appendix E.\footnote{In \cite{Upmeier}, Upmeier shows that every derivation of a reversible JC algebra $A$ acting on a complex Hilbert space $\cH$ extends to a $\ast$-derivation of the $C^{\ast}$ algebra of operators on $\cH$ generated by $A$. In finite dimensions, the hypothesis of reversibility is not needed. 
This is doubtless well known, but as we could find no reference to it, we have included a proof.}

\begin{proposition} \label{prop: canonical product composite}
If $(A, \M_A)$ and $(B, \M_B)$ are  
EJC-algebras, then their canonical tensor product $A \odot B$ is a 
composite in the sense of  Definition \ref{def: composites of EJAs}. 
\end{proposition}

\noindent{\em Proof:}   First we must show that $A \odot B$ is a dynamical composite, in the 
sense of Definition \ref{def: dynamical composites}.  In particular, it must 
be a composite of probabilistic models in the sense of Definition \ref{def: composites}. 
Conditions (a)-(c) of that definition are easily verified:   
That $u_{AB} = u_{A} \otimes u_{B}$ follows from the unitality of 
the embeddings $A \mapsto \M_A$, $B \mapsto \M_B$ and $AB \mapsto \M_{AB}$. 
For all $a, x \in (\M_A)_{\sa}$ and $b, y \in (\M_B)_{\sa}$, we have 
\[\langle a \otimes b | x \otimes y \rangle = \Tr(ax \otimes by) = \Tr(ax)\Tr(by) = \langle a |x \rangle \langle b | y \rangle\]
Thus,  for any states $\alpha = \langle a |$ and $\beta = \langle b |$, where $a \in A_+$ and $b \in B_+$, we have 
a state $\gamma = \langle a \otimes b |$ on $AB$ with $\gamma(x \otimes y) = \alpha(x) \beta(y)$ for 
all $x \in A_+$, $y \in B_+$.   

 That $A \odot B$ satisfies the additional conditions required
to be a \emph{dynamical} composite in the sense of Definition \ref{def: dynamical composites} is not trivial.  
Using Lemma \ref{lemma: extension of derivations} 
one can show that  any $\phi \in G(A)$ extends to an
 element $\hat{\phi} \in G((\M_A)_{\sa})$ that preserves $A$, and that
 $\hat{\phi} \otimes \1_{M_B}$ is an order-automorphism of $(\M_A
 \otimes \M_B)_{\sa}$ preserving $A \odot B$.  It follows that 
 $\hat{\phi} \otimes \hat{\psi} = (\hat{\phi} \otimes \1_{\M_B}) \circ (\1_{\M_A} \otimes \hat{\psi})$ 
 preserves $A \odot B$ as well.  {\awedit The details are presented in Appendix \ref{appendix: extending}.} 
It is not difficult to verify that 
 $(\hat{\phi} \otimes \hat{\psi})^{\dagger} = \hat{\phi}^{\dagger} \otimes \hat{\psi}^{\dagger}$. 
Thus $A \odot B$ satisfies
 conditions (a) and (b) of Definition \ref{def: composites of EJAs}.  Condition 
(c) is immediate from the definition of $A \odot B$.
$\Box$\\[0.3cm]
Combining this with Theorem \ref{thm: composites special}, we see that if $A$ and $B$ are simple reversible EJCs 
for which $A \hotimes B$ is also simple, then the {\redd canonical} and universal tensor products of $A$ and $B$ coincide. 
This covers all cases except those involving two factors of the form $C_n$ and $C_k$, and those involving 
$Q_2$ as a factor. In fact, many of the latter are covered by the following.

\begin{corollary}\label{cor: canonical composites with involutions}
Let $(A,\M_A)$ and $(B,\M_B)$ be {\hblue simple} reversible EJCs with $A$ generating
$M_A$ and $B$ generating $M_B$ as $\ast$-algebras.  Suppose $M_A$ and
$M_B$ carry involutions $\Phi$ and $\Psi$, respectively, with $\Phi$
fixing points of $A$ and $\Psi$ fixing points of $B$ (i.e.
  $A \subseteq \M_A^\Phi$ and $B \subseteq \M_B^\Phi$). Then $A \odot
B = (\M_{A} \otimes \M_{B})^{\Phi \otimes \Psi}_{\sa}$, the set of
self-adjoint fixed points of $\Phi \otimes \Psi$.
\end{corollary} 

\noindent{\em Proof:} By the preceding proposition,  
$A \odot B$ is a dynamical
composite of $A$ and $B$; hence, by Theorem \ref{thm: composites
  special} {\hblue and the simplicity of $A$ and $B$}, $A \odot B$ is universally reversible. Since $\Phi \otimes
\Psi$ fixes points of $A \otimes B$, it also fixes points of the
Jordan algebra this generates in $M_A \otimes M_B$, i.e., of $A \odot
B$. But then, by Proposition \ref{prop: UR fixed points}, $A \odot B$
is exactly the set of fixed points of $\Phi \otimes \Psi$. $\Box$\\[0.3cm]
 For example, the quaternionic bit $Q_2$ standardly embedded in $M_4(\C) = M_2(M_2(\C))$ as the set of 
block matrices of the form
\begin{equation}
	\begin{pmatrix} \hspace{0.23cm} a & b \\ -\bar{b} & \bar{a}
	\end{pmatrix}
\end{equation}
with $a$ self-adjoint and $b$ antisymmetric is exactly the fixed-point set of the involution $\Phi(x) = -J_{2}x^T\!J_{2}$, where with $\1_{2}$ the $2\times 2$ identity matrix
\begin{equation}
	J_{2}=\begin{pmatrix} \hspace{0.23cm}0 & \1_{2}\\ -\1_{2} & 0
	\end{pmatrix}\text{.}
\end{equation} 
Thus, $Q_2 \odot Q_2$ is 
the set of self-adjoint fixed points of $\Phi \otimes \Phi$ acting on $M_4(\C) \otimes M_4(\C) = M_{16}(\C)$. 
From this, it follows that $Q_2 \odot Q_2 \simeq R_{16} = M_{16}(\R)_{\sa}$. For details, see 
Appendix \ref{appendix: the quabit}. Notice that Corollary \ref{cor: canonical composites with involutions} 
applies equally to all simple, universally embedded EJCs, {\em and} to all {\em standardly} embedded 
EJCs {\em except} for $C_n$ (recalling 
here that no (complex-linear) involution on $M_n(\C)$ fixes all 
points of $M_n(\C)_{\sa}$).\\[0.3cm]
\noindent{\bf Canonical composites of standardly embedded, reversible EJAs} 
It follows from Proposition
\ref{prop: canonical product composite}, together with 
Theorem \ref{thm: simple composites ideals} that 
the canonical and universal tensor products coincide for simple, reversible EJCs whose universal 
tensor products are simple. This covers all cases except for those in which one factor is 
$M_2(\Q)_{\sa}$ (the quaternionic bit), and those in which both factors have the form $M_n(\C)_{\sa}$ for some $n$. 
Restricting attention to standardly embedded EJCs, these missing cases can be computed directly: see appendix \ref{appendix: the quabit} and \cite{Graydon16}. The results are summarized, {\mmagenta up to isomorphism}, in the following table.  (Only those products with a factor of $Q$ involve a nontrivial isomorphism; details of 
these will appear elsewhere.)  
Recall here the abbreviations $R_n = M_{n}(\R)_{\sa}$, $C_n = M_n(\C)_{\sa}$ and $Q_n = M_{n}(\Q)_{\sa}$. 
\[\begin{array}{c}
\begin{array}{c|ccc}
\odot & R_n & C_n & Q_n \\
\hline
R_m   & R_{mn} & C_{mn} & Q_{mn} \\
C_m   & C_{mn} & C_{mn} & C_{2mn} \\
Q_m   & Q_{mn} & C_{2mn} & R_{4mn} 
\end{array}\\
\\
\mbox{Table 3: $\odot${\hblue, up to isomorphism,} for standardly embedded, reversible EJCs}
\end{array} \]

\noindent{\bf Associativity} We now establish that the canonical
tensor product is associative, setting the stage for the construction
of symmetric monoidal categories of EJCs in Section \ref{sec: categories EJC}.

\begin{proposition}  \label{prop: associativity}
$\odot$ is associative. More precisely,  the associator mapping $\alpha : \M_{A} \otimes (\M_{B} \otimes \M_{C}) \rightarrow (\M_{A} \otimes \M_{B}) \otimes \M_{C}$ carries $A \odot (B \odot C)$ isomorphically onto 
$(A \odot B) \odot C$. \end{proposition}

The essence of the proof is the following Lemma.

\begin{lemma} \label{lemma: associativity}
Let $\M$ and $\N$ be $\ast$-algebras, and let $A$ and $B$ be subspaces  
of $\M_{\sa}$ and $\N_{\sa}$, respectively, {\mmagenta with $1 = 1_{\M} \in A$ and similarly for $B$.} Then $J(A \otimes J(B)) = J(A \otimes B)$\footnote{Note that on the left, $J(A \otimes J(B))$ refers to the Jordan subalgebra of $(A \otimes B)_{\sa}$ generated by $A \otimes J(B)$, where $J(B)$ refers to the Jordan subalgebra of $\N_{\sa}$ generated by {\mmagenta $B$}.}.\end{lemma}

\noindent{\em Proof of Lemma \ref{lemma: associativity}}: 
We make use of the following notation. If $X$
and $Y$ are subsets (note: not necessarily subspaces) of $\M$ and
$\N$, respectively, let
\[X  \jProd Y \ := \ \{\, a {\bblue \dot} b \ | \ a \in X, b \in Y \, \};\] 
similarly, $X + Y$ is the set of sums $a + b$ 
with $a \in X$, $b \in Y$, and, for $t \in \R$, $tX$ consists of multiples $ta$ of elements $a \in X$. 
Finally, we write $X \circledcirc Y$ for the set $\{ a \otimes b | a \in X, b \in Y\}$ of pure tensors 
of elements in $X$ and $Y$. Note that if $X$ and $Y$ happen to be subspaces of $\M$ and $\N$, respectively, 
then $J(X \potimes Y) = J(X \otimes Y)$. 

Now let $\cal N$ denote the set of all subsets $Y$ of $\N$ such that 
\begin{itemize} 
\item[(i)] $B \subseteq Y \subseteq J(B)$, and 
\item[(ii)]$A \potimes Y \subseteq J(A \potimes B)$.
\end{itemize} Note that $B \in {\cal N}$; hence, ${\cal N}$ is non-empty.  We are going to 
show first that $J(B) = \bigcup {\cal N}$. 

{\em Claim 1:} If $Y_1, Y_2 \in {\cal N}$, then the sets $Y_1  \jProd Y_2$, $Y_1 + Y_2$ and $tY_1$ ($t$ any real 
scalar) belong to ${\cal N}$. 

To see this, suppose that $y_1 \in Y_1 \in {\cal N}$ and $y_2 \in Y_2 \in {\cal N}$. 
Then for any $a \in A$, we have 
\begin{eqnarray*}
a \otimes (y_1 {\bblue \dot} y_2 ) & = & a \otimes \frac{1}{2} (y_1y_2 + y_2 y_1) \\
& = & \frac{1}{2} [(a \otimes y_1 y_2) + (a \otimes y_2 y_1)]\\
& = & \frac{1}{2} [(a \otimes y_1)(1 \otimes y_2) + (1 \otimes y_2)(a \otimes y_1) ]\\ 
& = & (a \otimes y_1) {\bblue \dot} (1 \otimes y_2).
\end{eqnarray*}
Since $A \potimes Y_1$ and $A \potimes Y_2$ are both, by assumption,
contained in $J(A \potimes B)$, the last expression defines an element
of $J(A \otimes B)$. Since $a$ is arbitrary here, all pure tensors
from $A \potimes (Y_1  \jProd Y_2)$ belong to $J(A \potimes B)$. 
{\redd Since
the latter is a subspace of $(\M \otimes \N)_{\sa}$, it follows that
linear combinations of pure tensors from $A \potimes (N_1  \jProd N_2)$
are contained in $J(A \potimes B)$ as well.} Finally, notice that $B
\subseteq Y_1$ and $B \subseteq Y_2$ implies that $B \subseteq Y_1
 \jProd  Y_2$ (for instance, express $b \in B$ as $b {\bblue \dot} 1$). The
corresponding claims for $Y_1 + Y_2$ and $t Y_1$ are straightforward,
since $a \otimes (y_1 + y_2) = (a \otimes y_1) + (a \otimes y_2)$ and
$a \otimes ty_1 = t(a \otimes y_1)$.

Now let $J = \bigcup {\cal N}$. {\em Claim 2:} $J = J(B)$. Since $B
\subseteq J \subseteq J(B)$, it's enough to show that $J$ is a Jordan
subalgebra of $\N_{\sa}$. Let $y_1, y_2 \in J$: then for some $Y_1, Y_2
\in {\cal M}$, we have $y_1 \in Y_1$ and $y_2 \in Y_2$.  By Claim 1,
$y_1 {\bblue \dot} y_2 \in Y_1  \jProd Y_2 \in {\cal N}$, $y_1 + y_2 \in {\mmagenta Y}_1
+ Y_2 \in {\cal N}$, and $t y_1 \in tY_1 \in {\cal N}$; hence, $y_1
{\bblue \dot} y_2$, $y_1 + y_2$ and $t y_1$ belong to $J$. In other words,
$J$ is closed under scalar multiplication, addition and the Jordan
product. This proves Claim 2.

We now use the fact that $J(B) = \bigcup {\cal N}$ to show that $J(A
\otimes J(B)) = J(A \otimes B)$. Let $\tau = \sum_i a_i \otimes y_i$
be an arbitrary element of $A \otimes J(B)$. Since $J(B) = \bigcup {\cal
  N}$, for each $i$, $y_i \in Y_i$ for some $Y_i \in {\cal N}$. But
then, by definition of ${\cal N}$,  {\mmagenta for every $i$} $a_i \otimes y_i \in A {\mmagenta \potimes} Y_i
\subseteq J(A \otimes B)$; since the latter is a subspace of $\M
\otimes \N$, $\tau \in J(A \otimes B)$ as well. Thus, $A \otimes J(B)
\subseteq J(A \otimes B)$.

It now follows that $A \otimes B \subseteq 
A \otimes
J(B) \subseteq J(A \otimes B)$, hence, that $J(A \otimes B) \subseteq
J(A \otimes J(B)) \subseteq J(A \otimes B)$. $\Box$\\[0.3cm]
\noindent{\em Proof of Proposition \ref{prop: associativity}:} Let $(A,\M_A)$,
$(B,\M_B)$ and $(C,\M_C)$ be EJC-algebras. We
need to show that the associator mapping $\alpha : \M_A \otimes (\M_B
\otimes \M_C) \rightarrow (\M_A \otimes \M_B) \otimes \M_C$ carries $A
\odot (B \odot C)$ onto $(A \odot B) \odot C$.  Applying Lemma {\mmagenta 5.6}, {\mmagenta 
(and noting that $A, B \odot C$, $A \odot B$ and $C$ all contain the relevant units)}, we
have
\[ A \odot (B \odot C) = J(A \otimes J(B \otimes C)) = J(A \otimes (B \otimes C))\]
and 
\[ (A \odot B) \odot C = J(J(A \otimes B) \otimes C) = J((A \otimes B) \otimes C).\] 
The associator mapping is a $\ast$-isomorphism, and carries $A \otimes (B \otimes C)$ to 
$(A \otimes B) \otimes C$. Hence, it also carries $J(A \otimes (B \otimes C))$ onto 
$J((A \otimes B) \otimes C)$. $\Box$\\[0.3cm] 
\noindent{\bf Direct sum of EJCs} If $(A,\M_A)$ and $(B,\M_B)$ are two embedded EJAs, define 
$(A,\M_A) \oplus (B,\M_B) = (A \oplus B, \M_A \oplus \M_B)$, where the embedding of $A \oplus B$ in $\M_A \oplus \M_B$ is the obvious one.  {\awedit Notice that by Proposition \ref{prop: Cu additive}, for universally embedded EJCs we 
have $(A,\Cu(A)) \oplus (B,\Cu(B)) \simeq (A \oplus B, \Cu(A \oplus B))$. }

{\mmagenta \begin{lemma}\label{Distributivity} For all EJCs $A$, $B$ and $C$, we have 
\[A \odot (B \oplus C) = (A \odot B) \oplus (A \odot C).\]
\end{lemma}

\noindent{\em Proof:} } One can easily check that 
for sets $X \subseteq \M_A$ and $Y \subseteq \M_B$, $J(X \oplus Y) = J(X) \oplus J(Y)$. Using this, and 
the distributivity of tensor products over direct sums in the contexts of vector spaces and $\ast$-algebras, we have 
\begin{eqnarray*}
A \odot (B \oplus C) = J(A \otimes (B \oplus C)) & = & J((A \otimes B) \oplus (A \otimes C)) \\
& = & J(A \otimes B) \oplus J(A \otimes C) \\
& = & (A \odot B) \oplus (A \odot C)
\end{eqnarray*} 
(where $J$ refers variously to generated Jordan subalgebras of $\M_{A}
\otimes (\M_{B} \oplus \M_{C})$, $(\M_{A} \otimes \M_{B}) \oplus
(\M_{A} \otimes \M_{C})$, $\M_{A} \otimes \M_{B}$, and $\M_{A} \otimes
\M_{C}$). {\mmagenta $\Box$}


\section{Monoidal Categories of EJC-algebras}
\label{sec: categories EJC}

As discussed in Section 2, a physical theory ought to be represented
by a category, in which objects correspond to systems and
morphisms to physical processes. An obvious candidate for a category
in which objects are embedded JC-algebras is the following. {\newawe Recall here that a linear map $\phi: \M_A \rightarrow \M_B$ is \emph{completely positive} 
if the map $\phi: M_A \otimes C_n \rightarrow M_B \otimes C_n$ is positive for each positive integer $n$.}

\begin{definition}{\em Let $(A,\M_A)$ and $(B,\M_B)$ be EJC-algebras. A linear mapping 
 $\phi : \M_A \rightarrow \M_B$ 
is {\em Jordan preserving} iff 
$\phi(A) \subseteq B$. The category $\EJC$ has, as objects, 
EJC-algebras, and, as morphisms, completely positive Jordan-preserving mappings. }
\end{definition}

In view of Proposition \ref{prop: associativity}, one might guess that
$\EJC$ is symmetric-monoidal under $\odot$. There is certainly a
natural choice for the monoidal unit, namely $I = (\R, \C)$. However,
the following examples show that tensor products of $\EJC$ morphisms
are generally not morphisms.

\begin{example}\label{ex: no states} {\em Let $(A,\Cu(A))$ and 
$(B,\Cu(B))$ be simple, nontrivial, universally embedded EJCs, and suppose 
that $B$ is not UR (e.g., the spin factor $V_4$). 
{\redd Suppose, further, that $A \hotimes B$ is irreducible --- 
for instance, let $A = R_n$ for any $n$.}
 Let
$\hat{B}$ be the set of fixed points of the canonical involution
$\Phi_B$ (cf. Definition \ref{def: canonical involution}). Then by Theorem \ref{thm: simple composites ideals}, $A \odot B = A \hotimes B$, the set of
fixed points of $\Phi_A \otimes \Phi_B$. In particular, $u_A \otimes
\hat{B}$ is contained in $A \odot B$. Now let $\alpha$ be a state on
$\Cu(A)$.  It is easily verified that every state is completely positive,\footnote{Letting $\alpha$ be a state on $A$ and therefore
on the complex matrix algebra $\M_A$, $\beta$ a positive functional on any complex matrix algebra $\M_B$, and $f \in 
(\M_A \otimes \M_B)_+$, 
we have $\beta((\alpha \otimes \id_{B})(f))  = (\alpha \otimes \beta)(f)$.  This is nonnegative since $\alpha \otimes \beta$ is 
a positive functional.  Since this holds for all positive $\beta$, $(\alpha \otimes \id_{B})(f)$ is positive, i.e. (since $\M_B$
was arbitrary) $\alpha$ 
is completely positive.}
and it is trivial that it is Jordan-preserving, and so, a
morphism in $\EJC$. But
\[(\alpha \otimes \id_{B})(u_A \otimes \hat{B}) = \alpha(u_A)\hat{B} = \hat{B},\]
which {\magenta by Proposition \ref{prop: UR fixed points} 
is larger than $B$ because $B$ is not UR}. So ${\hblue \alpha} \otimes \id_{B}$ isn't Jordan-preserving.}
\end{example}

The next example is similar:

\begin{example}\label{ex: no states 2} {\em Let $C_n = M_n(\C)_{\sa}$ and $R_k = M_k(\R)_{\sa}$, as usual. 
Consider these as standardly embedded, i.e., consider the EJCs $(C_n, M_n(\C))$ and $(R_k, M_k(\C))$, where 
in the latter, $R_k$ is embedded as the set $k \times k$ complex matrices with real entries, 
i.e., the set of self-adjoint symmetric $k \times k$ complex matrices. Then we have 
\[(C_n, M_n(\C)) \odot (R_k, M_k(\C)) = (C_{nk}, M_{nk}(\C))\]
where the embedding of $C_{nk}$ in $M_{nk}(\C)$ is the {\redd standard} one. Now let $\alpha$ be a state on 
$M_n(\C)$, and let $b$ be any self-adjoint matrix in $M_k(\C)$. Then $\1_n \otimes b$ is self-adjoint 
in $M_{n}(\C) \otimes M_{k}(\C) = M_{nk}(\C)$, and $(\alpha \otimes \id_{M_{k}}(\C))(\1_n \otimes b) = b$. Since 
$b$ needn't belong to $R_k$, the mapping $\alpha \otimes \id_{M_k}(\C)$ isn't Jordan-preserving. }
\end{example}

\subsection{Completely Jordan Preserving Maps}

Examples \ref{ex: no states} and \ref{ex: no states 2} suggest the following adaptation 
of the notion of complete positivity to our setting. 

\begin{definition} 
{\em 
A linear mapping $\phi : (A,\M_A) \rightarrow (B,\M_B)$ is 
\emph{completely Jordan preserving} (CJP)
iff, for any embedded JC-algebra $(C,\M_C)$, the mapping $\phi \otimes
\id_{\M_C}$ is positive, and takes $A \odot C$ into $B \odot C$.  
}
\end{definition}

{\redd Note that this implies that $\phi$ is both Jordan-preserving (take $C = \R$) and completely positive (take 
$C = C_n$ for any $n$). }

\begin{lemma}\label{lemma: CJP1}  If $\phi : \M_A \rightarrow \M_B$ is CJP, then for any $(C,\M_C)$, $\phi \otimes \id_{\M_{C}}$ is again CJP.\end{lemma}

\noindent{\em Proof:}  If $(D,\M_D)$ is another embedded JC-algebra, 
consider $(C \odot D, \M_{C} \otimes \M_{D})$. The associativity 
of $\odot$ gives us  
\[(\phi \otimes \id_{\M_{C}}) \otimes \id_{\M_D} = \phi \otimes (\id_{\M_C \otimes \M_D}).\]
Since $\phi$ is CJP, the latter {\redd carries} $A \odot (C \odot D)$ into $B \odot (C \odot D)$. 
Since $\odot$ is associative, this tells us that 
{\redd $(\phi \otimes \id_{C}) \otimes \id_{D}$ carries} $(A \odot C) \odot D$  into $(B \odot C) \odot D$. 
$\Box$\\[0.3cm]
While all CJP morphisms are CP, 
Example \ref{ex: no
  states} shows that the converse is false. On the other hand, the
class of CJP morphisms is still quite large.

\begin{example} {\em Let $\phi : \M_A \rightarrow \M_B$ be a $\ast$-homomorphism taking $A$ to $B$.  It
is easily verified that 
$\phi \otimes \id_{\M_C}$ is again a $\ast$-homomorphism.  Since  $*$-homomorphisms preserve the 
concrete Jordan product,\footnote{That is, a $*$-homomorphism $\varphi$ satisfies $\varphi((xy + yx)/2) = (\varphi(x)\varphi(y) + \varphi(y)\varphi(x))/2$.} $\phi \otimes \id_{\M_C}$ 
    takes the Jordan subalgebra generated by $A \otimes C$ to that
    generated by $B \otimes C$, i.e., sends $A \odot C$ into $B \odot
    C$. So $\phi$ is CJP. }\end{example}

\begin{example} 
{\em Let $(A,\M_A)$ be an embedded JC-algebra, and let $a \in A$.  The
  mapping $U_a : \M_A \rightarrow \M_A$ given by $U_a(b) = aba$ is 
  {\redd completely} positive;  recall that its restriction to $(\M_A)_{sa}$ can be expressed in terms of the Jordan product on
  $(\M_A)_{sa}$ by
\[U_a(b) = 2 a  \dot (a \dot b) - (a^2)  \dot b\]
Since $A$ is a Jordan subalgebra of
$(\M_A)_{\sa}$, this makes it clear that $U_a(b) \in A$ for all $a, b \in A$.  Thus, $U_a$ is a
morphism.  Now if $(C,\M_C)$ is another embedded JC-algebra, one easily sees\footnote{Observe that 
 that $(U_a \otimes U_c) (x \otimes y) = axa \otimes cyc = (a \otimes c)(x \otimes y)(a \otimes c) 
= U_{a \otimes c} (x \otimes y)$ and that the ``pure tensors" $x \otimes y$ span $\M_A \otimes \M_C$.}
that
\[U_{a} \otimes U_{c} = U_{a \otimes c}\]
for all $c \in \M_C$; in particular, 
\[U_a \otimes \id_{\M_C} = U_a \otimes U_{\1} = U_{a \otimes \1}.\]
Since $a \otimes \1 \in A \odot C$, it follows that $U_{a}$ is
CJP. }\end{example}

\begin{proposition} 
\label{prop: CJP composes and tensors}
Let $(A,\M_A)$, {\redd $(A', \M_{A'})$, $(B,\M_B)$, $(B', \M_{B'})$ and and $(C,\M_{C})$}  be EJC-algebras.
\begin{itemize} 
\item[(i)] If $\phi : \M_{A} \rightarrow \M_{B}$ and $\psi : \M_{B} \rightarrow \M_{C}$ 
are CJP, then so is $\psi \circ \phi$, and 
\item[(ii)] if $\phi : \M_{A} \rightarrow \M_{B}$ and $\psi : \M_{A'} \rightarrow \M_{B'}$ are CJP, 
then so is $\phi \otimes \psi : \M_{A \odot A'} = \M_{A} \otimes \M_{A'} \rightarrow \M_{B} \otimes \M_{B'} = \M_{B \odot B'}$. 
\[\phi \otimes \psi = (\phi \otimes \id_{\M_{B'}}) \circ (\id_{\M_{A}} \otimes \psi)\]
is again CJP. 
\end{itemize} 
\end{proposition}

\noindent{\em Proof:}  (i) is straightforward, and (ii) follows from (i) and Lemma \ref{lemma: CJP1}.  $\Box$\\[0.3cm] 
{\awedit As Example \ref{ex: no states} illustrates, for certain non-UR EJCs $A$, there are {\em no} 
non-zero CJP maps from $A$ to $I$. In particular, normalized states on such an $A$ will not count as 
morphisms in the category $\CJP$. 
Operationally, this means that  the preparation of states and registration of effects need not correspond to any physical process in $\CJP$. We now show that this defect can sometimes be remedied by relativizing the definition of CJP mappings to 
certain sub-categories of $\CJP$.} 
\subsection{Relatively CJP Mappings} 

\begin{definition}{\em Let $(A,\M_A)$ and $(B,\M_B)$ belong to $\Cat$. 
 A positive linear mapping $\phi : \M_A \rightarrow \M_B$ is {\em relatively CJP} with respect to $\Cat$ 
if, for all $(C,\M_C) \in \Cat$, the mapping $\phi \otimes \id_{\M_{C}}$ is positive and maps 
$A \odot C$ into $B \odot C$. We denote the set of all such maps by $\CJP_{\Cat}(A,B)$. } 
\end{definition}

\begin{proposition}\label{prop: Relatively CJP categories}
Let $\Cat$ be any class of EJCs closed under the formation of the canonical tensor product and containing 
the trivial EJC $I$. Then $\Cat$ is a symmetric monoidal category with completely positive relatively CJP mappings as morphisms 
and the canonical tensor product as the monoidal product.
\end{proposition}

\noindent{\awedit \em Proof:} Exactly as in the proof of Proposition \ref{prop: CJP composes and tensors}, 
 we see that if $A,B,C \in \Cat$ and $\phi \in \CJP_{\Cat}(A,B)$ and $\psi \in \CJP_{\Cat}(B,C)$, then $\psi \circ \phi \in \CJP_{\Cat}(A,C)$, and also that if $A,B,C,D \in \Cat$ and 
$\phi \in \CJP_{\Cat}(A,B)$ and $\psi \in \CJP_{\Cat}(C,D)$, 
then $\phi \otimes \psi \in \CJP_{\Cat}(A \otimes C,  B \otimes D)$. In other words, $\Cat$ becomes a symmetric 
monoidal category with relatively CJP mappings as morphisms. We denote this category by $\CJP_{\Cat}$. 
 Note that $\phi$'s being relatively CJP with respect to $\Cat$ 
implies that $\phi$ is CP.  For, either $\Cat$ contains an $A$ such that $\M_A$ is noncommutative, or it does not.
If all $\M_A$ are commutative, then all positive maps between them are CP.  On the other hand, if some $\M_A$ is noncommutative, 
it contains a summand isomorphic to $M_n(\C)$ for some $n \ge 2$.  By closure under $\odot$, it contains $C$ with a summand
equal to $M_m(\C)$ for arbitrarily large $m$ (just take $C = A \odot A \odot \cdots \odot A$ with sufficiently many factors of $A$).
Since the definition of ``completely Jordan preserving relative to $\Cat$" requires $\phi \otimes \id_C$ be positive for all $C$,
and positivity on a matrix algebra requires positivity on all summands, we have that for every 
positive integer $n$, there exists $m \ge n$ such that 
$\phi \otimes M_m(\C)$ is positive.  That is sufficient for $\phi$ to be completely positive.

\begin{example}[The category $\CQM$]\label{ex: CQM} 
{\em Let $\Cat$ be the class of hermitian parts of 
complex $*$-algebras with standard embeddings. Then $\phi$ 
belongs to $\mbox{CJP}_{\Cat}$ 
iff $\phi$ is CP.  Evidently, {\redd this category is} essentially orthodox, mixed-state QM with
  superselection rules. From now on, we shall call this category $\CQM$.
  }
\end{example} 

\begin{example}[The category $\RSE$] \label{ex: RSE}{\em 
Recall that the standard embeddings for simple, reversible Jordan algebras represent 
\begin{itemize} 
\item[(a)] $R_{n}$ as the set of self-adjoint elements of $M_{n}(\C)$ having real entries; 
\item[(b)] $C_{n}$ as {\hblue itself, that is,} the set of all self-adjoint elements of $M_{n}(\C)$; 
\item[(c)] $Q_{n}$ as the set of $2n \times 2n$ complex matrices having  symplectic $2 \times 2$ block structure.
\end{itemize} 
For $A = R_n, C_n$ or $Q_n$, let $\St(A)$ denote the matrix algebra, and $\sigma_A : A \rightarrow \St(A)$ the embedding, 
given above.  
Now let $\Cat_o$ denote the class of reversible, simple EJCs $(A,\M_A)$ that are {\em standardly embedded up 
to isomorphism}, in the sense that there exists a $\ast$-isomorphism $\phi : \M_A \simeq \St(A)$ acting on the 
identity on $A$ (or, more exactly, such that $\phi \circ \pi_A = \sigma_A$. )  
It is straightforward to check 
that $\Cat_o$ is closed under the canonical tensor product \cite{Graydon16}. Now 
let $\Cat$ consist of direct sums of EJCs belonging to $\Cat_o$. Since the canonical tensor product distributes over direct sums, 
by Lemma \ref{Distributivity}, $\Cat$ is also closed under $\odot$. By the remarks above, then, $\CJP_{\Cat}$ is a 
symmetric monoidal category, which we call $\RSE$.   
This represents a kind of unification of finite-dimensional real, complex and quaternionic quantum mechanics, in so far as its objects are the Jordan algebras associated with real, complex 
  and quaternionic quantum systems, and direct sums of these. Moreover, as restricted to complex (or real) systems, 
  its compositional structure is the standard one. However, Example \ref{ex: no states 2} shows that not every quantum-mechanical 
  process --- in particular, not even processes that prepare states --- will count as a morphism in $\RSE$, 
  so this unification comes at a high cost. }
\end{example} 

\begin{example}[The category $\URUE$]\label{ex: URUE} {\em 
Let $\Cat$ be the class of universally reversible EJCs with universal embeddings. {\mmagenta 
By  Proposition \ref{prop: canonical product composite} and Theorem \ref{thm: composites special}, if $A$ and $B$ are simple, then $A \odot B$ is UR. It is straightforward that 
$\Cstar(A \oplus B) = \Cstar(A) \oplus \Cstar(B)$ (Appendix \ref{appendix: direct sums}, Proposition A.6) and that 
$A \oplus B$ is UR iff $A$ and $B$ are UR (Corollary A.7). Finally, 
we know that the canonical tensor product distributes over direct sums (Lemma \ref{Distributivity}).  Putting these observations 
together, we see that $\Cat$ is closed under $\odot$. We denote the category $\CJP_{\Cat}$ by $\URUE$.}} \end{example}

{\mmagenta Morphisms in $\URUE$ can be characterized more 
concretely: they are precisely those CP maps that intertwine the canonical involutions 
on universal $\Cstar$ algebras. 

\begin{proposition}\label{URUEintertwiners} Let $A$ and $B$ be UR, universally embedded EJCs. 
Then a mapping $\phi : \Cstar(A) \rightarrow \Cstar(B)$ belongs to $\URUE$ iff it is 
CP, and satisfies $\phi \circ \Phi_{A} = \Phi_{B} \circ \phi$. \end{proposition} 
 
\noindent{\em Proof:} First, we show that CP intertwiners are relatively CJP. 
{\hblue Let $\phi : A \rightarrow B$ be a CP intertwiner.  We are to show that for every $C \in \Cat$, 
$\phi \otimes \id_C: A \odot C \rightarrow B \odot C$ is positive and Jordan-preserving.  It is 
positive because $\phi$ is CP.  By  Corollary \ref{cor: canonical composites with involutions}, we have $A \odot C = (\M_A \otimes \M_C)^{\Phi_A \otimes \Phi_C}$ 
and $B \odot C = (\M_B \otimes \M_C)^{\Phi_B \otimes \Phi_C}$.  Intertwining then immediately gives that 
$\phi \otimes \id_C$ is Jordan-preserving:  suppose $x \in A \odot C$, i.e. $(\Phi_A \otimes \Phi_C) (x) = x$.
Then $(\Phi_B \otimes \Phi_C) \circ (\phi \otimes \id_C (x)) = ((\Phi_B \circ \phi) \otimes \Phi_C )(x) = 
(\phi \circ \Phi_A) \otimes \Phi_C)(x) = ((\phi \otimes \id) \circ (\Phi_A \otimes \Phi_C))(x) = (\phi \otimes \id_C)(x)$, 
i.e. $(\phi \otimes \id_C)(x)$ is fixed by $\Phi_B \otimes \Phi_C$, i.e. it is in $B \odot C$.}

For the converse, suppose 
$\phi$ is relatively CJP for the class of UR, universally embedded EJCs. Note that 
$\Cstar(A)_{\sa}$ decomposes as the direct sum of $\Phi_A$'s $+1$ and $-1$ eigenspaces. 
Moreover, as $A$ is UR, the $+1$ eigenspace is precisely $A$. Now, 
$\phi \circ \Phi_A = \Phi_B \circ \phi$ trivially on $A$. We now claim 
that $\phi$ also intertwines $\Phi_A$ and $\Phi_B$ on the $-1$-eigenspace. 
To see this, note that $\phi_{A} \tensor \id_{C}$ is also relatively CJP 
for any object $C$ in $\URUE$. If $a$ and $c$ belong to the 
$-1$-eigenspaces of $\Phi_A$ and $\Phi_C$, respectively, then $a \otimes c$ belongs 
to the $+1$ eigenspace of $\Phi_A \otimes \Phi_C$, whence, 
\[(\Phi_{B} \otimes \Phi_{C})(\phi \otimes \id)(a \otimes c) = (\phi \otimes \id)(\Phi_{A} \otimes \Phi_C)(a \otimes c).\]
Since $\Phi_{C}(c) = -c$, this reduces to 
\[-(\Phi_{B}(\phi(a)) \otimes c) = -(\phi(\Phi_A(a)) \otimes c).\]
With $c$ non-zero, this implies that $\Phi_{B}(\phi(a)) = \phi(\Phi_A(a))$ for all $a$ in the 
$-1$ eigenspace of $\Phi_A$, as required. Thus, $\phi$ intertwines $\Phi_A$ and $\Phi_B$ on 
all of $\Cstar(A)_{\sa}$, and hence, by the Cartesian decomposition, on all of $\Cstar(A)$. $\Box$ }\\[0.3cm]
{\awedit As we will see below ( Corollary \ref{cor: things that are InvQM morphisms}), maps in $\URUE(A,I)$ and $\URUE(I,A)$ correspond exactly 
to states and effects, respectively, on $A$. In this respect, t}he category $\URUE$ is closer {\hblue than $\RSE$} to being a legitimate ``unified" 
  quantum theory, although it omits the quaternionic bit (which is reversible, but not 
  universally so).  Even as 
  restricted to complex quantum systems, however, it differs from orthodox QM 
  in two interesting ways. First, and most conspicuously, the tensor product 
  is not the usual one: $C_n \hotimes C_k = C_{nk} \oplus C_{nk}$, rather 
  than $C_{nk}$.   Secondly, it allows some processes that orthodox QM does not. 
In particular, the mapping  
  on $\Cu(C_n) = M_n(\C) \oplus M_n(\C)$ that swaps the two summands is a morphism {\mmagenta in $\URUE$}. Since 
  the image of $C_n$ in $\Cu(C_n)$ consists of pairs $(a,a^T)$, this mapping
  effects the transpose automorphism on $C_n$. This is not 
  permitted in orthodox QM, as the transpose is not a CP mapping on 
  $M_n(\C)$.  (This causes no difficulty with positivity in the context of $\URUE$, where the 
  tensor product is different.)

  In spite of its divergences from orthodoxy, $\URUE$ is in many respects a 
  well-behaved probabilistic theory. {\mmagenta {\hblue Proposition} \ref{URUEintertwiners} suggests a way in which 
  it can be improved, which we explore in the next section. }

\subsection[The Category InvQM]{\bf The Category $\InvQM$} 

Recall that we write $\M^{\Phi}$ for the set of fixed-points of an involution $\Phi$  on a complex $\ast$-algebra $\M$. 
Notice that the involution $\Phi$ on $M_n(\C)$ with  $R_n = M_n(\C)^{\Phi}_{\sa}$ (the transpose) and  
on $M_{2n}(\C)$ with $Q_n = M_{2n}(\C)^{\Phi}_{\sa}$, namely $\Phi(a) = -Ja^{T}\! J$ where $J$ is the unitary  in Eq.~\eqref{bigJ}, are both unitary with respect to the 
trace inner product $\langle a, b \rangle := \Tr(ab^{\ast})$\footnote{\redd We follow the convention that 
inner products are linear in the {\em first} argument.}. The canonical involution on $\Cu(C_n) = M_n(\C) \oplus M_n(\C)$, namely, 
the mapping $(a,b) \mapsto (b^T, a^T)$, {\redd is likewise unitary}. 

\begin{definition} {\em Call an EJC $(A,\M_A)$ {\em involutive} iff there exists an involution $\Phi$ on $\M_A$, 
{\awedit unitary with respect to the trace inner product on $\M_A$}, with $A = (\M_A)^{\Phi}_{\sa}$.  
Let $\InvQM$ denote the category in which objects are involutive EJC-algebras, 
and in which a morphism $\phi : A \rightarrow B$ is a CP mapping $\phi : \M_{A} \rightarrow \M_B$ intertwining 
$\Phi_{A}$ and $\Phi_{B}$, i.e, $\Phi_{B} \circ \phi = \phi \circ \Phi_{A}$.                                                              
}
\end{definition}

As pointed out earlier, the condition that $A$ be the set of
  self-adjoint fixed points of an involution makes $A$ reversible in
  $\M_A$. Thus, the class of involutive EJC-algebras 
  contains no ``higher'' ($n = 4$ or $n > 5$)
  spin factors. In fact, {\hblue up to isomorphism}  
  it contains exactly {\redd direct sums of} the universally
  embedded UR EJCs $(A,\Cu(A))$ where $A = R_n, C_n$, with $n$ arbitrary, 
  or $Q_n$ with $n  > 2$, {\em together with} the {\em standardly} embedded quabit, i.e.,
  $(Q_2,M_4(\C))$, with $Q_2$ the self-adjoint fixed points of the
  involution $\Phi(a) = -Ja^{T}J$ discussed in Section \ref{subsec:
    reps of EJAs}.  In other words, $\InvQM$ includes 
     exclusively {\em
    quantum} systems over the three division  algebras $\R, \C$ and $\Q$,
  albeit with the complex systems represented in their universal
  embeddings. Note that $(\R, M_1(\R))$ counts as an involutive EJC:
  since $M_1(\R) = \R$ is commutative, the identity map provides the
  necessary involution.

{\mmagenta {\hblue Proposition} \ref{URUEintertwiners} tells us that $\URUE$ is a full subcategory of $\InvQM$, obtained by 
omitting the quabit. }  {\mmagenta On the other hand, 
it} is easy to see that CP mappings $\M_A \rightarrow \M_B$ intertwining $\Phi_A$ and $\Phi_B$ are automatically 
relatively CJP for the class of involutive EJCs. By Corollary \ref{cor: canonical composites with involutions} 
the canonical tensor product of involutive
EJCs is again involutive, as 
\[A \odot B = (\M_{A} \otimes \M_{B})^{\Phi_A \otimes \Phi_B}_{\sa}\]
for all $A, B \in \InvQM$.  Proposition 5.3 implies 
that $A \odot B$ is a  composite of $A$ and $B$ {\hblue in the sense of Definition \ref{def: composites of EJAs}}. 
Composites and tensor products of intertwining CP maps 
are also intertwining CP-maps (as are associators, unit-introductions and the swap mapping), so $\InvQM$ is a symmetric monoidal category --- indeed, a monoidal subcategory of the category 
of involutive EJC-algebras and relatively CJP maps.  {\hblue In fact, these two categories are equal, because every step of the
proof of Proposition \ref{URUEintertwiners} is valid for $\InvQM$, indeed, for any category of the form 
$\CJP_\Cat$, where $\Cat$ is an $\odot$-closed subset of (the set of objects of) $\InvQM$.  So we have: 
\begin{proposition}
Let $\Cat$ be any set of involutive EJCs closed under $\odot$.  Then $\CJP_\Cat$ is a symmetric monoidal category, 
with hom-sets $\CJP_{\Cat}(A,B) = \{\phi: \M_A  \rightarrow \M_B: \phi \mbox{ is completely positive and } \phi \circ \Phi_A 
= \Phi_B \circ \phi \}$.
\end{proposition}

In other words, $\CJP_\Cat$ for such a $\Cat$ is a full monoidal subcategory of $\InvQM$.  In particular, 
$\InvQM = \CJP_{\text{ob}\mathbf{InvQM}}$.  
}

In the special case in which $A$ and $B$ are universally embedded complex quantum systems, say 
$A = C_n$ and $B = C_k$, we have $\M_{A} = \Cu(C_n) = M_n(\C) \oplus M_n(\C)$ and 
similarly $\M_{B} = M_k(\C)\oplus M_k(\C)$. The involutions $\Phi_A$  
are given by $\Phi_{A}(a,b) = (b^T, a^T)$, and similarly for $\Phi_B$. In 
this case, the interwining CP-maps are sums of mappings of the two forms: $(a, b) \mapsto (\phi(a), \phi^T(b))$ and $(a, b) \mapsto 
(\phi^T(b), \phi(a))$, where $\phi$ is a CP mapping $M_n(\C) \rightarrow M_k(\C)$ 
and $\phi^T$ is determined by the condition  
$\phi^T(x^T) = (\phi(x))^T$, i.e. $\phi^T := T \circ \phi \circ T$.

{\mmagenta Two} special cases of $\InvQM$-morphisms are worth emphasising:

{\hblue 
\begin{corollary}\label{cor: things that are InvQM morphisms} Let $(A,\M_A)$ belong to $\InvQM$. Then
\begin{itemize} 
\item[(a)] $\InvQM(I,A)$ is the set of linear mappings $\phi_a : \R \rightarrow \M_A$, for each $a \in A_+$, 
where $\phi_a$ is the map determined by $\phi_a(1) = a$; 
\item[(b)] $\InvQM(A,I)$ is the set of positive linear functionals (including, in particular, every state) on $\M_A$ of the form $| a \rangle$,  $a \in A_+$. 
\item[(c)] Each of $\InvQM(I,A)$ and $\InvQM(A,I)$ is a cone affinely isomorphic to $A_+$.
\end{itemize}	 \end{corollary}

\noindent{\em Proof:} A linear map $\phi: I \rightarrow M_A$ is determined by $\phi(1)$.  Let 
$\phi_a \in \InvQM(I,A)$ be the map with $\phi(1)=a$.  Complete positivity of $\phi_a$ is equivalent
to $a \in (\M_A)_+$,\footnote{{\hblue Complete positivity certainly implies $a \in (\M_A)_+$, since positivity 
is equivalent to $a \in (\M_A)_+$.  On the other hand, for any $C$, $\phi_a \otimes \id_C: \M_I  \otimes \M_C 
\rightarrow \M_A \otimes \M_C$ acting on $x \in (\M_I \otimes \M_C)_+ \iso (\M_C)_+$ just gives $a \otimes x$, which is positive iff $a \in (\M_A)_+$.  }} and intertwining, to the condition that $a \in A$.\footnote{{\hblue Noting that $\Phi_I = \id: \M_I \rightarrow \M_I$, we see that 
$\phi_a \circ \Phi_I = \phi_a$,  so intertwining says that $\Phi_A \circ \phi_a = \phi_a$.  Since a CP map intertwines
iff it does so on the Hermitian part of its domain, intertwining is equivalent to 
$\Phi_A \circ \phi_a (\lambda) = \phi_a(\lambda)$ holding for all $\lambda \in \R$, 
i.e. $\lambda \Phi_A(a) = \lambda a$, which by involutiveness
is equivalent to $a \in A$.}}
So $a  \in (\M_A)_+ \intersect A$, which by Proposition \ref{prop: embedded order} is equal to $A_+$. 
 
The completely positive maps $\M_A \rightarrow \M_I$ are the maps $\ket{a}$, $a \in (\M_A)_+$, 
i.e. $x \mapsto \langle x | a \rangle$, where the inner product is the canonical one on $\M_A$.
(In particular, $a$ is Hermitian.)   
Recall that $\Phi_I = \id_\C$ so intertwining, for such a map, means that $\langle \Phi_A(x) | a \rangle = \langle x | a \rangle$.  Unitarity of $\Phi_A$ 
implies 
$\langle \Phi_A(x), a \rangle = \langle x, \Phi_A(a) \rangle$, so
intertwining, for Hermitian $x$,
becomes  $\langle x , \Phi_A(a) \rangle = \langle x, a \rangle$, i.e. $\Phi_A(a) =  a$, i.e. $a \in A$.  So 
we have established that $a \in (\M_A)_+ \intersect A = A_+$.

Considered as maps $A \rightarrow \cL(\R , (\M_A)_{sa})$ and $\L( (\M_A)_{sa}, \R)$, the 
maps $\gamma: a \mapsto \phi_a$ and $\mu: a \mapsto \ket{a}$ induce isomorphisms of linear spaces between 
$A$ and their ranges, taking $A_+$ onto $\InvQM(I,A)$ and $\InvQM(A, I)$ respectively, which establishes the affine isomorphisms claimed in (c).   
$\Box$ \\[0.3cm]
}
The category $\InvQM$ provides a unification of
  finite-dimensional real, complex and quaternionic quantum mechanics,
  but with the same  important caveats that apply to $\URUE$: the representation of orthodox, complex
  quantum systems $C_n$ in $\InvQM$ is through the universal embedding
  $\psi : C_n \mapsto \Cu(C_n) = M_n(\C) \oplus M_n(\C)$, $a \mapsto
  (a,a^T)$.  As a consequence, the composite of two complex quantum
  systems in $\InvQM$ is a direct sum of two copies of the their
  standard composite --- equivalently, is the standard composite,
  combined with a classical bit.  Moreover, the mapping that swaps the direct summands of 
  $\Cu(C_n)$, a perfectly good morphism in $\InvQM$, acts as the transpose on 
  $\psi(a) = (a, a^T)$.

\subsection{Compact Closure} 
 
A {\em compact structure} on a symmetric monoidal category $\Cat$ is a choice, for every object $A \in \Cat$, of a 
{\em dual object}: A triple $(A',\eta_A,f_A)$ consisting of an object $A' \in \C$, a {\em co-unit} $\eta_A : A \otimes A' \rightarrow I$ and a {\em unit} $f_{A} : I \rightarrow A' \otimes A$
such that  
\[A \rightarrow A \otimes I \stackrel{\id_A \otimes f_{A}}{\longrightarrow}  A \otimes (A' \otimes A) \longrightarrow (A \otimes A') \otimes A \stackrel{\eta_{A} \otimes \id_A}{\longrightarrow} I \otimes A \rightarrow A\]
and 
\[A' \rightarrow I \otimes A' \stackrel{f_{A} \otimes \id_{A'}}{\longrightarrow} (A' \otimes A) \otimes A' \longrightarrow A' \otimes (A \otimes A') \stackrel{\id_{A'} \otimes \eta_{A}}{\longrightarrow} A' \otimes I \rightarrow A'\]
give the identity morphisms on $A$ and $A'$, respectively (and where the unlabeled arrows are the obvious {\magenta unit introductions and associators}).\footnote{Our usage is slightly perverse. The usual convention is to denote the {\em unit} by $\eta_A$ and the 
co-unit by $\epsilon_A$. Our choice is motivated in part by the desire to represent states as morphisms $A \rightarrow I$ 
and effects as morphisms $I \rightarrow A$, rather than the reverse, together with the convention that 
takes the unit to correspond to the maximally entangled state.}
The standard example is the category $\FinVec$ of finite-dimensional vector spaces (say, over $\C$) and linear mappings. Here there is a canonical linear functional $\eta_{A} : A \otimes A^{\ast} \rightarrow \C$, namely, the trace. A canonical unit is supplied by picking a basis $E$ for $A$, and setting $f_{A} = \sum_{x \in E} f_x \otimes f_x$, where $\{f_x\}$ is the dual basis for $A^{\ast}$; one then shows that this is independent of $E$, {\redd and that the  identities above hold.}

We say a symmetric monoidal category is \emph{compact closed} if it admits a compact structure. \footnote{This makes compact closure a \emph{property} of some symmetric monoidal categories (SMCs).  Some authors instead define 
a compact closed category as a distinct mathematical structure, namely an SMC equipped with a distinguished 
compact structure.} 
In \cite{Abramsky-Coecke}, it is shown that a large number of
information-processing protocols, including 
{\magenta in particular} conclusive
teleportation and entanglement-swapping, hold in any compact closed
symmetric monoidal category, if we interpret objects as systems and
morphisms as physically allowed processes. In this section, we shall
see that our category $\InvQM$ is compact closed. More exactly, we
shall show that it inherits a compact structure from the {\magenta
  natural compact structure on the} category $\stalg$ of
finite-dimensional complex $\ast$-algebras,  which we now review.

If $\M$ is a finite-dimensional complex $\ast$-algebra, let $\Tr$ denote the
canonical trace on $\M$, regarded as acting on itself by left
multiplication (so that $\Tr(a) = \tr(L_a)$, $L_a : \M \rightarrow \M$
being $L_a(b) = ab$ for all $b \in \M$). This induces an inner product\ffootnote{\red double check!! Checked.}
on $\M$, given by $\langle a , b \rangle_{\M} = \Tr(ab^{\ast})$\footnote{Again, we are following the convention 
that complex inner products are linear in the first argument.}.  Note
that this inner product is self-dualizing, i.e,. $a \in \M_+$ iff 
$\langle a , b \rangle \geq 0$ for all $b \in \M_+$. 

Now let $\bar{\M}$ be the conjugate algebra, writing $\bar{a}$ for $a \in \M$ when regarded as belonging to $\bar{\M}$ (so that $\bar{ca} =
\bar{c}~\bar{a}$ for scalars $c \in \C$ and $\bar{a} \bar{b} =
\bar{ab}$ for $a, b \in \M$).  Note that $\langle \bar{a}, \bar{b} \rangle = \langle b ,
a \rangle$.  Now define
\[f_{\M} = \sum_{e \in E} e \otimes \bar{e} \in \M \otimes \bar{\M}\]
where $E$ is any orthonormal basis for $\M$ with respect to
$\IP_{\M}$. {\redd Then straightforward computations show that $f_{\M} \in (\M \otimes \bar{\M})_{+}$, and that, 
for all $a, b \in \M$, 
\[\langle a \otimes \bar{b}, f_{\M} \rangle = \langle a, b \rangle = \Tr(ab^{\ast}),\]
where the inner product on the left is the trace inner product on $\M \otimes \M^{\ast}$. Now define 
$\eta_{\M} : \M \otimes \bar{\M} \rightarrow \C$ by $\eta_{\M} = | f_{\M} \rangle$, noting that this functional 
is positive (so, up to normalization, a state) since $f_{\M}$ is positive in $\M \otimes \bar{\M}$. }

A final
computation  shows that, for any states $\alpha$ and
$\bar{\alpha}$ on $\M$ and $\bar{\M}$, respectively, and any $a \in
\M, \bar{a} \in \bar{\M}$, we have
\[(\eta_{\M} \otimes \alpha)(a \otimes f_{\bar{\M}}) = \alpha(a) \ \ \mbox{and} \ \ (\bar{\alpha} \otimes \eta_{\M})(f_{\bar{M}} \otimes \bar{a}) = \bar{\alpha}(\bar{a}).\]
Thus, $\eta_{\M}$ and $f_{\bar{\M}}$ define a compact structure on
$\stalg$, for which the dual object of $\M$ is given by $\bar{\M}$.

\begin{definition} {\em The {\em conjugate} of a EJC-algebra $(A,\M_{A})$ is $(\bar{A},\bar{\M}_{A})$, where 
$\bar{A} = \{ \bar{a} | a \in A\}$. We write $\eta_{A}$ for $\eta_{\M_A}$ and $f_A$ for $f_{\M_A}$. }\end{definition}

Any linear mapping $\phi : \M \rightarrow \N$ between $\ast$-algebras $\M$ and $\N$ gives rise to a 
linear mapping $\bar{\phi} : \bar{\M} \rightarrow \bar{\N}$, given by $\bar{\phi}(\bar{a}) = \bar{\phi(a)}$ 
for $a \in \M$. It is straightforward {\redd to show} that if $\Phi$ is a unitary involution on $\M_A$ with $A = {\M_{A}}^{\Phi}_{\sa}$, then $\bar{\Phi} : \bar{\M} \rightarrow \bar{\M}$ is also a unitary involution with ${\bar{\M}_{A}}^{\bar{\Phi}}_{\sa} = \bar{A}$. Thus, the class of involutive EJCs is closed under the formation of conjugates.

\begin{lemma} Let $(A,\M_A)$ belong  {\red to $\InvQM$.} Then 
$f_{A} \in A \odot \bar{A}$.
\end{lemma}

\noindent{\em Proof:}  {\red By assumption,} there is a unitary involution 
$\Phi$ on $\M_A$ such that $A = (\M_{A})^{\Phi}_{\sa}$; by Corollary \ref{cor: canonical composites with involutions}, 
$A \odot \bar{A}$ is then the set of self-adjoint fixed points of $\Phi \otimes \bar{\Phi}$. 
Since $\Phi$ is unitary,  if $E$ is an orthonormal basis for $\M_A$, then so 
is $\{\Phi(e) | e \in E\}$; hence, as $f_A$ is independent of the choice of 
orthonormal basis, $f_A$ is invariant under $\Phi \otimes \bar{\Phi}$. 
Since $f_A$ is also self-adjoint, it belongs to $(\M_A \otimes \M_{\bar{A}})^{\Phi \otimes \bar{\Phi}}_{\sa}$, 
i.e., to $A \odot \bar{A}$. $\Box$ \\[0.3cm]
It follows now from part (b) of Corollary \ref{cor: things that are InvQM morphisms} that the 
functional  
\[{\redd \eta_{A}} = | f_{\bar{A}} \rangle : \M_{\bar{A}} \otimes \M_{A} \rightarrow \R\]
is an $\InvQM$ morphism.  Hence, $\InvQM$ inherits the compact structure from $\stalg$, as promised. We have arrived at the following.

\begin{theorem}\label{thm: InvQM is compact closed} $\InvQM$ is compact closed. \end{theorem} 
 
\noindent{\bf Dagger compactness} In fact, we can do a bit better. {\redd A {\em dagger} on a category $\Cat$ 
is an involutive contravariant functor $\dagger : \Cat \rightarrow \Cat$ that fixes objects; that is, $A^{\dagger} = A$ 
for all $A \in \Cat$, and $f^{\dagger} \in \Cat(B,A)$ for all $f \in \Cat(A,B)$ with $f^{\dagger \dagger} = f$.  
If $\Cat$ is a symmetric 
monoidal category  {\hblue equipped with a dagger satisfying }$(f \otimes g)^{\dagger} = f^{\dagger} \otimes g^{\dagger}$ for all 
morphisms $f$ and $g$, and also $\sigma_{A,B}^{\dagger} = \sigma_{B,A}$, where $\sigma_{A,B} : A \otimes B 
\rightarrow B \otimes A$ is the ``swap" morphism, then $\Cat$ is said to be 
{\em dagger-monoidal.}  Finally, if $\Cat$ {\hblue admits} a compact structure such that $\eta_{A}^{\dagger} = f_{A}$, 
then $\Cat$ is said to be {\em dagger-compact.} }

It is not difficult to show that $\stalg$ is dagger-compact, where, if $\M$ and $\N$ are finite-dimensional $\ast$-algebras and $\phi : \M \rightarrow \N$ is a linear mapping, $\phi^{\dagger}$ is the hermitian adjoint 
of $\phi$ with respect to the natural trace inner products on $\M$ and $\N$. If $(A,\M_A)$ and $(B,\M_B)$ 
are involutive 
EJC-algebras with given unitary involutions $\Phi_A$ and $\Phi_B$, then for any intertwiner $\phi : \M_A \rightarrow \M_B$, 
$\phi^{\dagger}$ also intertwines $\Phi_A$ and $\Phi_B$. Hence, we have

\begin{corollary}\label{cor: InvQM is dagger compact} $\InvQM$ is dagger-compact.\end{corollary} 

\section{Conclusion} 
\label{sec: conclusion}

We have constructed 
two categories of probabilistic models --- the categories {\mmagenta $\RSE$} and
$\InvQM$ --- that, in different ways, unify finite-dimensional real,
complex and quaternionic quantum mechanics.  In each case, there is a
price to be paid for this unification.  For {\mmagenta $\RSE$}, this price is 
steep: $\RSE$ is a monoidal category, but one in which states (for
instance) on complex systems don't count as physical processes. 
In particular, {\mmagenta $\RSE$} is
very far from being compact closed.\footnote{\redd In \cite{BGraW}, we erroneously claimed 
that the category $\URSE$ of  universally reversible, but standardly embedded, EJC algebras, 
with relatively CJP mappings, is compact closed. That this is not the case is clear from {\hblue Example}
\ref{ex: no states 2}.}

{\magenta In contrast, 
$\InvQM$ is clearly 
a well-behaved --- indeed, dagger-compact --- probabilistic theory, in which the states, as well
as the effects, of real, 
complex, and quaternionic Euclidean Jordan algebras appear as morphisms. 
  On the other hand, $\InvQM$ 
 admits the transpose
  automorphism on the complex Hermitian Jordan algebra,} and requires
complex quantum systems to compose in a non-standard way.
Nevertheless, by virtue of being dagger compact, ${\redd \InvQM}$ 
continues
to enjoy many of the information-processing properties of standard
complex QM, e.g., the existence of conclusive teleportation and
entanglement-swapping protocols \cite{Abramsky-Coecke}. {\magenta Also, composites in $\InvQM$
satisfy the Cirel'son bound on correlations owing to the way that,
by construction, these composites are embedded within a tensor product
of complex matrix algebras.}

All of this is in spite of the fact that composites in $\InvQM$ are 
not locally tomographic: 
the canonical composite $A \odot B$ is larger than the vector 
space tensor product $A \otimes B$.
Local tomography is well known to separate complex QM from its 
real and quaternionic variants, so its failure in $\URUE$ and $\RSE$ 
is hardly surprising,  but it is noteworthy that we are able to construct
(non-locally tomographic) composites in $\URUE$ in all of the non-real cases, and
certain composites involving quaternions even in $\RSE$.     
Even more interesting is the fact that, for 
quaternionic systems $A$ and $B$, the  information capacity --- 
the number of sharply distinguishable states --- of $A \odot B$ is {\em larger} than the  
product of the capacities of $A$ and $B$. A related point is that, for 
 quaternionic quantum systems $A$ and $B$, the product of a pure state 
of $A$ and a pure state of $B$ will generally be a {\em mixed} state 
in $A \odot B$.

The category $\InvQM$ contains interesting compact closed
subcategories.  In particular, real and quaternionic quantum systems,
taken together, form a (full) monoidal sub-category of $\InvQM$ closed under
composition.  We  {\magenta conjecture} 
that this is exactly what one gets by applying Selinger's CPM construction \cite{Selinger} to Baez'
(implicit) category of pairs $(\Hilb,J)$, $H$ a finite-dimensional
Hilbert space and $J$ an anti-unitary with $J^2 = \pm 1$ \cite{BaezDivAlg}.

Another compact-closed subcateory of $\InvQM$, which we might call $\InvCQM$, 
consists of universally embedded {\em complex} quantum systems $C_n$. 
It is interesting to note that, 
in an hypothetical universe described by $\InvQM$, the subcategory $\InvCQM$ 
acts as a kind of ``ideal", in that
if $A \in \InvQM$ and $B \in \InvCQM$, then $A \odot B \in \InvCQM$ as well.
This is provocative, as it suggests that such a universe, initially
consisting of many systems of all three types, {\redd might} eventually evolve
into one in which complex systems greatly predominate. 

Although it is not compact closed, the 
category {\mmagenta $\RSE$} of reversible, standardly embedded EJCs remains of interest. This is 
still a monoidal category, and contains, in addition to real and quaternionic quantum systems, orthodox complex 
quantum systems in their standard embedding (and composing in the normal way). 
Indeed, these form a monoidal subcategory, $\CQM$, which, again, 
functions as an ``ideal".

It is worth noting that the set of quaternionic quantum systems 
does {\em not} form a monoidal subcategory of either {\mmagenta $\RSE$} or 
$\InvQM$, as the composite of two quaternionic systems is {\em real}. 
Efforts to construct a free-standing quaternionic quantum theory 
have had to contend with the absence of a suitable {\em quaternionic} 
composite of quaternionic systems. For instance, as pointed out by Araki 
\cite{Araki80}, the obvious candidate for the composite 
of $A = M_m(\Q)_{\sa}$ and $B = M_{n}(\Q)_{\sa}$, $M_{mn}(\Q)_{\sa}$, 
does not have a large enough dimension to accommodate the 
real tensor product $A \otimes B$, 
{\magenta causing difficulty for}  the 
representation of product effects.\footnote{Attempts to interpret
  the quaternionic ``Hilbert space'' $\Q^{mn}$ as a tensor 
  product of $\Q^m$
  and $\Q^n$ raise at least the possibility of signaling via the
  noncommutativity of scalar multiplication.  This noncommutativity
  underlies the the argument in \cite{McKague} that
  stronger-than-quantum correlations are achievable in such a model.}
 In our approach, the issue simply doesn't arise. It seems that ``quaternionic quantum mechanics" is best seen as an 
inextricable part of a larger theory. Essentially 
the same point has also been made by Baez \cite{BaezDivAlg}. 
 
The category $\InvQM$ is somewhat mysterious. It encompasses 
real and quaternionic QM in a completely natural way; 
however, while it also contains complex quantum systems, these compose 
in an exotic {\awedit manner}: {\redd as pointed out above}, the composite of two complex quantum systems 
in $\InvQM$ comes with an extra classical bit ---
equivalently, $\{0,1\}$-valued superselection rule. 
 This functions to make the transpose automorphism of 
$M_n(\C)_{sa}$ count as a morphism.   
The extra classical bit is flipped by the Jordan transpose 
(swap of $C^*$ summands) on either factor of {\redd such} a composite, but unaffected 
if {\em both} parties {\redd implement} the Jordan transpose (which does, of course, 
effect a Jordan transpose on the composite).
The precise physical significance of this is a subject for
further study. 

As Example \ref{ex: no states} shows, there is no way to enlarge
  $\InvQM$ so as to include higher spin factors, without either
  sacrificing compact closure {\magenta (and even rendering the set
  $\cc(A,I)$, which might naturally be thought to represent states,
  trivial)} or venturing outside the ambient category of
  EJC-algebras, to make use of morphisms that are not {\redd (relatively)} completely
  Jordan-preserving maps.  Example \ref{ex: no states 2} shows, more strikingly, that
  there is no way to construct a category {\redd of the form $\CJP_{\Cat}$} that contains
  {\em standardly embedded} complex quantum systems {\em and} real 
  systems, without, again, sacrificing compact closure
  (indeed, the representation of states by morphisms).


\appendix

\section{Direct Sums of EJAs}
\label{appendix: direct sums}

The first part of this Appendix collects some basic facts about direct sums of EJAs that are used in the body 
of the paper. The second part contains a proof that 
$\Cu(A \oplus B) = \Cu(A) \oplus \Cu(B)$, and that the direct sum of universally reversible EJAs is again UR.\\[0.3cm]
\noindent{\bf Direct Summands and Central Projections} The {\em direct sum} of EJAs $A$ and $B$ is $A \oplus B := A \times B$, equipped with the slotwise operations, 
so that the canonical projections $\pi_1 : A \times B \rightarrow A$ and $\pi_2 : A \times B \rightarrow B$ 
are unital Jordan homomorphisms. 
Identifying  $A$ and $B$ with $A \times \{0\}$ and $\{0\} \times B$, respectively, we  write 
$a + b$ for $(a,0) + (0,b)$.  Note that $A$ and $B$ are then ideals in $A \oplus B$, and 
that $B = A^{\perp} := \{ z \in A \oplus B | \langle a, z \rangle = 0 \ \forall a \in A\}$.  Conversely, 
we will show that if $E$ is an EJA and $A$ is an ideal in $A$, then $A^{\perp}$ is also an ideal, and $a\jProd b = 0$ for 
all $a \in A, b \in A^{\perp}$; hence, $E \simeq A \oplus A^{\perp}$. 

Suppose $E$ is an EJA and $A \leq E$ is an ideal: let $B = A^{\perp}$. Then for all $z \in E$, 
$a \in A$ and $b \in B$, 
\[\langle a, z\jProd b \rangle = \langle a\jProd z, b \rangle = 0\]
since $a\jProd z \in A$. Thus, $B$ is also an ideal, and $E = A \oplus B$ as a vector space. Finally, 
if $a \in A$ and $b \in B$, then $a\jProd b \in A \cap B = \{0\}$. Hence, 
if $a, x \in A$ and $b, y \in B$ then $(a + b)\jProd (x + y) = a\jProd x + b\jProd y$, i.e., in the 
representation of $A \oplus B$ as $A \times B$, operations are slotwise. 
{\newawe Note that $u = p_A + p_B$ for a unique $p_A \in A$ 
and $p_B \in B$. For $a \in A$ we have $a \jProd p_B = 0$, so 
$a = a \jProd u = a \jProd p_A$, whence $p_A = p_{A}^2$ and 
 $A = A \jProd p_A$. In other words, we have}


\begin{lemma} Let $A$ be an ideal in an EJA $E$. Then there exists a projection $p \in E$ 
such that $p\jProd a = a$ for every $a \in A$. Thus, $A = p\jProd A$, and $E = p\jProd A \oplus p'\jProd A$, where $p' = 1 - p$. \end{lemma}

\tempout{
For a proof in the more general setting of JBW algebras, see  \cite{AS}, Propositions 2.7, 2.39 
and 2.41. 
}
 
The {\em center} of an EJA $E$ is the set of elements operator-commuting with all other elements. Denote this by $C(E)$. If $E = A \oplus B$, and $p$ is the unit of $A$, so that $A = p\jProd A$, then it's easy to check 
that $p  \in C(E)$.  Conversely, if $p$ is a central projection, then $p\jProd A$ is an ideal, with unit element $p$. 
If $p$ is a {\em minimal} central projection, then $p\jProd A$ is a minimal direct summand of $E$.  If $E$ is simple, 
then its only central projections are $0$ and $1$, and conversely.

One can show that for every projection $p$ in an EJA $E$, there exists a unique minimal projection $c(p) \in C(E)$, 
the {\em central cover} of $p$, such that $p \leq c(p)$. Then $A := c(p)E$ is an ideal of $E$, in which 
$c(p)$ is the unit. If $A$ is a minimal ideal, then  
elements of $A$ are exactly those with central cover $c(p)$ (see 2.37 and 2.39 in \cite{AS}.) More generally, 
two elements of $E$ having the same central cover have nonzero components in exactly the same ideals of $E$.

Recall that a {\em symmetry} of $A$ is an element $s \in A$ with $s^2 = u$. 
Projections $e, f$ in $A$ are {\em exchanged by a symmetry} $s$ iff $U_{s}(e) = f$. If there 
exists a sequence of symmetries $s_1,...,s_n$ with $U_{s_n} \circ \cdots \circ U_{s_1} (e) = f$, then 
$e$ and $f$ are {\em equivalent}. 

\begin{lemma}[\cite{AS}, Lemma 3.9] 
Equivalent projections have the same central cover. 
\end{lemma}

\noindent{\bf The universal $C^{\ast}$-algebra of a direct sum} 
Recall that a sequence of vector spaces and linear maps, or of Jordan algebras and Jordan homomorphisms, or of 
$C^{\ast}$ algebras and $\ast$-homomorphisms 
\[ A \stackrel{\alpha}{\longrightarrow} B \stackrel{\beta}{\longrightarrow} C\]
is said to be {\em exact at $B$} iff the image of $\alpha$ is the kernel of $\beta$. A {\em short exact sequence} is 
one of the form 
\[0 \longrightarrow A \stackrel{\alpha}{\longrightarrow} B \stackrel{\beta}{\longrightarrow} C \longrightarrow 0\]
that is exact at $A$, $B$ and $C$ (with the maps on the ends being the only possible ones). This means that 
$\alpha$ is injective (its kernel is $0$), while $\beta$ is surjective (its image is the kernel of the zero map, 
i.e., all of $C$). 

Let $\EJA$ and $\Cstar$ be the categories of EJAs and Jordan homomorphisms, and of $C^{\ast}$-algebras and 
$\ast$-homomorphisms, respectively.

\begin{theorem}[\cite{HO}] 
\label{HOET}
$A \mapsto \Cu(A)$ is an exact functor from $\EJA$ to $\Cstar$. 
In other words, if $A \stackrel{\alpha}{\longrightarrow} B \stackrel{\beta}{\longrightarrow} C$ is an exact sequence in $\EJA$, then $\Cu(A) \stackrel{\Cu(\alpha)}{\longrightarrow} B \stackrel{\Cu(\beta)}{\longrightarrow} \Cu(C)$ is an exact sequence in $\Cstar$. 
\end{theorem}

We are going to use this to show that $\Cu(A \oplus B) = \Cu(A) \oplus \Cu(B)$. We need some preliminaries. 
The following is standard: 

\begin{lemma}\label{exact sequences} Let 
\[0 \longrightarrow A \stackrel{\alpha}{\longrightarrow} C \stackrel{\beta}{\longrightarrow} B \longrightarrow 0\]
be a short exact sequence of vector spaces. Then the following are equivalent:
\begin{itemize} 
\item[(a)] There is an isomorphism $\phi : A \oplus B \simeq C$ such that $\alpha$ and $\beta$ are respectively the canonical injection and surjection given by  
\[\alpha(a) = \phi(a,0) \ \ \mbox{and} \ \ \beta(\phi(a,b)) = b\]
\item[(b)] The sequence is {\em split} at $B$: there exists a linear mapping $\gamma : B \rightarrow C$ such 
that $\beta \circ \gamma = \id_{B}$.
\end{itemize}
\end{lemma}

The idea is that, given $\phi$, we can define $\gamma$ by $\gamma(b) = \phi(0,b)$ and, given $\gamma$, we can 
define $\phi$ by $\phi(a,b) = \alpha(a) + \gamma(b)$.

If $A$, $B$ and $C$ are Jordan algebras or $C^{\ast}$ algebras, the implication from (a) to (b) is obviously valid, 
but the converse requires additional assumptions. 

\begin{lemma}\label{exact sequences and ideals} 
Let \[0 \longrightarrow A \stackrel{\alpha}{\longrightarrow} C \stackrel{\beta}{\longrightarrow} B \longrightarrow 0.\]
be a short exact sequence of $\ast$-algebras and $\ast$-homomorphisms, which is split by a $\ast$-homomorphism 
$\gamma :  B \rightarrow C$ with $\beta \circ \gamma = \id_{B}$. Let $\phi : A \oplus B \rightarrow C$ be 
as defined above, {\redd i.e, $\phi(a,b) = \alpha(a) + \gamma(b)$ for $a \in A$, $b \in B$.}  Then the following are equivalent: 
\begin{itemize} 
\item[(a)] $\gamma(B)$ is a (2-sided) $\ast$-ideal in $C$;
\item[(b)] $\phi$ is multiplicative, and thus a $\ast$-isomorphism;  
\item[(c)] There exists a $\ast$-homomorphism $\delta : C \rightarrow A$ with 
\[0 \longleftarrow A \stackrel{\delta}{\longleftarrow} C \stackrel{\gamma}{\longleftarrow} B \longleftarrow 0\]
exact.
\end{itemize} 
\end{lemma} 

\noindent{\em Proof:} (a) $\Rightarrow$ (b). It is easy to see that a $C^{\ast}$-algebra $C$ is the direct sum of two $\ast$-ideals $A, B \leq C$ iff $A \oplus B = C$ and $A \cap C = \{0\}$, i.e., iff $C$ is the vector-space 
direct sum of $A$ and $B$.  We already know that $\alpha(A) + \beta(B) = C$ (since $\phi$ is a linear 
isomorphism). It therefore suffices to show that $\alpha(A)$ and $\gamma(B)$ are $\ast$-ideals with 
zero intersection. We are assuming that $\gamma(B)$ is a $\ast$-ideal. As it's the kernel of a $\ast$-homomorphism, 
$\alpha(A)$ is automatically a $\ast$-ideal. To see that $\alpha(A) \cap \gamma(C) = \{0\}$, 
let $c \in C$ with $c = \alpha(a) = \gamma(b)$ for some $a \in A$ and $b \in B$. Then we have 
\[b = \beta(\gamma(b)) = \beta(\alpha(a)) = 0\]
whence, $c = \gamma(0) = 0$. 

(b) $\Rightarrow$ (c). If $\phi$ is a $\ast$-isomorphism, then let $\delta = \pi_{A} \circ \phi^{-1}$ where 
$\pi_{A} : A \oplus B \rightarrow A$ is the projection $\pi_{A}(a,b) = a$. Note that $\delta$ is the composition 
of two $\ast$-homomorphisms, and thus, a $\ast$-homomorphism. To verify exactness, 
note that as $\phi(a,b) = \alpha(a) + \gamma(b)$, 
we have $\phi(0,b) = \gamma(b)$, whence, $\phi^{-1}(\gamma(b)) = (0,b)$. Thus, 
$\delta(\gamma(b)) {\redd = \pi_{A}(0,b)} = 0$. 

(c) $\Rightarrow$ (a). By exactness, $\gamma(C)$ is the kernel of the $\ast$-homomorphism $\delta$, and hence, 
a $\ast$-ideal. $\Box$ \\[0.3cm]
Now let $E = A \oplus B$. Then we have a short exact sequence 
\[0 \longrightarrow A \stackrel{j}{\longrightarrow} A \oplus B \stackrel{p}{\longrightarrow} B \longrightarrow 0.\]
where $j(a) = (a,0)$ and $p(a,b) = b$. This is split by the homomorphism $k : A \rightarrow A \oplus B$ given 
by $k(b) = (0,b)$. Hanche-Olsen's exactness theorem, that is Theorem \ref{HOET}, gives us a short exact sequence 
\[0 \longrightarrow \Cu(A) \stackrel{C^{\ast}(j)}{\longrightarrow} \Cu(A \oplus B) \stackrel{C^{\ast}(p)}{\longrightarrow} \Cu(B) \longrightarrow 0.\]
By functoriality, $\Cu(p) \circ \Cu(j) = \id_{\Cu(B)}$, so this is again split. Thus, {\em regarded as a vector space}, $\Cu(A \oplus B)$ is canonically isomorphic to 
$\Cu(A) \oplus \Cu(B)$. On the other hand, we {\em also} have an exact sequence 
\[0 \longleftarrow A \stackrel{q}{\longleftarrow} A \oplus B \stackrel{k}{\longleftarrow} B \longleftarrow 0\]
where $q(a,b) = a$. By the same argument, then, we have a short exact sequence 
\[0 \longleftarrow \Cu(A) \stackrel{\Cu(q)}{\longleftarrow} \Cu(A \oplus B) \stackrel{\Cu(k)}{\longleftarrow} \Cu(B) \longleftarrow 0.\]
Applying the preceding Lemma, we have

\begin{proposition}\label{prop: Cu additive} If $A$ and $B$ are EJAs, then 
\[ \Cu(A \oplus B) \simeq \Cu(A) \oplus \Cu(B).\]
\end{proposition}

Notice that if $\Phi_A$ and $\Phi_B$ are, respectively, the canonical involutions on $\Cu(A)$ and $\Cu(B)$ fixing points of $A$ and $B$, then $\Phi_{A} \oplus \Phi_{B}$ is a $\ast$-involution on $\Cu(A) \oplus \Cu(B)$ fixing points of 
$A \oplus B$. But there is only one such  $\ast$-involution on $\Cu(A \oplus B)$, the canonical one. In other words, 
in identifying $\Cu(A \oplus B)$ with $\Cu(A) \oplus \Cu(B)$, we also identify $\Phi_{A \oplus B}$ with $\Phi_{A} \oplus \Phi_{B}$. 

Recalling now the fact (\cite{HO}, Lemma 4.2) that an EJA
$A$ is universally reversible (UR) iff $A$ coincides with the set of
self-adjoint fixed points in $\Cu(A)$ of the canonical
$\ast$-involution $\Phi_A$, we have the following.

\begin{corollary}\label{prop: UR composites}{\mmagenta $A \oplus B$ is UR iff $A$ and $B$ are UR.} 
\end{corollary} 

\noindent{\em Proof:} Let $\Phi = \Phi_{A} \oplus \Phi_{B}$ be the canonical involution on $\Cu(A \oplus B) = \Cu(A) \oplus \Cu(B)$. 
For $(a,b) \in \Cu(A) \oplus \Cu(B)$, we have 
 $\Phi(a,b) = (\Phi_{A}(a), \Phi_{B}(b)) = (a,b)$ iff $\Phi_{A}(a) = a$ and $\Phi_{B}(b) = b$. Since $A$ and $B$ 
are UR, this holds iff $a \in A$ and $b \in B$, i.e., iff $(a,b) \in A \oplus B$. Thus, $A \oplus B$ is exactly the 
set of fixed-points of $\Phi$, and so, is UR. 

{\mmagenta Conversely, } {\hblue let $A \oplus B$ be UR.  Suppose for a contradiction that one of $A$ or $B$, say 
$B$, is not UR: then there exists $b \in C^*(B)$ such that $\Phi_B(b) = b$ but $b \notin B$.   Let $a$ be in 
$A$.  Then $\Phi_{A \oplus B}((a,b)) \equiv (\Phi_A(a),\Phi_B(b)) = (a,b)$, whence by the fact that $A \oplus B$
is UR, $(a,b) \in A \oplus B$, but since $b \notin B$ this is in contradiction with the definition of $A \oplus B$ as $A \times B$ equipped with a product.}
$\Box$ 

 (One {\awedit can} also easily prove the above Corollary directly from the definition of universal reversibility, 
without using the canonical involutions.)  

\section{The Quabit}
\label{appendix: the quabit}

In this appendix, we show that the canonical tensor product $Q_2 \odot Q_2$ of two quabits in their standard representation is $R_{16}$, the self-adjoint part of the real matrix algebra $M_{16}(\R)$. \\[0.3cm]
\noindent{\bf The symplectic representation} A quaternion $q = a + bi + cj + dk$ can alternatively be expressed in the form $(a + bi) + (c + di)j$, i.e., 
$z + wj$ where $z, w \in \C$, and also has a standard representation as a $2 \times 2$ complex matrix, 
namely 
\[ [q] := \left [ \begin{array}{cc} z & w \\ -\bar{w} & \bar{z} \end{array} \right ].\]
Treating $\Q$ as a $\ast$-algebra over $\C$, the mapping $q \mapsto [q]$ is a $\ast$-homomorphism from $\Q$ into $M_2(\C)$.  
This  {\hblue yields} a natural representation --- that is, $\ast$-homomorphism --- $\pi_{o} : M_n(\Q) \rightarrow M_{n}(M_{2}(\C)) \simeq M_{2n}(\C)$,  {\hblue given by}  
\[ \pi_{o}(a)_{p,q} = [a_{p,q}]\]
for $a \in M_n(\Q)$ and $p, q = 1,...,n$.  An equivalent,  but for our purposes, more useful, representation is given by 
\[\pi(a)_{\hblue {p,q}} = F \pi_{o}(a)_{\hblue {p,q}} F\]
where 
\begin{equation}
F = \left(\begin{array}{cccc} 1 & 0 & 0 & 0 \\ 0 & 0 & 1 & 0 \\ 0 & 1 & 0 & 0 \\ 0 & 0 & 0 & 1 \end{array}\right)\text{.}\nonumber
\end{equation}
 This is called the {\em symplectic} representation of $M_n(\H)$. 

If we express each entry of $a$ in the form $a_{p,q} = z_{p,q} + w_{p,q} j$, then $a = \Gamma_{1} + \Gamma_{2} j$ 
where $(\Gamma_1)_{p,q} = z_{p,q}$ and $(\Gamma_{2})_{p,q} = w_{p,q}$. Computing, one finds that 
\[\pi(a) = \left [ \begin{array}{cr} \Gamma_{1} & \Gamma_{2} \\ - \bar{\Gamma}_{2} & \bar{\Gamma}_{1} \end{array} \right ].\]
Notice that $\pi(a)$ is self-adjoint iff $\Gamma_1$ is self-adjoint and $\Gamma_{2}$ is anti-symmetric, and that 
this is the case iff $a$ is self-adjoint in $M_2(\Q)$.

From now on, we identify $Q_2$ with $\pi(M_2(\Q)_{\sa}) \leq M_{4}(\C)_{\sa}$. Regarded as an embedded EJA in this way, 
the canonical tensor product $Q_2 \odot Q_2$ is defined to be the Jordan subalgebra of the self-adjoint part of $M_{4}(\C) \otimes M_{4}(\C) = 
M_{16}(\C)$ generated by $Q_2 \otimes Q_2$. Our main goal in this Appendix is to prove 

\begin{proposition}\label{universal C* for two quabits}
{\em $\Cu(Q_2 \odot Q_2) \simeq M_{16}(\C)$.} 
\end{proposition}

Since $Q_2 \odot Q_2$ is UR, this will follow from (\cite{HO}, Theorem 4.4) 
if we can show that (i) $Q_2 \odot Q_2$ generates $M_{16}(\C)$ as a $\ast$-algebra, and (ii) there is a $\ast$-involution on $M_{16}(\C)$ fixing elements of 
$Q_2 \odot Q_2$ pointwise. \\[0.3cm]
\noindent{\bf Quaternionic Pauli Matrices} In order to show that $Q_2
\odot Q_2$ generates $M_{16}(\C)$, we begin by writing down some
useful elements of $Q_2$. The analogues of the Pauli matrices
$\sigma_x, \sigma_y, \sigma_z \in M_2(\C)$ are the following
\emph{quaternionic Pauli matrices}:
\[
\begin{array}{ccccc} 
\sigma_0 = \left [ \begin{array}{cc} 1 & 0 \\ 0 & 1 \end{array} \right ] & & 
\sigma_1 = \left [ \begin{array}{cr} 1 & 0 \\ 0 & -1 \end{array} \right ] & & 
\sigma_2 = \left [ \begin{array}{cc} 0 & 1 \\ 1 &  0 \end{array} \right ]\\ 
& & & & \\
\sigma_3 = \left [ \begin{array}{cr} 0 & -i \\  i & 0 \end{array} \right ] & & 
\sigma_4 = \left [ \begin{array}{cr} 0 & -j \\  j & 0 \end{array} \right ] & &  
\sigma_5 = \left [ \begin{array}{cr} 0 & -k \\  k & 0 \end{array} \right ] 
\end{array} 
\]
Evidently, $\sigma_1, \sigma_2$ and $\sigma_3$ are the standard Pauli matrices 
$\sigma_z$, $\sigma_x$ and $\sigma_y$, respectively. Note that these are all traceless and self-adjoint,  
{\awedit and satisfy 
$\sigma_{a} \dot \sigma_{b} = \delta_{a,b} \sigma_0$ 
where $a, b \in \{1,...,5\}$ --- that is, $\sigma_a$ and $\sigma_b$ anti-commute if $a \not = b$, and $\sigma_{a}$ 
squares to the identity. }
Applying the representation $\pi$ gives us six elements of $Q_2$, 
$s_0 = \pi(\sigma_0)$, $s_1 = \pi(\sigma_1)$, ..., $s_5 = \pi(\sigma_5)$. 
Direct computation reveals that 
\[
\begin{array}{ccccc} 
s_0 = \left [ \begin{array}{cc} \1 & 0 \\ 0 & \1 \end{array} \right ] = \sigma_0 \otimes \sigma_0 
& & 
s_1 = \left [ \begin{array}{cr} \sigma_z & 0 \\ 0 & \sigma_z \end{array} \right ] = \sigma_0 \otimes \sigma_z 
& & 
s_2 = \left [ \begin{array}{cc} \sigma_x & 0 \\ 0 & \sigma_x \end{array} \right ] = \sigma_0 \otimes \sigma_x \\ 
& & & & \\
s_3 = \left [ \begin{array}{cc} \sigma_y & 0 \\ 0 & -\sigma_y \end{array} \right ] = \sigma_z \otimes \sigma_y 
& & 
s_4 = \left [ \begin{array}{cr} 0 & -i\sigma_y \\ i\sigma_y & 0 \end{array} \right ] = \sigma_y \otimes \sigma_y 
& & 
s_5 = \left [ \begin{array}{cc} 0 & \sigma_y \\ \sigma_y & 0 \end{array} \right ] = \sigma_x \otimes \sigma_x \\ \end{array} 
\]
These again obey the Pauli-like identities mentioned above. 
Using these, we can compute (associative) products of these matrices, 
e.g., 
\[s_3 s_4 = (\sigma_z \otimes \sigma_y)(\sigma_y \otimes \sigma_y) = \sigma_z \sigma_y \otimes \sigma_y \sigma_y 
= - i \sigma_x \otimes \sigma_0.\]

\begin{lemma}\label{Two quabits generate}
$Q_2 \otimes Q_2$ generates $M_{16}(\C)$ as a $\ast$-algebra.
\end{lemma}

\noindent{\em Proof:} Begin by noting that the elements 
\[\sigma_{a} \otimes \sigma_{b} \otimes \sigma_{c} \otimes \sigma_{c},\]
where $a,b,c,d \in \{0,x,y,z\}$, are a basis for $M_{16}(\C)$. 
For each  $a \in \{0,x,y,z\}$, let 
\[x_1(a) = \sigma_a \otimes \sigma_0 \otimes \sigma_0 \otimes \sigma_0\] 
\[x_2(a) = \sigma_0 \otimes \sigma_a \otimes \sigma_0 \otimes \sigma_0\] 
\[x_3(a) = \sigma_0 \otimes \sigma_0 \otimes \sigma_a \otimes \sigma_0\] 
\[x_4(a) = \sigma_0 \otimes \sigma_0 \otimes \sigma_0 \otimes \sigma_a\]
Then $x_1(a) x_2(b) x_3(c) x_4(d) = \sigma_{a} \otimes \sigma_{b} \otimes \sigma_{c} \otimes \sigma_{d}$. These 
last form a basis for $M_{16}(\C)$, so it will suffice to show that, for $a \in \{x,y,z\}$, the elements $x_{p}(a)$ can be manufactured from elements of $Q_2 \otimes Q_2$ by forming (associative) products.  

For a start, notice that 
\[s_1 \otimes s_0 = \sigma_0 \otimes \sigma_{z} \otimes \sigma_0 \otimes \sigma_0 = x_2(z) \ \ \mbox{and} \ \ 
  s_0 \otimes s_1 = \sigma_0 \otimes \sigma_0 \otimes \sigma_0 \otimes \sigma_z = x_4(z).\]
Similarly, with $s_2$ in place of $s_1$, we have $x_2(x)$ and $x_4(x)$ in $Q_2 \odot Q_2$. 

As noted above, $s_3 s_4 =  -i \sigma_{x} \otimes \sigma_0$. 
Similarly, 
\[s_3 s_5 = i \sigma_{y} \otimes \sigma_{o} \ \mbox{and} \ s_4 s_5 = -i \sigma_{z} \otimes \sigma_{o}.\]
Hence, 
\[(s_3 \otimes s_4)(s_{4} \otimes s_{4}) = s_{3}s_{4} \otimes s_{4} s_{4} = -i \sigma_y \otimes \sigma_0 \otimes \sigma_0 \otimes \sigma_0 = -i x_1(y)\]
and similarly 
\[((s_3 \otimes s_5)(s_5 \otimes s_5) = i x_1(y) \ \ \mbox{and} \ \ (s_4 \otimes s_5)(s_5 \otimes s_5) = -i x_1(z).\]
Thus, we have $x_1(a)$ for all $a$. In an entirely similar way, we find that considering 
\[(s_3 \otimes s_3)(s_3 \otimes s_4) = -i x_3(y), \ (s_3 \otimes s_3)(s_3 \otimes s_5) = i x_3(x), \ 
(s_4 \otimes s_4)(s_4 \otimes s_5) = -i x_3 (z).\]
So we have $x_3(a) \in (Q_2 \odot Q_2)^2$ for all $a$. 

It remains to obtain $x_2(y)$ and $x_4(y)$. But now we have 
\[(s_3 \otimes s_0)x_1(z) = (\sigma_z \otimes \sigma_y \sigma_0 \otimes \sigma_0)(\sigma_z \otimes \sigma_0 \otimes \sigma_0  \otimes \sigma_0) = x_{2}(y)\]
and similarly 
\[x_3(z) (s_0 \otimes s_3) = x_3(y).\]
Hence, $x_2(y)$ and $x_4(y)$ also belong to $(Q_2 \odot Q_2)^2$, completing the proof. $\Box$\\[0.3cm]
\noindent{\bf An Involution} We now wish to find an involution on $M_{16}(\C)$ fixing $Q_2 \odot Q_2$ pointwise. 
If $\phi : \M \rightarrow \M$ is a $\ast$-involution on a complex $\ast$-algebra $\M$, 
recall that we write 
$\M^{\phi}_{\sa}$ for the set of self-adjoint fixed-points of $\phi$. In finite dimensions, this is 
always an EJA. Suppose $\phi$ and $\psi$ are $\ast$-involutions on complex $\ast$-algebras $\M$ and $\N$, 
respectively: if $A$ is a Jordan subalgebra of $\M^{\phi}_{\sa}$ and $B$ is a Jordan subalgebra of $\N^{\psi}_{\sa}$, 
then it's easy to see that $A \odot B$ is a Jordan subalgebra of $(\M \otimes \N)_{\sa}^{\phi \otimes \psi}$. 

Let $J = \left [\begin{array}{cc} \0 & -\1 \\ \1 & \0 \end{array} \right ]$. Then 
\[\phi(a) := -Ja^{T}J = -(J a J)^{T}\]
is a $\ast$-involution (a $\ast$-antiautomorphism of period two) on $M_n(\C)$. This fixes $Q_2$ pointwise, 
i.e., $\phi(\pi(a)) = \pi(a)$ for every $a \in M_2(\H)_{\sa}$. Identifying $M_{16}(\C)$ with 
$M_{4}(M_{4}(\C)) = M_{4}(\C) \otimes M_{4}(\C)$, we then have an involution 
\[\Phi = \phi \otimes \phi : M_{16}(\C) \rightarrow M_{16}(\C).\] 
By the comments above, we have 

\begin{lemma}\label{Two quabits fixed}
$\Phi = \phi \otimes \phi$ fixes every element of $Q_2 \odot Q_2$. 
\end{lemma}

This completes the proof of Proposition {\hblue \ref{universal C* for two quabits}}. Moreover, as a consequence of (\cite{HO},  Theorem 4.4), we have

\begin{corollary}\label{standard tp of two quabits is set of fixed points}
$Q_2 \odot Q_2 = M_{16}(\C)^{\Phi}$.
\end{corollary}

Since $\Cu(A) \simeq \M \simeq \Cu(B)$ (as $\ast$-algebras 
with involution) implies $A \simeq \M^{\phi}_{\sa} \simeq B$, and since 
$M_{16}(\C) \simeq \Cu(M_{16}(\R)_{\sa})$, we have

\begin{corollary}\label{standard tp of two quabits is R16}
$Q_2 \odot Q_2 {\hblue \iso } M_{16}(\R)_{\sa} =: R_{16}$.
\end{corollary}

\section{Spin factors}
\label{appendix: spin factors} 
The Jordan-von Neumann-Wigner Classification Theorem \cite{JNW} singles out exactly three classes of finite dimensional simple euclidean Jordan algebras: the matrix algebras $R_{n}, C_{n}, Q_{n}$, the exceptional Jordan algebra $M_{3}(\mathbb{O})_{\text{sa}}$, and one further type. The remaining type were dubbed \textit{Spin Factors} by Topping in \cite{Topping65}. For each finite $\mathbb{N}\ni n>1$, there exists a unique spin factor of dimension $1+n$ (up to Jordan isomorphism \cite{HO-Stormer}) denoted by $V_{n}$. Abstractly, $V_{n}$ is generated as a Jordan algebra by a \textit{spin system} of cardinality $n$: a collection of $2 \leq n \in \mathbb{N}$ symmetries (\textit{i.e.} self-adjoint unitaries) $s_{p}$ in a unital JB algebra $\mathcal{A}$, with $s_{p}\neq\pm u_{\mathcal{A}}$ such that $s_{p}\dot s_{q}=u_{\mathcal{A}}\delta_{p,q}$. It follows that $V_{n}\cong\mathbb{R}\oplus\mathbb{R}^{n}$ as a real inner product space, 
and also as a euclidean Jordan algebra with 
\begin{equation}
\big(\lambda_{0}\oplus\vec{\lambda}\big)\dot\big(\mu_{0}\oplus\vec{\mu}\big)=\lambda_{0}\mu_{0}+\langle \vec{\lambda},\vec{\mu}\rangle\oplus \lambda_{0}\vec{\mu}+\mu_{0}\vec{\lambda}\text{.}
\end{equation}
Concretely, the usual complex Pauli matrices can be used to define the spin factors. We recall the usual complex Pauli matrices as follows
\begin{equation}
u_{\mathbb{C}_{2}}=\sigma_{0}=\begin{pmatrix}1&\hspace{0.2cm}0\\0&\hspace{0.2cm}1\end{pmatrix}\qquad
\sigma_{1}=\begin{pmatrix}1&\hspace{0.2cm}0\\0&-1\end{pmatrix}\qquad \sigma_{2}=\begin{pmatrix}0&\hspace{0.2cm}1\\1&\hspace{0.2cm}0\end{pmatrix}\qquad\sigma_{3}=\begin{pmatrix}0&-i\\i&\hspace{0.2cm}0\end{pmatrix}\text{.}
\end{equation}
Following \cite{HO-Stormer}, we define for each finite $\mathbb{N}\ni n>1$ and $\forall 1\leq p\leq n$, with $\left\lfloor{\cdot}\right\rfloor$ and $\left\lceil{\cdot}\right\rceil$ the usual floor and ceiling functions
\begin{eqnarray}
t_{p}=\begin{cases}
\begin{cases}
\sigma_{3}^{\otimes^{\left\lceil{\frac{p}{2}}\right\rceil-1}}\otimes\sigma_{1}\otimes\sigma_{0}^{\otimes^{\left\lfloor{\frac{n}{2}}\right\rfloor-\left\lceil{\frac{p}{2}}\right\rceil}} & p \text{ odd} \\ 
\sigma_{3}^{\otimes^{\left\lceil{\frac{p}{2}}\right\rceil-1}}\otimes\sigma_{2}\otimes\sigma_{0}^{\otimes^{\left\lfloor{\frac{n}{2}}\right\rfloor-\left\lceil{\frac{p}{2}}\right\rceil}} & p \text{ even}
\end{cases} & n \text{ even} \\[1cm]
\begin{cases}
\sigma_{3}^{\otimes^{\left\lceil{\frac{p}{2}}\right\rceil-1}}\otimes\sigma_{1}\otimes\sigma_{0}^{\otimes^{\left\lceil{\frac{n}{2}}\right\rceil-\left\lceil{\frac{p}{2}}\right\rceil}} & p \text{ odd} \\ 
\sigma_{3}^{\otimes^{\left\lceil{\frac{p}{2}}\right\rceil-1}}\otimes\sigma_{2}\otimes\sigma_{0}^{\otimes^{\left\lceil{\frac{n}{2}}\right\rceil-\left\lceil{\frac{p}{2}}\right\rceil}} & p \text{ even}
\end{cases} & n \text{ odd}
\end{cases}
\end{eqnarray}
where our notation is such that $x^{\otimes^{0}}=1\in\mathbb{R}$, $x\otimes 1=x=x^{\otimes^{1}}$, $x^{\otimes^{2}}=x\otimes x$, and so on. One can easily check that for each $n>1$, $\{t_{1},\dots,t_{n}\}$ generates a spin factor of dimension $1+n$ with $t_{p}\dot t_{q}=(t_{p}t_{q}+t_{q}t_{p})/2$. It turns out \cite{HO}, with finite $k\in\mathbb{N}$, that the maps
\begin{eqnarray}
\psi_{2k}:V_{2k}\longrightarrow M_{2^{k}}(\mathbb{C})_{\text{sa}}::s_{p}\longmapsto t_{p} \\
\psi_{2k+1}:V_{2k+1}\longrightarrow M_{2^{k}}(\mathbb{C})_{\text{sa}}\oplus M_{2^{k}}(\mathbb{C})_{\text{sa}}::s_{p}\longmapsto t_{p} 
\end{eqnarray}
are precisely the canonical injections of $V_{n}$ into their universal C$^{*}$\!-algebras (\textit{i.e.} their \textit{universal} representations). Our \textit{standard} representation $\pi_{n}$ of $V_{n}$ differs from the universal representation when $n$ is odd. Specifically, when $n$ is even we define $v_{p}=t_{p}$, and when $n$ is odd we define $\forall 1\leq p <n$
\begin{eqnarray}
v_{p}&=&\begin{cases}
\sigma_{3}^{\otimes^{\left\lceil{\frac{p}{2}}\right\rceil-1}}\otimes\sigma_{1}\otimes\sigma_{0}^{\otimes^{\left\lfloor{\frac{n}{2}}\right\rfloor-\left\lceil{\frac{p}{2}}\right\rceil}} &  p  \text{ odd} \\ 
\sigma_{3}^{\otimes^{\left\lceil{\frac{p}{2}}\right\rceil-1}}\otimes\sigma_{2}\otimes\sigma_{0}^{\otimes^{\left\lfloor{\frac{n}{2}}\right\rfloor-\left\lceil{\frac{p}{2}}\right\rceil}} & p  \text{ even}
\end{cases}\\
v_{n}&=&\sigma_{3}^{\otimes^{\left\lfloor{\frac{n}{2}}\right\rfloor}}
\end{eqnarray}
and we embed $V_{n}$ into its \textit{standard} C$^{*}$\!-algebra via the following Jordan monomorphism with $v_{p}\dot v_{q}=(v_{p}v_{q}+v_{q}v_{p})/2$
\begin{equation}
\pi_{n}:V_{n}\longrightarrow M_{2}(\mathbb{C})_{\text{sa}}^{\otimes^{\left\lfloor{\frac{n}{2}}\right\rfloor}}::s_{p}\longmapsto v_{p}\text{.}
\end{equation}
For example, the universal and standard representations for the qubit --- \textit{i.e}.\ $V_{3}$ --- differ as follows
\begin{eqnarray}
t_{1}=\sigma_{1}\otimes \sigma_{0}\qquad v_{1}=\sigma_{1}\\
t_{2}=\sigma_{2}\otimes \sigma_{0}\qquad v_{2}=\sigma_{2}\\
t_{3}=\sigma_{3}\otimes \sigma_{1}\qquad v_{3}=\sigma_{3}
\end{eqnarray}
hence the name \textit{standard} representation. Incidentally, note that $V_{3}\cong {C}_{2}$ as a euclidean Jordan algebra. Furthermore, $V_{2}\cong {R}_{2}$, $V_{5}\cong {Q}_{2}$, and $V_{9}\cong M_{2}(\mathbb{O})_{\text{sa}}$; the ambient spaces for the real, quaternionic, and octonionic quantum bits.

{
\section{Weak Composites} 
\label{appendix: weak composites}

By a {\em weak composite} of EJAs $A$ and $B$, we mean an EJA $AB$ and a bilinear mapping 
 $\pi : A \otimes B \rightarrow AB$  satisfying parts (a) and (b), but not necessarily 
part (c), of Definition \ref{def: composites of EJAs}. {\redd That is, we suspend the requirement 
that $AB$ be generated, as a Jordan algebra, by $\pi(A \otimes B)$.} We are going to show that 
the Jordan subalgebra of $AB$ generated 
by the image of $A \otimes B$ in $AB$ also satisfies these conditions, and hence, is a composite in the strict 
sense.  

Observe, first, that we made no use of Condition (c) in proving any of the results leading up to, and including, 
Corollary \ref{cor: cor to main equation}. So all of these are also satisfied by weak composites. 

\begin{proposition} Let $AB$ be a weak composite  of EJAs $A$ and $B$. Then the Jordan subalgebra of $AB$ generated by 
$A \otimes B$ is also a composite. 
\end{proposition}

\noindent{\em Proof:} {\redd Identifying $A \otimes B$ with its image, $\pi(A \otimes B)$, in $AB$,} 
let $A \odot B = J(A \otimes B)$ denote the Jordan subalgebra of $AB$ generated by $A \otimes B$. That 
the co-restriction $\pi : A \otimes B \rightarrow A \odot B$ satisfies the conditions for a composite 
(Definition \ref{def: composites}) is straightforward. 
  It is also straightforward that $A \odot B$ will 
satisfy the conditions for a composite of Jordan models (Definition \ref{def: composites of EJAs}), 
 provided that it satisfies the conditions for a dynamical composite (Definition \ref{def: dynamical composites}).
To see that it does,     
let $\phi \in G(A)$ and $\psi \in G(B)$. We need to show that $\phi \otimes \psi$ preserves $A \odot B$. 
We can expand $\phi$ as $U_{a} \circ g$ and $\psi$ as $U_b \circ h$ for some interior elements $a \in A_+$, $b \in B_+$, and Jordan homomorphisms $g \in G(A)$ and  
$h \in G(B)$ (\cite{FK}, III.5.1). Now 
\[\phi \otimes \psi = (U_a \circ g) \otimes (U_b \circ h) = (U_{a} \otimes U_{b}) \circ (g \otimes h).\]
Since $(g \otimes h)(u_{AB}) = g(u_A) \otimes h(u_B) = u_{A} \otimes u_{B} = u_{AB}$, 
$g \otimes h$ is a symmetry, hence, a Jordan automorphism of $AB$ (\cite{AS}, Theorem 2.8). As it maps $A \otimes B$ to itself, 
it also preserves $J(A \otimes B)$, i.e., $A \odot B$. 

It remains to show that $U_{a} \otimes U_{b}$ also preserves $A \odot B$. 
Begin with the observation that if $A$ is an EJA and $X \subseteq A$, then for all $x \in X$, 
$U_{x}(J(X)) \subseteq J(X)$. This is evident from the fact that, as $x \in J(X)$ and the latter is a Jordan 
subalgebra of $A$, $U_{x}(y) = 2x\dot (x\dot y) - (x^2) \dot y$ for all $y \in J(X)$. 

Now consider that the proof of the identities $(a \otimes u)\dot (x \otimes y) = ( a\dot x ) \otimes y$ and 
$(u \otimes b)\dot (x \otimes y) = x \otimes ( b \dot y )$ relies only on (a) and (b), and so, holds in our context. 
By Corollary \ref{cor: cor to main equation},  $U_{a \otimes u} = U_{a} \otimes \id$ and $U_{u \otimes b} = \id_{A} \otimes U_{b}$. 
If $a, b$ are interior elements, then $U_a \in G(A)$ and $U_b \in G(B)$, so by (b), we have 
\[U_{a} \otimes U_{b} = (U_{a} \otimes \id_{B}) \circ (\id_{A} \otimes U_{b}) = U_{a \otimes u_B} \circ U_{u_A \otimes b}.\]
Since $U_{a \otimes u_B}$ and $U_{u_A \otimes b}$ preserve $J(A \otimes B) = A \odot B$, by the remark above, 
so does $U_{a} \otimes U_{b}$, and the proof is finished. $\Box$

\section{Extending order automorphisms} 
\label{appendix: extending}

{\awedit

In this appendix, for the reader's convenience, we collect facts concerning extensions of derivations used in the body of this paper. 
Throughout, $A$ is an EJC algebra, that is, a Jordan subalgebra of $\M_{\sa}$, where $\M$ is a finite-dimensional complex matrix algebra. We begin by recalling some facts about order-automorphisms and Jordan derivations 
(\cite{AS}, Chapter 6).

If $\{\phi(t)\}_{t \in \R}$ is a one-parameter
group of order-automorphisms of $A$, then for every $a \in A$, 
$\phi(t)(a) = e^{t\delta}a$ where $\delta = \phi'(0)$ is a linear operator on $A$. The linear operators $\delta$ arising in this way are called {\em order derivations} of $A$.  Order-derivations come in two basic types:
those having the form $\delta = L_a$ for some $a \in A$ 
and those having the property $\delta(u) =
0$. The latter are exactly the Jordan derivations of $A$, that is, the linear maps $A \rightarrow A$ satisfying 
the Leibniz law $\delta(a \dot b) = \delta(a) \dot b + a \delta(b)$ for all $a, b \in A$. Order derivations 
of the form $L_a$ are said to be {\em self-adjoint}; those that are Jordan derivations are said to be {\em skew}. 
The former are self-adjoint with respect to the canonical inner product on $A$,  
by the definition of a euclidean Jordan algebra, while the latter are skew-adjoint.   
Every order-derivation has the form $\delta = L_{a} + \delta'$ where
$\delta'$ is skew (\cite{AS}, Proposition 1.60).  It follows that $\phi := e^{t\delta}$ fixes the Jordan unit $u_A$ 
iff $\delta$ is skew.

A mapping $\delta : \M \rightarrow \M$ is said to be an order-derivation of $\M$ iff it preserves $\M_{\sa}$ and its restriction to $\M_{\sa}$ is an order derivation in the sense discussed above. 
If $a \in \M$, define $\delta_{a} : \M \rightarrow \M$ by 
\[\delta_{a}(x) = \frac{1}{2}(a x + x a^{\ast}).\]
Then $\delta_{a}$ is an order derivation in this broader sense. Moreover, every order-derivation of $\M$ has this 
form for some $a \in \M$ (\cite{AS}, Appendix 183.) 
Note that $\delta_{a}$ is self-adjoint precisely when $a$ is self-adjoint and 
skew precisely when $a$ is skew-adjoint.  In the latter case, a direct computation shows that 
$\delta_{a}$ is actually a {\em $\ast$-derivation} 
of $\M$, that is, $\delta_{a}(xy) = \delta_{a}(x)y + x \delta_{a}(y)$ and $\delta_{a}(x^{\ast}) = \delta_{a}(x)^{\ast}$. 

The following essentially restates Lemma \ref{lemma: extension of derivations}:

\begin{lemma}\label{lemma: extension of derivations-appendix}  Any Jordan derivation of $A$ extends to a $\ast$-derivation of $\M$. Hence, any order-derivation of $A$ extends to an order-derivation of $\M$.
\end{lemma}

\begin{proof}
The set of derivations of a finite-dimensional real or complex algebra is a Lie algebra, and in the case
of a Jordan algebra $A$, it is the linear span of the elements $[L_a, L_b]$, for $a,b \in A$, in fact every
derivation is a sum of such elements.\footnote{See e.g. \cite{FK}, Proposition II.4.1 for the fact that these are 
derivations, called inner derivations.  In \cite{Upmeier}, for example, it is said to be well-known that all 
derivations of a finite-dimensional JB algebra (i.e. an EJA) are inner.}  Since $L_a$ and $L_b$ are linear operators
$A \rightarrow A$, $[L_a, L_b]$ {\awedit belongs to} the real associative algebra of such operators.  Since $A$ is a subalgebra of
$(\M_A)_{sa}$, $L_a$ and $L_b$ extend to linear operators $(\M_A)_{sa} \rightarrow (\M_A)_{sa}$, namely  
the Jordan multiplication operators corresponding to $a$ and $b$ viewed as elements of $(\M_A)_{sa}$.   
But as noted above, every Jordan derivation of $\M_A$ is also a $\ast$-derivation for the associative product, 
establishing the first claim. A $\ast$-derivation on $\M$ preserves $\M_{\sa}$, and is also a Jordanderivation on the latter, so this also establishes that all skew order derivations on $A$ extend to 
order-derivations of $\M$.   A self-adjoint order derivation of $A$ is simply a map $L_a : A \rightarrow A$ 
with $a \in A$, which extends to $\delta_{a} : \M \rightarrow \M$ since $a$ is self-adjoint in $\M$. Thus, 
all order-derivations on $A$ extend to order-derivations on $\M$. $\Box$
\end{proof}

}

\begin{lemma} \label{lemma: derivations extend}
Any one-parameter group of order automorphisms of $A$ extends to a
one-parameter group of order-automorphisms of $\M$.  
\end{lemma}

\noindent{\em Proof:} If $\{\phi(t)\}_{t \in \R}$ is a one-parameter
group of order-automorphisms of $A$, then $\phi(t)(a) = e^{t\delta}a$ where
$\delta = \phi'(0)$ is an order-derivation of $A$.  Thus, we have  
$\delta = L_{a} + \delta'$ where $a \in A$ and 
$\delta'$ is skew (\cite{AS}, Proposition 1.60). $L_a$ obviously extends from $A$ to $\M$, simply because $a \in \M$
and the Jordan product on $A$ is the restriction of that on $\M$. By 
 Lemma \ref{lemma: extension of derivations}, 
\ffootnote{\red But it looks,on reading his Example 2.3, as though
  finite-dimensional spin factors also have the extension
  property. Can we find a better (f.d.) reference?} $\delta'$
also extends to a Jordan derivation $\delta''$ on $\M$. Thus, we have
an extension $\hat{\delta} = L_a + \delta''$ on $\M$. In particular,
$\delta''(A) \subseteq A$.
We now have an order-automorphism $\hat{\phi}(t) = e^{t \hat{\delta}}$ of $\M_+$. Note that this preserves $A$, since 
\[\hat{\phi}(t)x = \sum_{k=1}^{\infty} \frac{t^k}{k!} \hat{\delta}^{k} x\]
and $\hat{\delta}x = (L_a + \delta'')x = a\dot x + \delta''(x)$, 
which belongs to $A$ if $x$ does. $\Box$
\begin{corollary}\label{cor: automorphisms extend} 
Every element of $G(A)$ extends to an element of
$G((\M_{A})_{\sa})$. \end{corollary}

\begin{lemma}\label{lemma: fixing} Let {\awedit $A$ and $B$ be EJC-algebras. If $\delta$ is} any $\ast$-derivation of $\M_A$ fixing $A$, then $\delta \otimes \1$ is a $\ast$-derivation of $\M_A \otimes \M_B$ fixing $A \odot B$. \end{lemma} 

\noindent{\em Proof:} Let $\M$ and $\N$ be $\ast$-algebras, and let $a,
b \in \M$ and $x,y \in \N$. Then
\[(a \otimes x) {\bblue \dot} (b \otimes y) = \frac{1}{2} ( ab \otimes xy + ba \otimes yx).\]
If $\delta$ is a $\ast$-derivation of $\M$, then it is straighforward to check that $\delta \otimes \1$ is a $\ast$-derivation of $\M \otimes \N$, 
and that for all $a, b \in \M$ and $x, y \in \N$,  
\[(\delta \otimes \1)((a \otimes x) {\bblue \dot} (b \otimes y)) = (a \otimes x) {\bblue \dot} (\delta(b) \otimes y) + (\delta(a) \otimes x) {\bblue \dot} (b \otimes y).\]
In particular, if $A \subseteq \M$ and $\delta(A) \subseteq A$, it follows that $(\delta \otimes \1)(A \otimes B) \subseteq (A \otimes B) \dot (A \otimes B)$ 
for any $B \subseteq \N$. It follows easily that, 
{\redd where $A$ and $B$ are EJCs and $\M = \M_A$ and $\N = \M_B$,} $\delta \otimes \1$ preserves $A \odot B$.
\footnote{The details: let $\delta$ be a $\ast$-derivation on a $\ast$-algebra $\M$, and let $X \subseteq M_{\sa}$ with $\delta(X) \subseteq X$. Let $Y = \{ a \in J(X) | \delta(a) \in J(X)\}$. Evidently $X \subseteq Y$. Now if $a, b \in Y$ and $t \in \R$, then $\delta(ta + b) = t\delta(a) + \delta(b) \in J(X)$, so $J(X)$ is a subspace of $M$. If $a, b \in Y$ then $\delta(a \dot b) = a \dot \delta(b) + \delta(a) \dot b \in J(X)$. Thus, $Y$ is a Jordan subalgebra of $\M_{\sa}$, containing 
$X$, and contained in $J(X)$. Ergo, $Y = J(X)$, and $\delta(J(X)) \subseteq J(X)$.} 

\begin{proposition} If $\phi$ and $\psi$ are order-automorphisms  in {\awedit $G(A)$ and $G(B)$}, 
respectively, then $\phi \otimes \psi {\awedit : A \otimes B \rightarrow A \otimes B}$ extends to an order-automorphism {\awedit in $G(A \odot B)$}. 
\end{proposition}

\noindent{\em Proof:} By Corollary \ref{cor: automorphisms extend}, we
can assume that $\phi$ is an order-automorphism of $M_{\sa}$ fixing
$A$.  Since $\phi \in G(A)$, it occurs as part of a one-parameter group 
$\phi(t) = e^{t\delta}$ of order-automorphisms, say as $\phi = \phi(1) = e^{\delta}$, 
where $\delta$ is an order-derivation of $A$.  By Lemma \ref{lemma: extension of derivations-appendix}, $\delta$ extends to an order derivation 
of $\M$ fixing $A$.
It follows that
\[\phi \otimes \1 = e^{t\delta} \otimes \1  = \sum_{n=0}^{\infty} \frac{t^n}{n!} \delta^n \otimes \1 = \sum_{n=0}^{\infty} \frac{t^{n}}{n!} (\delta \otimes \1)^n = e^{t(\delta \otimes \1)}.\]
By Lemma \ref{lemma: fixing}, $\delta \otimes \1$ fixes $A \odot B$; thus, so does the
series at right, whence, so does $\phi \otimes \1$.  It follows that
if $\phi$ is an order-automorphism of $(M_A)_{\sa}$ fixing $A$, {\awedit so} $\phi
\otimes \1$ is an order-automorphism of $A \otimes B$ fixing $J(A
\otimes B) = A \odot B$. Hence, if $\phi$ and $\psi$ are
order-automorphisms of $M_A$ and $M_B$, respectively fixing $A$ and
$B$, then $\phi \otimes \psi = (\phi \otimes \1) \circ (\1 \otimes
\psi)$ fixes $A \odot B$. $\Box$

\tempout
{\awedit 
\setcounter{section}{8}
\section{The inner product on a composite}
\label{appendix: Inner}

This appendix collects two currently homeless facts about inner products that we may want  (or need) in later work. 

\begin{proposition}\label{prop: inner product on a sub-EJC}
Let $(A,\M_A)$ be a reversible EJC (i.e., reversibly embedded). Then 
the canonical inner product on $A$ is equal, up scalar multiples on each irreducible factor (direct summand) of $A$, 
to the {\em restriction of} the inner product on $(M_{A})_{\sa}$.
\end{proposition}

\noindent{\em Proof:} Suppose first that $A$ is simple.  Since there is a symmetry {\awcomment [which must be unitary?]} exchanging any two 
primitive idempotents, all primitive idempotents in $A_i$ have the same norm. 
Accordingly, we can normalize 
the inner product on $A$ so that primitive idempotents have norm $1$. 
By the spectral decomposition, if $a \in A$ then $a = \sum_{i} \lambda_i e_i$, where the $e_i$ are pairwise orthogonal primitive idempotents in $A$; hence $\langle a, a \rangle = \sum_i \lambda_i$. Since the idempotents $e_i$ remain idempotent and orthogonal (though perhaps not primitive) in $\M_{\sa}$, we have 
$\Tr(a^2) = \sum_i \lambda_i r_i$ where $r_i$ is the rank of $x_i$ in $\M$ (a positive integer).  Proposition 5.3
(Howard's FD improvement of Upmeier's result) tells us that 
if $A_i$ is embedded in $\M$, then automorphisms of $A_i$ extend to automorphisms of $\M$. Hence, 
every primitive idempotent $e \in A_i$, regarded as an element of $\M$, has the same rank, call it $r$, and we 
have $\Tr(e) = r$ for every such idempotent.  If $e, f \in A_i$ are two primitive idempotents,  
let $a = e + f$. Then we have $\Tr(a^2) = r \langle a, a \rangle$, i.e., 
\[2r + 2 \Tr(e,f) = r(2 + \langle e, f \rangle),\]
whence, $\Tr(e,f) = r \langle e, f \rangle$.  It now follows from spectral theorem plus the bilinarity of the inner products that $Tr(ab) = r \langle a, b \rangle$ for all $a, b \in A$. 

Now consider the case where $A = \bigoplus_i A_i$ is a direct sum of simple EJCs. Notice that since the central projections 
associated with each summand are Jordan orthogonal, they are also orthogonal with respect to the 
inner product on $A$; hence, the inner product on $A$ is the sum of the inner products on the summands. 
Moreover, since Jordan orthogonal elements of $A$ remain Jordan orthogonal as embedded in $\M_A$, 
elements of distinct summands also orthogonal in $\M_A$ with respect to the trace inner product. 
Applying the previous argument separately to each summand (separately normalizing the inner product on 
each $A_i$ so as to give primitive idempotents norm $1$) completes the proof. $\Box$

\begin{proposition}  Let $AB$ be a composite of simple EJAs $A$ and $B$. Then 
for all $a, x \in A$ and all $b, y \in B$, 
\[\langle a \odot x , b \odot y \rangle = c \langle a , x \rangle \langle b , y \rangle.\]
where $c = \|a \otimes b\|^2/\|a\|^2 \|b\|^2$. Moreover, if inner products on $A$, $B$ and $AB$ 
are normalized so that order units are unit vectors, then $c = 1$. 
\end{proposition}

\noindent{\em Proof:} We begin with the case in which $a$ and $b$ are minimal projections. Since $a \odot b$ is then 
also a projection, the discussion in Section 3.2 tells us that 
\[\hat{a \odot b} = \|a \odot b \|^{-2}(a \odot b)\]
defines a state. Now define a positive bilinear form $\omega : A \times B \rightarrow \R$ 
by setting 
\[\omega(x,y) = \langle \hat{a \odot b} , x \odot y \rangle = \langle a \odot b , x \odot y \rangle\]
for all $ x \in A$, $y \in B$. Note that $\omega$ is normalized, i.e., a state in the maximal tensor product 
$A \otimes B$. 
Now evaluate the first marginal of $\omega$ at $a$: 
\begin{eqnarray*}
\omega_{1}(a) = \omega(a \otimes u_B) & = & \langle \hat{a \odot b} , a \otimes u_B\rangle \\
& = & \langle \hat{a \odot b} , a \odot b + a \odot b' \rangle\\
& = & (\langle \hat{a \odot b} , a \odot b \rangle + \langle \hat{a \odot b} , a \odot b' \rangle. 
\end{eqnarray*}
Since $\langle \hat{a \odot b} , a \odot b \rangle = 1$, the second summand at the end is $0$, and we have 
$\omega_{1}(a) = 1$. Since $a$ is minimal, there is only one such state: $\omega_1(a) = \langle \hat{a} |$, 
i.e., the functional $b \mapsto \langle a, b \rangle$. Moreover, 
this is a {\em pure} state. The same argument shows that $\omega_2(b) = 1$, so that $\omega_2 = \langle \hat{b} |$. 
As is well known, if a non-signaling state has pure marginals, then it's the product of these marginals \cite{BBLW}.
 Thus, $\omega = \langle \hat{a} | \otimes \langle \hat{b} |$. This gives us 
 \[\langle a \odot b , x \odot y \rangle = c \langle a , x \rangle \langle b , y \rangle.\]
where $c := \frac{\|a \odot b \|^2}{\|a\|^2 \|b\|^2}$. 

Now suppose that $\|u_A\| = \|u_B \| = \|u_{AB}\| = 1$. We want to show that $c = 1$.  As a first step, note that $c$ is independent of the choice of minimal projections $a$ and $b$ (Argue by symmetry: since $A$ and $B$ are simple, there are symmetries $\phi$ and $\psi$ taking any given 
pair $a,b$ to any other such pair $a', b'$; hence, there's a symmetry taking $a \odot b$ to $a' \odot b'$ 
 by Lemma \ref{lemma: exchange}.) 
 Extend $a$ and $b$ to 
orthogonal decompositions
 of $u_A$ and $u_B$ as sums of projections:  $u_A = \sum_{a_i} a_i$ with $a = a_1$, and $u_B = \sum_j b_j$ with $b = b_1$ Then we have 
 $u_{AB} = \sum_{i,j} a_i \odot b_j$,  so 
 \[\sum_{i,j, k, l} \langle a_i \odot b_j , a_k \odot b_l \rangle = \langle u_{AB} , u_{AB} \rangle = 1\]
 but also, noting that all $a_i$ and $b_j$ here are minimal projections, so that the constant $c$ above is 
 the same for all choices of $a_i \odot b_j$, 
 \[\sum_{i,j,k,l} \langle a_i \odot b_j , a_i \odot b_l \rangle = \sum_{i,j,k,l} c \langle a_i \odot b_j , a_k \odot b_l \rangle = c.\] 
 Hence, $c = 1$. 

Now suppose $a$ and $b$ are arbitrary elements of $A$ and $B$, respectively. Spectrally decomposing $a$ and $b$ as 
\[a \ = \ \sum_i t_i a_i \ \mbox{and} \  b \ = \ \sum_j s_j b_j\]
where $a_i$ and $b_j$ are pairwise orthogonal families of minimal projections, we have 
\[\langle a \odot b|  = \langle \sum_{i, j} t_i s_j a_i \odot b_j |  = \sum_{i,j} t_i s_j \langle a_i \odot b_j |.\]
Hence, for all $x, y$,
\begin{eqnarray*} 
\langle a \odot b , x \odot y \rangle & = & \sum_{i,j} t_i s_j \langle a_i \odot b_j , x \odot y \rangle \\
& = & \sum_{i,j} t_i s_j \langle a_i | x \rangle \langle b_j , y \rangle \\
& = & \langle a , x \rangle \langle b , y \rangle
\end{eqnarray*}
as advertised. $\Box$\\ %
}
\end{document}